%% file: gravSK-spin0-arXiv-v3.tex
\documentclass[11pt]{article}
\pdfoutput=1
\usepackage{jheppub}

\usepackage[svgnames,table]{xcolor}
\usepackage{graphicx}
\usepackage{amsfonts,amsmath,amssymb,amsthm}
\usepackage{mathrsfs,bm,bbm,esint}
\usepackage[mathscr]{eucal}
\usepackage{physics}
\usepackage{slashed,soul}
\usepackage{rotating,pdflscape}
\usepackage{enumerate}
\usepackage{multirow,array}
\usepackage[numbers,sort&compress]{natbib}
\usepackage{tikz}
\usetikzlibrary{cd}
\usetikzlibrary{decorations.pathmorphing}
\usetikzlibrary{decorations.pathreplacing}
\usetikzlibrary{decorations.markings}
\tikzset{snake it/.style={decorate, decoration=snake}}
\tikzset{->-/.style={decoration={
  markings,
  mark=at position .5 with {\arrow{>}}},postaction={decorate}}}
\usepackage{subcaption}
\usepackage{float}
\usepackage{afterpage}
\usepackage{cleveref}
\crefname{section}{\S\!\!}{\S\S\!\!}
\Crefname{section}{\S}{\S\S}
\crefname{appendix}{Appendix}{Appendices\!}
\crefname{figure}{Fig.\!}{Figs.\!}

\input{gravsk-macros}

\title{Open quantum systems and Schwinger-Keldysh holograms}
 
 \author[a]{Chandan Jana}
 \author[a]{, R. Loganayagam}
\author[b]{, Mukund Rangamani}
\affiliation[a]{
International Centre for Theoretical Sciences (ICTS-TIFR), \\ 
Tata Institute of Fundamental Research, Shivakote, Hesaraghatta, Bangalore 560089, India.}
\affiliation[b]{Center for Quantum Mathematics and Physics (QMAP)\\
Department of Physics, University of California, Davis, CA 95616 USA}

\emailAdd{chandan.jana@icts.res.in} 
\emailAdd{nayagam@icts.res.in}
\emailAdd{mukund@physics.ucdavis.edu}

\abstract{
We initiate the study of open quantum field theories using holographic methods. Specifically, we consider a quantum field theory (the system) coupled to a holographic field theory at finite temperature  (the environment). We investigate the effects of integrating out the holographic environment with an aim of obtaining an effective dynamics for the resulting open quantum field theory. The influence functionals which enter this open effective action are determined by the real-time (Schwinger-Keldysh) correlation functions of the holographic thermal environment. To evaluate the latter, we exploit recent developments, wherein the semiclassical gravitational Schwinger-Keldysh saddle geometries were identified as complexified black hole spacetimes. We compute real-time correlation functions using holographic methods in these geometries, and argue that they lead to a sensible open effective quantum dynamics for the system in question, a question that hitherto had been left unanswered. In addition to shedding light on open quantum systems coupled to strongly correlated thermal environments, our results also provide a principled computation of  Schwinger-Keldysh observables in gravity and holography. In particular, these influence functionals we compute capture both the dissipative physics of black hole quasinormal modes, as well as that of the fluctuations encoded in outgoing Hawking quanta, and interactions between them. We obtain  results for these observables at leading order in a low frequency and momentum expansion in general dimensions, in addition to determining explicit results for two dimensional holographic CFT environments. 
}

\begin{document}
\maketitle


\section{Introduction}
\label{sec:intro}

The study of quantum fields in curved spacetimes, especially in geometries with horizons, such as black holes or cosmological spacetimes, has been an immensely valuable window into  the semiclassical aspects of gravitational dynamics. Such investigations have been instrumental not only for understanding the effects of vacuum polarization, particle production, etc., but  also  have played an important role in the holographic AdS/CFT correspondence. In the latter context the semiclassical gravitational computations give us insight into observables e.g., correlation functions, von Neumann entropy, etc., of the dual strongly coupled QFT.

One natural set of observables in a quantum system are thermal correlation functions, capturing real-time response to perturbations and the attendant thermal fluctuations. These are especially interesting in strongly correlated systems where analytic techniques to compute  response functions are limited. Here, holography provides a valuable avenue: one can extract dynamical response of strongly correlated thermal plasma by performing classical calculations in the dual black hole geometry. This approach has paid rich dividends over the past two decades: ranging from understanding thermalization \cite{Horowitz:1999jd}, linear response and hydrodynamics \cite{Policastro:2002se}, to real-time transport computations \cite{Herzog:2007ij}. Much of this success owes to the fact that black holes in asymptotically AdS spacetimes are dual to thermal QFTs. 

We wish to argue here that black holes in fact provide a playground for the exploration of a much richer set of dynamics, viz., that of an open quantum field theory, where the degrees of freedom of the quantum system are non-trivially entangled with  some external environment (or bath) degrees of freedom. The set-up we have in mind is the following: consider a QFT say with one bosonic degree of freedom $\Psi(t,{\bf x})$ (for simplicity) which is our system of interest. We will model the environment  by another field theory, now with many degrees of freedom $X_i(t,{\bf x})$. The unitary microscopic theory is of the form: 
\begin{equation}\label{eq:sysenv}
	S_\text{s}[\Psi] + S_\text{e}[X_i] + S_\text{s-e}[\Psi, X_i]
\end{equation}	
The combined system and environment is prepared in some initial state, which we may even take to be factorized between the respective degrees of freedom. Integrating out the environment degrees of freedom $X_i$ we end up with a non-unitary evolution of our system. 

The basic paradigm for such an effective theory was described by Feynman and Vernon \cite{Feynman:1963fq} who noticed that the natural way to describe the system is in terms of a doubled set of degrees of freedom for the system, together with a non-trivial interaction between them, which they dubbed \emph{influence functionals}. Heuristically,
\begin{equation}\label{eq:feynver}
	\int [D\Psi] \int [DX_i] e^{i\left( S_\text{s}[\Psi] + S_\text{e}[X_i] + S_\text{s-e}[\Psi, X_i]\right)} 
	\mapsto \int [D\Psi_\skL]  [D\Psi_\skR] e^{i\left( S_\text{s}[\Psi_\skR] - S_\text{s}[\Psi_\skL] + S_\text{IF}[\Psi_\skR, \Psi_\skL]\right)} 
\end{equation}	
where $S_\text{IF}[\Psi_\skR, \Psi_\skL] $ is the aforementioned influence functional, induced onto the system owing to the coupling with the environment. This paradigm is well understood and tested for Gaussian dynamics in quantum mechanics, as exemplified by the Caldeira-Leggett description of quantum Brownian motion \cite{Caldeira:1982iu}. For an overview of developments in the study of open quantum systems see  \cite{Breuer:2002pc,Schlosshauer:2003zy,Sieberer:2015svu}.

One major question in this scenario is to find a set of sufficient conditions for a local effective field theory to emerge (for our system degree of freedom $\Psi$). This question is quite difficult to address within weakly coupled systems for the following reason: often a local description emerges at time scales longer than the `environmental memory' time scale $\tau_m$. Here, $\tau_m$ should be thought of as the time that the environment fields take to forget the information of the initial state; it is inversely proportional to the interactions within  the environment. Consequently, one has to often wait for a non-perturbatively long time for a local description to be valid. This necessarily means that derivation  of open quantum field theories is inevitably a non-perturbative question. This explains to some extent why to date there are no simple microscopic models from which a local non-unitary open quantum effective  field theory has been  systematically derived. In particular, as far as the authors are aware, currently there are no microscopic QFTs from which a local  open EFT with interactions can be derived.\footnote{ See \cite{Lombardo:1995fg} for early work on the subject and \cite{Agon:2014uxa} for recent attempts in this direction, in addition to \cite{Avinash:2017asn,Agon:2017oia,Gao:2018bxz,Avinash:2019qga} for technical issues regarding renormalization.} Our goal in this work is to address this lacuna by using holography.

The set-up we have in mind is semi-holographic, cf., \cite{Faulkner:2010tq}. Say we wish to understand the dynamics of a single bosonic degree of freedom which we continue to call $\Psi(x)$ in $d$ spa{}cetime dimensions. We imagine coupling this to a strongly coupled thermal environment comprising of some intrinsic microscopic degrees of freedom. For concreteness, one can imagine the environment to be the thermal large $N$, strongly coupled, $\mathcal{N}=4$ Super Yang-Mills (SYM) theory (gauge group $SU(N)$) in $d=4$, or a large $c$ thermal CFT in $d=2$.  Following common practice, we will often refer to the system and the environment as  the probe and  the bath, respectively. 

The coupling of the probe/system degree of freedom $\Psi$ to the thermal bath is via a local coupling $\int d^dx \, \Psi(x) \, \mathcal{O}(x)$where $\mathcal{O}(x)$ is a simple operator in the bath/environment theory. In the aforementioned examples $\mathcal{O}$ could  be  a low lying single trace conformal primary operator.  Crucially, the thermal bath theory is assumed to have a holographic dual. This will enable us  model the environment by a dual black hole geometry. The influence phases of our system are then encoded in the  real-time or  Schwinger-Keldysh (SK) correlation functions of the environmental degrees of freedom. The computation of the latter, thanks to the holographic map, is one which can be carried out using classical fields propagating in a classical black hole background. From the holographic standpoint, the problem therefore boils down to developing a formalism for computing real-time observables in black hole backgrounds (or more generally in spacetimes with horizons).

Before commenting on the real-time computation let us first recall a well-known, but remarkable, fact of the Euclidean gravitational path integral. Thermal boundary conditions require that the Euclidean time ($\tE$) direction be compact with period given by the inverse temperature asymptotically. For the gravitational path integral these boundary conditions pick out the Wick rotated black hole solution as the Gibbons-Hawking saddle point solution \cite{Gibbons:1976ue}.  This is unlike any non-gravitational system, where the Euclidean thermal circle is more of a computational aid, the background geometry being non-dynamical. Said differently, gravitational dynamics interplays non-trivially with thermal boundary conditions.

 Given this solution, one way to pass to a real-time description is to slice open the Euclidean solution at some instant of real-time, say at $t=0$, which exposes quite naturally two copies of the asymptotic region at ($\tE =0$ and $\tE = \frac{\beta}{2}$, respectively). This initial data can then be evolved in real Lorentzian time to give the (future half) of the eternal black hole solution.  This is the gravitational preparation of the thermofield double state in the doubled field theory Hilbert space, which in energy eigenbasis is expressed as 
\begin{equation}\label{eq:}
\ket{\text{TFD}} = \frac{1}{\sqrt{Z(\beta)}} \, \sum_n \, e^{-\frac{1}{2} \, \beta\, E_n} \, \ket{E_n^{\skR}} \otimes \ket{E_n^{\skL}} 
\end{equation}	
where the $R(L)$ refers to the asymptotic boundaries at $\tE =0 \ (\tE =\frac{\beta}{2})$.

Insofar as  thermal equilibrium properties are concerned, the thermofield double construction proves ample. One gets to ask questions about correlation functions with operators inserted in either copy of the doubled system. One can therefore view the gravitational path integral as computing the following generating function for asymptotic observers (in asymptotically AdS spacetimes) 
\begin{equation}\label{eq:Ztfd}
\mathcal{Z}_\text{TFD}\left[J_\skR, J_\skL\right] = 
\Tr{\mathcal{U}(J_\skR) \, \rho_\beta^\frac{1}{2} \, \left(\mathcal{U}[J_\skL]\right)^\dagger \, \rho_\beta^\frac{1}{2}}
\end{equation}	
The fact that we slice open the functional integral midway is what is responsible for the fractionation of the thermal density matrix $\rho_\beta$.

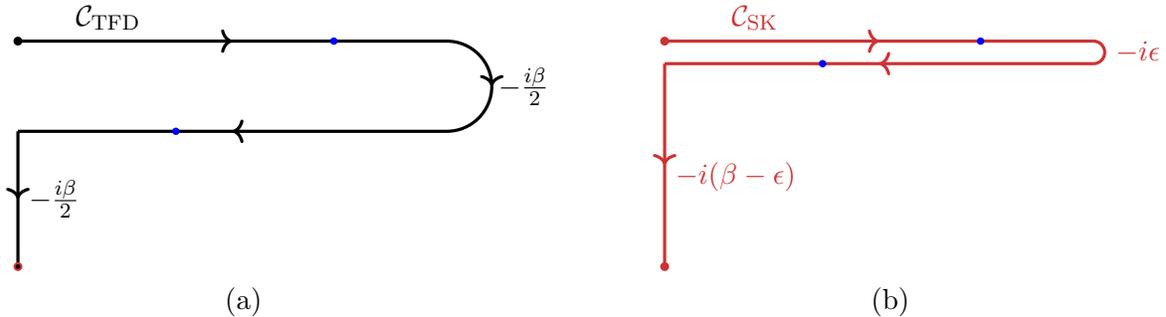
\begin{figure}[t!]
\begin{center}
\begin{tikzpicture}[scale=0.6]
\draw[very thick,color=black,->-] (-5,1) -- (-3,1) node [above] {$\color{black}\mathcal{C}_{\text{TFD}}$} -- (4.5,1);
\draw[very thick,color=black,->-] (4.5,-1)  -- (-5,-1);
\draw[very thick, color=black,->-] (-5,-1) -- (-5, -2.5) node [right]  {$ -\frac{i\beta}{2}$} --  (-5,-4);
\draw[very thick,color=black,->-] (4.5,1) arc (90:-90:1);
\draw[very thick,color=black] (5.4,0.0) node [right] {$ -\frac{i\beta}{2}$};
\draw[thick,color=black,fill=black] (-5,1) circle (0.45ex);
\draw[thick,color=rust,fill=black] (-5,-4) circle (0.45ex);
\draw[thick,color=blue,fill=blue] (2,1) circle (0.35ex);
\draw[thick,color=blue,fill=blue] (-1.5,-1) circle (0.35ex);
\draw[very thick,color=black] (0,-4.2) node [below] {(a)};
\end{tikzpicture}
\hspace{1cm}
\begin{tikzpicture}[scale=0.6]
\draw[very thick,color=rust,->-] (-5,1) -- (-3,1) node [above] {$\color{rust}\mathcal{C}_{\text{SK}}$} -- (4.5,1);
\draw[very thick,color=rust,->-] (4.5,0.5)  -- (-5,0.5);
\draw[very thick, color=rust,->-] (-5,0.5) -- (-5, -2) node [right]  {$ -i (\beta-\epsilon)$}-- (-5,-4);
\draw[very thick,color=rust,] (4.5,0.5) arc (-90:90:0.25);
\draw[very thick,color=rust] (4.75,0.75) node [right] {$ -i\epsilon$};
\draw[thick,color=rust,fill=rust] (-5,1) circle (0.45ex);
\draw[thick,color=rust,fill=rust] (-5,-4) circle (0.45ex);
\draw[thick,color=blue,fill=blue] (2,1) circle (0.35ex);
\draw[thick,color=blue,fill=blue] (-1.5,0.5) circle (0.35ex);
\draw[very thick,color=black] (0,-4.2) node [below] {(b)};
\end{tikzpicture}
\caption{A comparison of the (a) thermofield double  and (b) Schwinger-Keldysh complex  time contours  for a system prepared in a thermal state. The starting and end points of the contour are identified. The associated Euclidean (imaginary time) periodicity is set by the inverse temperature $\beta$.}
\label{fig:tfdsk}
\end{center}
\end{figure}

To obtain real-time response one then has to analytically continue these results to the real-time domain. In the absence of any sources deforming the state away from equilibrium there is in principle no obstacle to carrying out this analytic continuation. However, once one moves to the physics of systems in local equilibrium, or more generally out-of-equilibrium, which more pertinently applies to  our discussion of computing influence functionals, the thermofield double state proves less useful. In this context the Schwinger-Keldysh formalism provides a cleaner and more natural way for computing real-time observables by keeping manifest causality and unitarity, without relying on the aforesaid analytic continuation. The Schwinger-Keldysh generating function does not fractionate the density matrix, but rather computes the generating function:
\begin{equation}\label{eq:Zsk}
\mathcal{Z}_\text{SK}\left[J_\skR, J_\skL\right] = 
\Tr{\mathcal{U}(J_\skR) \, \rho_\beta \, \left(\mathcal{U}[J_\skL]\right)^\dagger}\,.
\end{equation}	
This fact is well-known in  non-equilibrium QFTs where real-time observables, say linear response captured by viscosity, conductivity, etc.,  are always computed (using Kubo formulae) from the Schwinger-Keldysh formalism. The pictorial representation of the path integral contour shown in \cref{fig:tfdsk} provides a quick way to see the difference between the two constructions (cf., \cite{Haehl:2016pec} for further discussion). 

Given this fact, it is natural to ask how the Schwinger-Keldysh construction can be adopted to gravitational theories. We will primarily focus on situations where the system is prepared in a thermal state. We wish to know how the gravitational dynamics fills in an asymptotic Schwinger-Keldysh contour shown in \cref{fig:tfdsk}(b). This question has been considered by many authors in the AdS/CFT context over the years. In \cite{Son:2002sd} the first proposal for computing real-time correlation functions was given. These authors posited that one should consider the future half of the domain of outer communication of a Lorentzian black hole spacetime, and impose ingoing boundary conditions on the future horizon to extract causal observables (retarded Green's functions). In addition they also argued for the absence of any boundary contribution from the horizon. This prescription was justified shortly thereafter in \cite{Herzog:2002pc} using the maximal Kruskal extension of the black hole geometry (the logic was to exploit the analytic structure taking inspiration from the Euclidean thermofield double construction). One limitation of this approach was that it was well adapted to the computation of two-point functions, but it left implicit how to obtain higher point functions. Nevertheless, over the years, various authors have attempted to use this prescription for various applications \cite{Barnes:2010jp,Son:2009vu,CaronHuot:2011dr,Chesler:2011ds,Botta-Cantcheff:2018brv,Botta-Cantcheff:2019apr}. 

In order to explain the subtlety let us remind the reader how Euclidean  $n$-point correlation functions in AdS/CFT  are computed via Witten diagrams \cite{Witten:1998qj}. One first computes the appropriate bulk-boundary propagators for  various external operator insertions. These enable us to `evolve' the corresponding fields into the bulk with sources set by the boundary conditions. One  convolves the bulk fields  thus obtained using the bulk interaction vertices which are dictated by gravitational dynamics. Since the fields interact locally, one has to integrate the position of the vertex over the entire Euclidean  bulk manifold. Modulo the choice of temporal boundary conditions one expects something similar for the computation of real-time retarded observables. However, it is was a-priori unclear what domain of the bulk geometry one ought to integrate the bulk interaction vertex over, even assuming that the ingoing boundary conditions serve to pick out the appropriate bulk-boundary propagator.

This issue was addressed  in \cite{Skenderis:2008dh,Skenderis:2008dg} who gave a more  detailed prescription for real-time computations, arguing that one should fill in asymptotic Schwinger-Keldysh contours with piecewise smooth geometries: real-time evolution sections of the boundary contour get filled in with Lorentzian geometries, and imaginary-time segments with Euclidean geometries. These geometries are glued together continuously along codimension-1 spacelike slices. The authors  developed a robust holographic renormalization scheme for asymptotically AdS geometries \cite{Skenderis:2008dg}. Furthermore, the ingoing boundary condition for retarded correlation functions was derived quite cleanly using this  prescription in \cite{vanRees:2009rw}.  This prescription was employed in the derivation of covariant holographic entanglement entropy proposal \cite{Dong:2016hjy}.

Per se, it is then clear that in order to carry out the computation of real-time correlation function of probe operators in a fixed state, one could simply use the prescription of  \cite{Skenderis:2008dh,Skenderis:2008dg}. However, if one were to ask questions about dynamically evolving geometries, one realizes that the piecewise smooth geometries pose a potential issue in the presence of horizons.  Physically, the  question essentially becomes one of coming up with a prescription ensuring that effects of the outgoing Hawking quanta are correctly accounted for (see \cite{CaronHuot:2011dr, Chesler:2011ds} for attempts in this direction). While a complete answer to this question is still unclear, an interesting prescription was recently given  by \cite{Glorioso:2018mmw} to address this lacunae in the probe limit  (see also \cite{deBoer:2018qqm}). 

\begin{figure}[h]
\centering
\includegraphics[width=.3\textwidth]{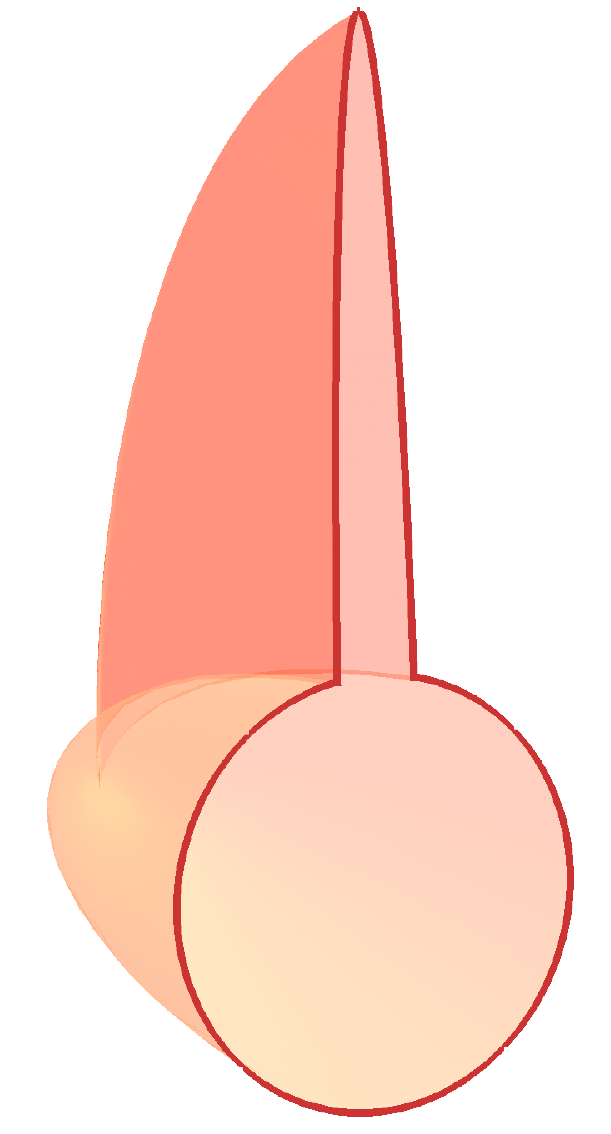}
\hspace{1cm}
\includegraphics[width=.45\textwidth]{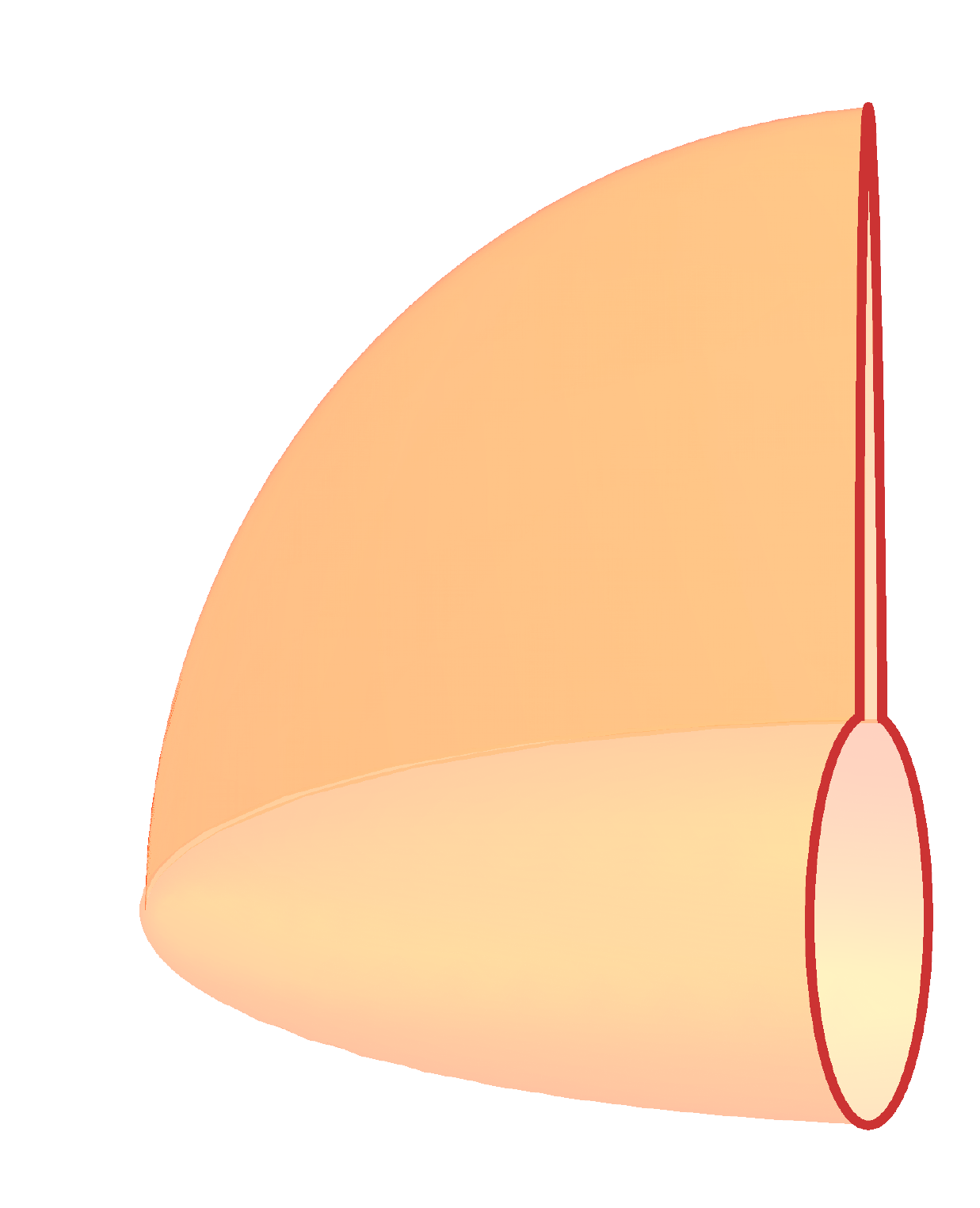}
\includegraphics[width=0.42\textwidth]{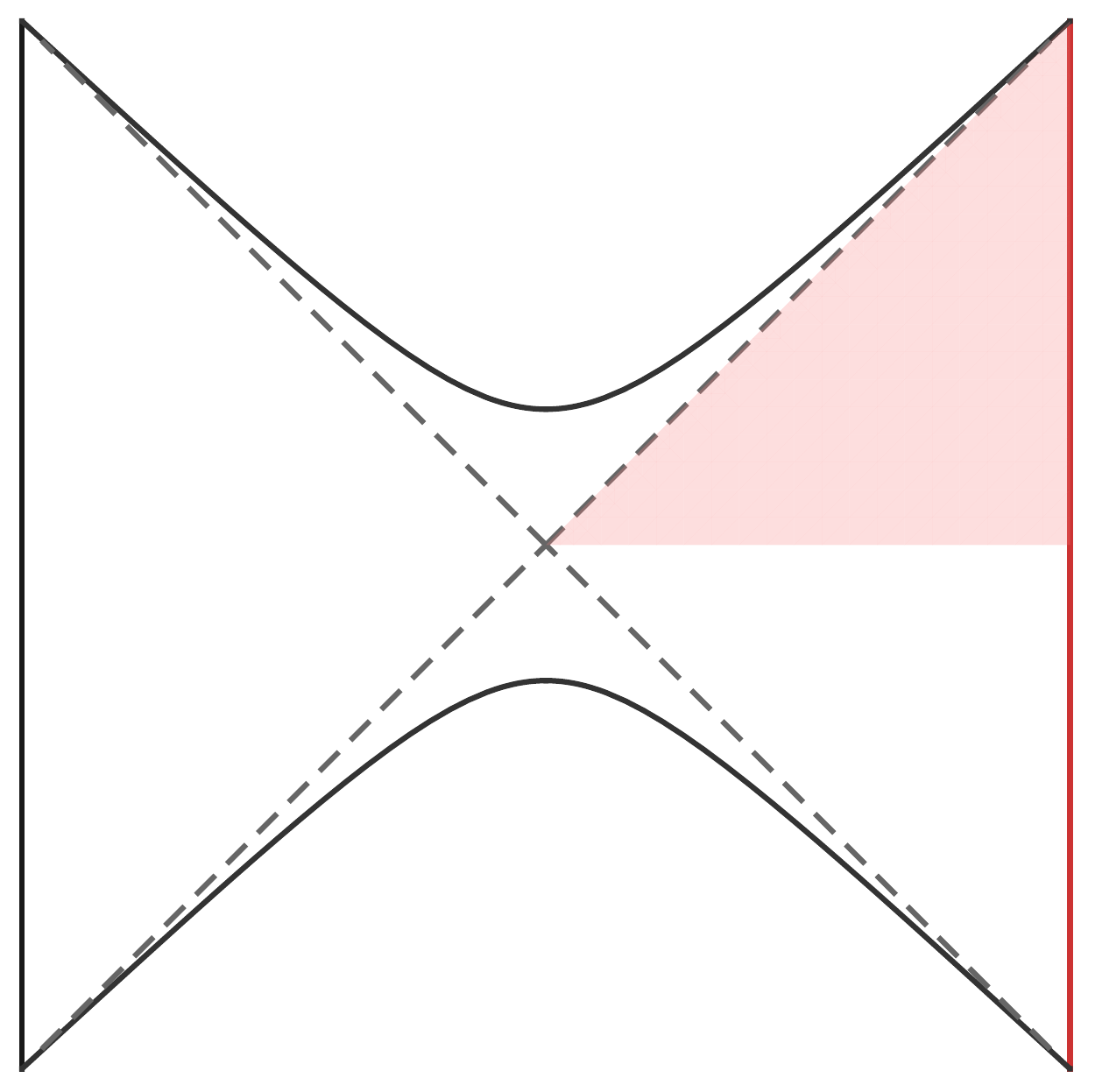}
\hspace{1cm}
\includegraphics[width=0.42\textwidth]{figures/sadspd.pdf}
\begin{picture}(0,0)
\setlength{\unitlength}{1cm}
\put (-11.,7) {$\circlearrowleft\scriptscriptstyle{\tE}$}
\put (-1.6,7) {$\circlearrowleft\scriptscriptstyle{\tE}$}
\put (-10.5,12) {$\uparrow\scriptscriptstyle{t}$}
\put (-11.5,12) {$\downarrow\scriptscriptstyle{t}$}
\put (-12,11) {$\skL$}
\put (-10.8,11) {$\skR$}
\put (-2.8,11) {$\skL$}
\put (-1.2,12.5) {$\skR$}
\put (-1.2,13) {$\uparrow\scriptscriptstyle{t}$}
\put (-1.85,11) {$\downarrow\scriptscriptstyle{t}$}
\put (-9.5,4) {$\searrow\skL$}
\put (-2,4) {$\nwarrow\skR$}
\end{picture}
\caption{The two-sheeted complex grSK geometry shown from two different perspectives. On the top left we display the boundary thermal SK contour which is filled in the Euclidean portion by the Euclidean black hole geometry (the cigar) and in the Lorentzian section by two copies of the domain of outer communication of the Lorentzian black hole spacetime. The top right panel displays the bulk perspective to emphasize the smooth join of the two sheets of the Lorentzian section. On the bottom panel we illustrate the Lorentzian sections of the geometry on the Schwarzschild-AdS{d+1} Penrose diagram. with the regions pertaining to the $L$ and $R$ sheets of the grSK spacetime shaded.}
\label{fig:grsk}
\end{figure}

The prescription of \cite{Glorioso:2018mmw} postulates that the gravitational dual of the asymptotic Schwinger-Keldysh contour is given by a complex two-sheeted spacetime. This geometry is made of  two copies of the black hole exterior  smoothly glued together across the future horizon, with a particular monodromy condition. One may view this as the statement that the asymptotic Schwinger-Keldysh contour gets filled in by a complex geometry. This can be motivated by the slicing open the Euclidean Gibbons-Hawking saddle \cite{Gibbons:1976ue} whilst incorporating the Schwinger-Keldysh boundary conditions, as we have attempted to illustrate in \cref{fig:grsk}.
The authors of \cite{Glorioso:2018mmw}  already demonstrated the efficacy of their prescription by obtaining the quadratic influence  functional in a low frequency and momentum expansion for a bulk scalar field in the probe limit.\footnote{ The  authors of
 \cite{Glorioso:2018mmw,deBoer:2018qqm} also addressed the problem of Maxwell gauge field in the probe limit. However, as pointed out in  \cite{Glorioso:2018mmw} many aspects of how the prescription works for gauge theories are still unclear.}

Subsequently, \cite{Chakrabarty:2019aeu} explored how this prescription may be employed to study non-linear Langevin dynamics of a single particle degree of freedom in quantum mechanics. Their analysis subjected the prescription to the test of dealing with self-interacting fields in the bulk and demonstrated that it continues to give reasonable answers. Specifically, the authors of \cite{Chakrabarty:2019aeu} modeled the Brownian particle using  the holographic construction described in \cite{deBoer:2008gu,Son:2009vu}.  Using the complex holographic geometry of \cite{Glorioso:2018mmw} to compute influence functionals, they  demonstrated that general expectations of non-linear Langevin dynamics explored in earlier  \cite{Chaudhuri:2018ihk,Chakrabarty:2018dov} was borne out.  We will demonstrate in the course of our analysis that the prescription correctly captures real-time observables and can be used in the semi-holographic setting to model open effective field theories. 

We have motivated the discussion in terms of invoking the holographic duality to learn about influence functionals in an open quantum system. It is instructive to also keep in mind that these influence functionals for holographic thermal baths in fact encode interesting semiclassical gravitational information about black holes. In the semi-holographic set-up we motivated above, the system degree of freedom $\Psi$ gets imprinted upon by the characteristics of the environment. Since our thermal environment is provided by a black hole, we would effectively be encoding not only the dissipative behaviour of the horizon, which we know to be characterized linearly by quasinormal modes, but also the fluctuations of the horizon. The latter are nothing but the outgoing Hawking radiation. Since the quasinormal modes refer to  the physical response of infalling matter we would effectively be capturing, in the non-Gaussian influence functionals, the interaction between infalling matter and the Hawking radiation.\footnote{ The non-Gaussian correlations we compute holographically using the semiclassical gravity approximation are suppressed in the planar expansion by powers of $1/N$, as indeed are all higher point functions in a large $N$ (or large central charge)  environment.} 

The coupling of AdS black holes to external systems has recently been of active interest in the context of the black hole information paradox \cite{Penington:2019npb,Almheiri:2019psf} (cf., \cite{Rocha:2008fe}). In these examples the external system is treated as a passive reservoir wherein one captures the Hawking radiation. Our discussion applies in this context as well; the external system's observables will faithfully be able to diagnose the interaction of Hawking radiation with infalling matter. The structure we get to probe however is only the leading semiclassical pieces of the interaction.  We are considering a single gravitational saddle point configuration, and not including contributions from non-trivial replica saddles which have been important in understanding the purification of the Hawking radiation and the reproduction of the Page curve for AdS black holes \cite{Penington:2019kki,Almheiri:2019qdq}.

In this paper, we will be considering the coupling of our probe/system (modeled by a single bosonic field) to a scalar operator in the holographic thermal system. Most of the technical computations we report will  involve computing Schwinger-Keldysh correlation functions of a scalar field in an asymptotically AdS black hole background using the holographic Schwinger-Keldysh geometry of \cite{Glorioso:2018mmw}. This work had already considered the computation of the two point function in a low-energy gradient expansion, i.e., perturbatively at low frequencies and momenta. We will extend this to higher point functions, but also show how to get results outside the gradient expansion in two dimensional CFTs.

The outline of the paper is as follows. We will review the gravitational prescription for filling in the Schwinger-Keldysh contour in \cref{sec:gravgeom}, illustrating in the process the generalizations to an arbitrary spacetime with a Killing horizon. In \cref{sec:openscalar} we outline the specifics of the open quantum system we wish to study and describe the holographic thermal environment we are coupling it to.
In \cref{sec:sprops} we then turn to the task of solving the scalar wave equation (which we generically do at long wavelengths), and describe various propagators of interest that appear in the computation of the Schwinger-Keldysh Witten diagrams. We compute the influence functionals using holographic methods in \cref{sec:influence} and furthermore demonstrate in \cref{sec:stochastic} that we can use our results to provide a stochastic description of the effective open quantum field theory. We end with a brief discussion in \cref{sec:discuss}.

Several technical steps are outlined in various appendices. In \cref{sec:eogradient} we explain various aspects of the long-wavelength gradient expansion we work in, while \cref{sec:gradexpmass} gives specifics of the Green's function in Schwarzschild-\AdS{d+1} geometries obtained in  this approximation. We review in \cref{sec:wittendia} why the standard Witten diagram technique continues to work for computing influence functionals.  In \cref{sec:2dcubic} we  describe how non-Gaussian influence functions in 2d CFTs can be computed. Finally, in \cref{sec:counterterns} contains details of the divergence structure of bulk Witten diagrams and a desciption of counterterms that enter into the influence functionals.

\section{grSK: the gravitational Schwinger-Keldysh saddle}
\label{sec:gravgeom}

The  Schwinger-Keldysh contour for a thermal state  is a complex time path running from $t=0$ to $t= T$ and thence to $t = 0 - i\,\beta$ as depicted in \cref{fig:tfdsk}(b). Thus we have a contour in a complex time plane for the temporal part of the action at the boundary for a holographic field theory. The proposal of \cite{Glorioso:2018mmw} is to extend this  contour to a codimension-1 hypersurface in the complexified bulk spacetime in gravity.
To be specific, let us first introduce this geometry for  stationary configurations with a timelike Killing field, such as the planar \SAdS{d+1} black hole. 

We start with the metric written in ingoing Eddington-Finkelstein coordinates, which are regular at the future horizon, viz., 
\begin{equation}\label{eq:efads}
ds^2 = -r^2\, f(r) \, dv^2 + 2\, dv\, dr + r^2\, d{\bf x}^2 \,, \qquad f(r) = 1 - \frac{r_h^d}{r^d} \,.
\end{equation}	
The coordinate $v$ is identified with the time coordinate $t$ on the boundary of the spacetime $r \to \infty$, which as we have argued, is to be interpreted as a curve in the complex plane.  The idea is to also upgrade the radial coordinate to the complex domain and pick a codimension-1 slice through the resulting complex spacetime.

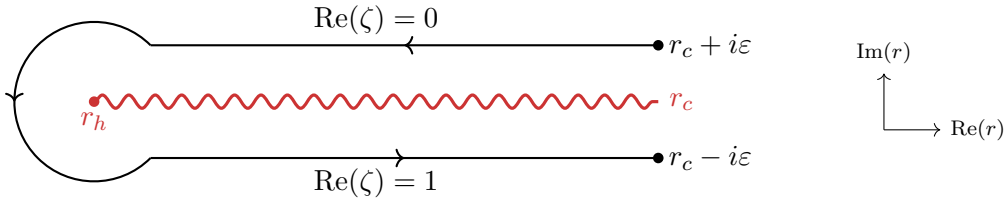
\begin{figure}[t!]
\begin{center}
\begin{tikzpicture}[scale=0.75]
\draw[thick,color=rust,fill=rust] (-5,0) circle (0.45ex);
\draw[thick,color=black,fill=black] (5,1) circle (0.45ex);
\draw[thick,color=black,fill=black] (5,-1) circle (0.45ex);
\draw[very thick,snake it, color=rust] (-5,0) node [below] {$r_h$} -- (5,0) node [right] {$r_c$};
\draw[thick,color=black, ->-] (5,1)  node [right] {$r_c+i\varepsilon$} -- (0,1) node [above] {$\Re(\ctor) =0$} -- (-4,1);
\draw[thick,color=black,->-] (-4,-1) -- (0,-1) node [below] {$\Re(\ctor) =1$} -- (5,-1) node [right] {$r_c-i\varepsilon$};
\draw[thick,color=black,->-] (-4,1) arc (45:315:1.414);
\draw[thin, color=black,  ->] (9,-0.5) -- (9,0.5) node [above] {$\scriptstyle{\Im(r)}$};
\draw[thin, color=black,  ->] (9,-0.5) -- (10,-0.5) node [right] {$\scriptstyle{\Re(r)}$};  
\end{tikzpicture}
\caption{ The complex $r$ plane with the locations of the two boundaries and the horizon marked. The grSK contour is a codimension-1 surface in this plane (drawn at fixed $v$). The direction of the contour is as indicated counter-clockwise encircling the branch point at the horizon.}
\label{fig:mockt}
\end{center}
\end{figure}

Operationally, we in fact upgrade radial tortoise coordinate to a complex variable, which we refer to as the \emph{mock tortoise} coordinate, $\ctor$. We define this coordinate by the differential relation:
\begin{equation}\label{eq:ctordef}
\frac{dr}{d\ctor} = \frac{i\,\beta}{2} \, r^2\, f(r) \,,
\end{equation}	
where $\beta = \frac{4\pi}{d \,r_h}$ is the inverse temperature of the black hole.
The rationale for introducing this coordinate is that $\ctor$ picks up a logarithmic branch cut from the integral about the zero of the emblackening function $f(r)$ of the black hole. The choice of normalization is such that the monodromy around this cut is set to unity. 
The coordinate $\ctor$ can be viewed as parameterizing a two-sheeted surface, each of which can be thought of as the bulk extension of the Schwinger-Keldysh contour. On each sheet $\ctor$ has an imaginary part running from $0$ at the AdS boundary to $\infty$ at the horizon. In addition it has a real part which differentiates the two sheets and is given by the monodromy around the horizon. By convention we will choose one of the sheets to have vanishing real part, and the other to have unit real part (this is based on our choice of normalization). We will also cut-off the AdS geometry at a radial cut-off $r=r_c$ for computational ease. With this choice, we have the two branches on which the mock tortoise coordinate asymptotes to
\begin{equation}\label{eq:ctorbc}
\ctor(r_c+i\,\varepsilon) = 0 \,, \qquad \ctor(r_c-i\,\varepsilon) = 1\,.
\end{equation}	
A section of geometry in the mock tortoise complex plane is illustrated in \cref{fig:mockt}.

The metric  then takes the form
\begin{equation}\label{eq:sadsct}
ds^2 = -r^2\, f(r) \, dv^2 +  i\, \beta\, r^2 \, f(r)\,  dv\, d\ctor + r^2\, d{\bf x}^2 \,, \qquad f(r) = 1 - \frac{r_h^d}{r^d}
\end{equation}	
where we treat $r(\ctor)$ using \eqref{eq:ctordef}. We will refer to  this geometry as the \emph{gravitational Schwinger-Keldysh (grSK) saddle or geometry}. One can integrate \eqref{eq:ctordef} explicitly to find that the mock tortoise coordinate for Schwarzschild-\AdS{d+1} geometries in terms of a hypergeometric function, viz.,  
\begin{equation}\label{eq:ctoradsd}
 \ctor +\ctor_c=  
\frac{i\, d}{2\pi\, (d-1)}  \left( \frac{r}{r_h}\right)^{d-1}\, {}_2F_1\left(1,\frac{d-1}{d};2-\frac{1}{d};\frac{r^d}{r_h^d}\right) ,
\end{equation}	
where $\ctor_c$ is chosen to make $\ctor=0$ at $r=r_c+i\,\varepsilon$. The branch-cut of the hypergeometric function is taken to run from $r=r_h$ to $\infty$.\footnote{  It is actually convenient in explicit computations to work with a redefined radial coordinate $\varrho = \frac{r^d}{r_h^d}$, which is the argument of the hypergeometric function.}

One can motivate the grSK geometry by recalling the Schwinger-Keldysh generating function \eqref{eq:Zsk} which computes casual response of the thermal state (as well as the fluctuations thereabout). In particular, it requires that we do not fractionate the thermal density matrix. Since the excursion into the complex time domain between the two segments of the contour on the boundary in \cref{fig:tfdsk}(b) is infinitesimal, the gravity saddle point should respect this separation.  One way to achieve this is to  prepare the thermal state using the Euclidean path integral. The real-time part of the contour can thence be obtained by  slicing open the Euclidean black hole solution around $t = i\, \varepsilon$ and $t = -i\, \beta + i \, \varepsilon$ and continue the geometry into the Lorentzian section. The evolution of the slice at $t= i\, \varepsilon$ will give a section of the domain of outer communication of the Lorentzian black hole geometry. The slice at $t = -i\, \beta + i \, \varepsilon$ will lead to something similar with a reversed temporal direction. These are the two slices of the geometry living at $\ctor = 0, 1$, respectively. 

This basic picture was first espoused in \cite{vanRees:2009rw}, but the prescription of \cite{Glorioso:2018mmw} has the added advantage of smoothly connecting the two slices across a `horizon-cap'. The resulting geometry is smooth and is coordinatized by \eqref{eq:sadsct}. 
Pictorially, the construction is depicted in \cref{fig:grsk}.

Now that we have identified the grSK geometry we can explain how to study dynamics thereupon. One should think of all the fields as residing on a complex $\ctor$ contour and upgrade the classical bulk action to a contour integral over the mock tortoise coordinate. We  write:
\begin{equation}\label{eq:}
S_\text{bulk} = \oint d\ctor \int d^dx \, \sqrt{-g} \; \mathcal{L}[g_{AB}, \Phi]\,,
\end{equation}	
where $x^\mu$ are the boundary coordinates. We will use this form of the action for computations of influence functionals. Before we get to those however, we describe how to generalize our construction covariantly to spacetimes with a Killing horizon, and also explain some useful properties of the grSK geometries.

\paragraph{Covariant grSK spacetime:}
While we described the construction in a coordinate dependent manner, one can give a more covariant presentation of the same. We can describe this for any spacetime with a smooth future horizon. Pick some intrinsic coordinates on the spatial sections of the horizon, call them ${\bf x}$. One can let the temporal evolution be determined by the affine parameter, $v$, along the horizon generators. A natural radial coordinate $r$ can be chosen by demanding that it be generated by the null normal to the horizon, normalized with respect to the horizon generators, i.e., $\partial_v \cdot \partial_r = 1$. For a non-degenerate horizon one has $\partial_v$ being timelike outside the codimension-1 null hypersurface (the future horizon). In a local neighbourhood of the horizon one would then end up with a metric of the form 
\eqref{eq:efads} with $f(r)$ having a simple zero. The details of the rest of the geometry will depend on the asymptotics etc., but the part of the construction that matters for us is indeed the neighbourhood of the future horizon. One can now convert this classical geometry to a two-sheeted geometry with a gluing condition across the horizon as described above by replacing $r \to \ctor$.

To illustrate the construction more generally, it is sufficient to consider the near-horizon region of the spacetime. For non-degenerate black holes this is given by the Rindler geometry.  In this case we have the familiar form of the geometry as well as the ingoing coordinatization to be given as:\ 
\begin{equation}\label{eq:rindler}
ds^2  = -r^2\, dt^2 + dr^2  =-r^2 dv^2+ 2r dv dr= 2 dv d\rho - 2\rho dv^2 \,.
\end{equation}	
Here we have defined $\rho\equiv \frac{1}{2}\,r^2$. It is then easy to explicitly identify the mock tortoise coordinate, $\ctor = \frac{1}{2\pi i}\, \log \rho$ on the primary branch (we work with the Rindler temperature normalized to be $2\pi$). The grSK Rindler geometry would then take the form 
\begin{equation}\label{eq:}
ds^2 = 2e^{2\pi i \ctor}\ dv (2\pi i  d\ctor -  dv) .
\end{equation}

\paragraph{grSK time reversal:} Before proceeding further, it is useful to note one useful feature of the grSK geometries. These geometries are not time reversal invariant as is indeed appropriate for the Schwinger-Keldysh dual. However, the Schwinger-Keldysh construction has a $\mathbb{Z}_2$ involution that can be thought of as time-reversal (cf., \cite{Haehl:2016pec} for a discussion). As described in \cite{Chakrabarty:2019aeu} the geometry does indeed have an involution which can be used to map ingoing solutions to outgoing ones.  In the coordinates used in \eqref{eq:sadsct}  the transformation takes the form:
\begin{equation}\label{eq:trevA}
v \to i\,\beta \, \ctor  - v \,, \qquad  \omega \to - \omega
\end{equation}	
where $\omega$ is the frequency conjugate to $v$. More generally, on tensor valued fields the map acts via an idempotent $(1,1)$ tensor:
\begin{equation}\label{eq:Treverse}
\mathcal{T}_{A}^{\ B} \equiv 
\begin{pmatrix}
-1 &0 & 0\\
 i\,\beta & 1 & 0 \\
0 & 0 & \delta_{ij}
\end{pmatrix} \,, 
\qquad \mathcal{T}_A^{\ B} \, \mathcal{T}_B^{\ C} = \delta_A^{\ C} \,.
\end{equation}	
Time reversal on tensors by contracting indices appropriately, which can be inferred from the action on one-forms and vectors, respectively. These are given to be:
\begin{equation}\label{eq:trevb}
\begin{split}
\mathfrak{W}_A(v,\ctor, {\bf x}) &\mapsto \mathcal{T}_A^{\ B} \, \mathfrak{W}_B(i\beta\, \ctor - v, \ctor, {\bf x}) \,, \\
\mathfrak{V}^A(v,\ctor, {\bf x}) &\mapsto  \mathfrak{V}^B(i\beta\, \ctor - v, \ctor, {\bf x}) \, \mathcal{T}_B^{\ A} \,. 
\end{split}
\end{equation}
It is useful to write these equations in terms of ingoing and outgoing modes explicitly, which we denote with superscripts `$\pm$'. So one has 
\begin{equation}\label{eq:trevc}
\mathfrak{f}^-(v,\ctor, {\bf x}) = \mathfrak{f}^+(i\,\beta\, \ctor -v , \ctor, {\bf x})
\end{equation}	
leading to the general tensor transformation:
\begin{equation}\label{eq:trevten}
\left(\mathfrak{T}^+\right)_{A_1\cdots A_n}^{B_1 \cdots B_m}\;\;  \mapsto \;\;  \mathcal{T}_{A_1}^{\ C_1} \, \cdots \mathcal{T}_{A_n}^{\ C_n} \; \left(\mathfrak{T}^-\right)_{C_1\cdots C_n}^{D_1 \cdots D_m} \;  \mathcal{T}_{D_1}^{\ B_1} \, \cdots \mathcal{T}_{D_m}^{\ B_m} \,.
\end{equation}	
We will exploit this symmetry to construct solutions of the scalar wave equations in an asymptotically AdS grSK geometry to obtain the corresponding boundary-bulk Green's functions.

\section{Open scalar field theory and holographic baths}
\label{sec:openscalar}

As described in \S\ref{sec:intro}, our goal is to construct the open effective field theory of a single scalar degree of freedom coupled to a holographic thermal field theory. We will start by describing the general set-up and then specialize to the case of two dimensional theories. We will first begin our description by focusing on the `bath/environment' theory with a holographic 
dual and how its real-time correlators can be computed via AdS/CFT. We will then describe how this computation amounts to deriving the open effective theory for the probe.

\subsection{General set-up}
\label{sec:gsetup}

Let us consider a scalar probe $\Psi(x)$ coupled to a $d$-dimensional field theory with fields denoted collectively by $X$. The latter is taken to be in a thermal state and we let $\mathcal{O} \equiv \mathcal{O}[X(x)]$ be a local gauge invariant operator in this theory. The action for our system is  then 
\begin{equation}\label{eq:SpsiO}
S = \int d^dx\,  \bigg( \mathcal{L}[\Psi] + \mathcal{L}[X] + \Psi(x) \, \mathcal{O}(x) \bigg) .
\end{equation}  
We wish to integrate out the thermal degrees of freedom characterized by $X$ and derive an effective action for $\Psi$.  
The couplings in $S_\text{eff}[\Psi]$ are determined by standard sum rules in terms of the Schwinger-Keldysh thermal correlators of the environment variables $X$ \cite{Feynman:1963fq,Caldeira:1982iu} (see also \cite{Avinash:2017asn}). So in what follows we will focus on computing thermal Schwinger-Keldysh observables for the environment, with the understanding that these feed
into the effective action of our open quantum system. The implications for the open effective theory will be described below in \cref{sec:holopen}.

The thermal field theory we consider will be taken to be holographic. For example we can consider $S[X]$ to refer to a strongly coupled, planar gauge theory in $d>3$ (like  $\mathcal{N} =4$ SYM) or a large $c$ 2d CFT.  Concretely, in the familiar duality between $SU(N)$ $\mathcal{N}=4$ Super Yang-Mills (SYM) and string theory on \AdS{5} $\times {\bf S}^5$, the map between parameters is 
\begin{equation}
g_{YM}^2\,N  \sim \left(\frac{\ell_\text{AdS}}{\ell_s}\right)^4 \,, \qquad  N \sim \left(\frac{\ell_\text{AdS}}{\ell_{_P}}\right)^4 \,.
\label{eq:N4map}
\end{equation}  
where $\ell_s$ is the string length scale, and $\ell_{_P}$ the five-dimensional Planck scale. We will work in the regime $N\gg1$ and $g_{YM}^2\, N \gg 1$ whence the holographic system can be described by classical gravitational dynamics on \AdS{5} $\times {\bf S}^5$.

The gravitational dual description is in terms of a planar \SAdS{d+1} black hole, whose metric in ingoing coordinates was presented in \eqref{eq:efads}. The scalar operator, $\mathcal{O}$, is characterized by its conformal dimension $\Delta$ and maps, under the AdS/CFT dictionary, to a scalar field  $\Phi$ propagating on this black hole background with mass $m^2\, \lads^2 = \Delta (\Delta-d)$. In particular, the marginal case $\Delta=d$ corresponds to $m^2=0$ in AdS. The correlation functions of the operator $\mathcal{O}$ can be computed by studying the dynamics of $\Phi$ in the gravitational theory. For the purposes of our discussion we will model the scalar dynamics by a minimally coupled scalar with a contact self-interaction. For the most part the self-interacting scalar action we work with takes the form:
\begin{equation}
\label{eq:sphi4a}
S_{\Phi} = - \int d^{d+1}x \, \sqrt{-g}\,\left[ \frac{1}{2} \, g^{AB} \, \partial_{A}\Phi \partial_{B}\Phi +\frac{1}{2}\, m^2\, \Phi^2+  \frac{\lambda_n}{n!}\, \Phi^n \right]\,.
\end{equation}
While this will be sufficient to illustrate the general features, it should be borne in mind that one can obtain in  
actual (top-down) holographic models, such effective action for the bulk fields using dimensional reduction from 10 or 11 dimensional supergravity.  

The above discussion describes the basic set-up for any AdS/CFT computation. However, for the purposes of our real-time computation we would need to upgrade this action to reside on the gravitational  Schwinger-Keldysh black hole geometry \eqref{eq:sadsct}.  This is readily done, for we simply change coordinates and rewrite the scalar action as a contour integral over the grSK geometry. To wit, 
\begin{equation}
\label{eq:sphi4}
S_{\Phi} = - \oint d\ctor \int d^d x \, \sqrt{-g}\,\left[ \frac{1}{2} \, g^{AB} \, \partial_{A}\Phi \partial_{B}\Phi +\frac{1}{2}\, m^2\, \Phi^2+  \frac{\lambda_n}{n!}\, \Phi^n \right]\,.
\end{equation}
We will study this problem in general dimensions, constructing first the boundary-bulk propagators on the grSK geometry. This involves only the quadratic part of the action; we essentially need to invert the kinetic terms on the grSK geometry.

 The boundary to bulk propagators will be specified by suitable boundary conditions around the horizon-cap in the geometry \eqref{eq:sadsct} and non-normalizable boundary conditions characterizing sources on the AdS boundary. We will find that there are two types of propagators: \emph{retarded (ingoing)} and \emph{advanced (outgoing)}. They will be related in a simple manner by the  time-reversal involution identified at the end of \S\ref{sec:gravgeom}. These Green's functions can be obtained by solving the linear scalar wave equation on the grSK geometry. Working in momentum space variables allows mode decoupling as usual.  For a general field $\mathfrak{f}$ on the grSK geometry we adopt the following notational contrivance for Fourier transforms:
\begin{equation}\label{eq:}
\mathfrak{f}(v,\ctor, {\bf x})  = \int \frac{d\omega}{2\pi}\, \frac{d^{d-1}{\bf k}}{(2\pi)^{d-1}} \, \mathfrak{f}(\omega, \ctor, {\bf k}) \, e^{-i\, \omega\, v + i\, {\bf k} \cdot {\bf x}} \equiv \int_k\, \mathfrak{f}_k\, e^{i\,k\,x} \,.
\end{equation}  
In general $d$ dimensions we will not be able to solve for the propagator in an explicit  analytic manner, as the wave equation on the \SAdS{d+1} black hole is not known to admit closed form solutions. Hence we will resort to a gradient expansion in  the frequencies and momenta, aiming for a low-frequency, long-wavelength expansion. In $d=2$ however we will be able to obtain closed form expressions for the propagators.

Once we have obtained the boundary to bulk propagators we can compute higher point functions using standard Witten diagram technology in the grSK geometry. Consider the computation of the 4-point function of the operator $\mathcal{O}$ in the boundary theory on the Schwinger-Keldysh contour. Of the $4!$ Wightman functions with various time-orderings, $8$ are computed by the Schwinger-Keldysh time-ordering. This is clear from the generating functional \eqref{eq:Zsk}; on the Schwinger-Keldysh contour operators can be inserted either in the forward (R) or backward (L) segments, effectively doubling the number of correlators. These correlation functions are simply related via the Keldysh rules to sequences of nested commutators and anti-commutators of the operator $\mathcal{O}$ with suitable time-ordering step-functions, see \cite{Chou:1984es,Haehl:2016pec}. Furthermore, the thermal KMS relations group the correlation functions into orbits of 4 elements (coming from cyclic symmetry of the thermal trace) \cite{Haehl:2017eob}. 

It is convenient to introduce a couple of different basis of operators and sources which  are convenient in the Schwinger-Keldysh formalism for various computations. First, we introduce the \emph{average-difference} or \emph{Keldysh} basis
\begin{equation}\label{eq:avdif}
\begin{split}
\text{operators}: &  \quad \mathcal{O}_a = \frac{1}{2} \left( \mathcal{O}_\skR + \mathcal{O}_\skL\right) \,, \qquad \mathcal{O}_d = \mathcal{O}_\skR - \mathcal{O}_\skL\,, \\
\text{sources}: & \quad J_a = \frac{1}{2} \left( J_\skR + J_\skL\right) \,, \qquad\quad J_d = J_\skR - J_\skL\,.
\end{split}
\end{equation}  
Within the Schwinger-Keldysh literature, the average/difference fields are also sometimes referred to as classical/quantum components respectively.   The Keldysh basis naturally 
separates out an averaged mean field vs the corrections due to quantum/statistical fluctuations.
Another useful basis which manifests the KMS relations is the \emph{retarded-advanced} (RA) basis\footnote{  Per se, our definition of the retarded-advanced basis differs mildly  from what is commonly used in the literature say in \cite{Chou:1984es,Haehl:2016pec}. The choice we make is inspired by certain simplifications for the spectral decomposition, not only of Schwinger-Keldysh observables, but also of the more general out-of-time-order correlation functions as described in \cite{Chaudhuri:2018ymp}.} 
\begin{equation}\label{eq:advret}
\begin{split}
\bar{J}_F(\omega,{\bf k}) &\equiv - \bigg((1+n_\omega)\, J_{\skR}(\omega, {\bf k})- n_\omega \, J_{\skL}(\omega, {\bf k}) \bigg) , \\
\bar{J}_P(\omega, {\bf k}) &\equiv -n_\omega\bigg( J_{\skR} (\omega, {\bf k})- J_{\skL}(\omega, {\bf k}) \bigg) \,,
\end{split}
\end{equation}
where $n_\omega$ is the Bose-Einstein statistical factor:
\begin{equation}\label{eq:benw}
n_\omega \equiv \frac{1}{e^{\beta\omega}-1}\,.
\end{equation}  
As we shall see below the distinction between the $P$ and $F$  combinations naturally shows up on the holographic side as the distinction between the ingoing modes and
the outgoing modes.

The structure of the influence functionals we wish to extract can be now succinctly summarized as follows. Having solved for the field $\Phi$ in terms of sources $J_{\skR}$ and $J_{\skL}$ on the asymptotic boundaries of the AdS grSK geometry the influence functionals are obtained from the generating functional, assuming an $n$-point bulk contact interaction:
\begin{equation}\label{eq:Sgen}
\begin{split}
S_{(n)} & \propto \prod_{i=1}^n\, \int_{ k_i} \, \delta(\sum_{i=1}^n k_i) \; \oint \, d\zeta\,\prod_{i=1}^n \, \Phi(\zeta, k_i)\\
& \equiv   \int \, \prod_{i=1}^n\, \frac{d^dk_i}{(2\pi)^d} \left( 2\pi\right)^d \delta(\sum_i\, k_i) \; 
\oint \, d\zeta\,\prod_{i=1}^n \, \Phi(\zeta, k_i)\\
&=  \int \, \prod_{i=1}^n\, \frac{d^dk_i}{(2\pi)^d} \left( 2\pi\right)^d \delta(\sum_i\, k_i) \; 
 \bigg[ \ifad_{p,n-p}(k_1,k_2,\cdots,k_n) \,  \prod_{i=1}^p\, J_a(k_i) \,  \prod_{j=p+1}^n \, J_d(k_j) \bigg]\\
 &=  \int \, \prod_{i=1}^n\, \frac{d^dk_i}{(2\pi)^d} \left( 2\pi\right)^d \delta(\sum_i\, k_i) \; 
 \bigg[ \ifra_{p,n-p}(k_1,k_2,\cdots,k_n) \,  \prod_{i=1}^p\, \bar{J}_F(k_i) \,  \prod_{j=p+1}^n \, \bar{J}_P(k_j)
\bigg].
\end{split}
\end{equation}  
In case there are lower-point contact interactions in the bulk, then we should also include tree level Witten diagrams where we have bulk-bulk propagators between the vertices of such lower order contact terms. While self-evident we nevertheless provide a brief argument for this prescription in Appendix~\ref{sec:wittendia}. 

We have given the influence functionals both in the average-difference basis, denoted by $\ifad$ as well as  in the retarded-advanced basis, denoted  $\ifra$. We will later write down expressions for them directly in terms of scalar Green's functions. However, there are some general statements that one can make regarding their structural properties prior to any explicit computation, which follow directly from the generating function \eqref{eq:Zsk}. In the average-difference basis the fact that $\mathcal{Z}_\text{SK}[J,J] = \Tr(\rho_\beta)$ implies that 
\begin{equation}\label{eq:}
\ifad_{a\cdots a}(k_1, k_2, \cdots, k_n) = 0 \,.
\end{equation}  
This is reflection of the Schwinger-Keldysh collapse rule: the difference operators (equivalently the average source) cannot be futuremost.

In the retarded-advanced basis the fact that we have folded in the statistical factors makes not only the Schwinger-Keldysh collapse rule manifest, but also incorporates the KMS condition. We have 
\begin{equation}\label{eq:}
\ifra_{P\cdots P}(k_1, k_2, \cdots, k_n) = 0 = \ifra_{F\cdots F}(k_1, k_2, \cdots, k_n)  \,.
\end{equation}  
We shall see the holographic SK geometry naturally incorporates these relations through the smoothness  of the solution across the future horizon cap region that glues together the two sheets of grSK geometry.

\subsection{Deriving an open EFT from holography}
\label{sec:holopen}

In \cref{sec:gsetup} we have described how one could get a generating functional for Schwinger-Keldysh correlators from holography. This generating functional is evaluated in the presence of a source $J$ for an operator $\mathcal{O}$ in the theory with holographic dual. This selfsame generating functional has another physical interpretation as emphasized by Feynman and Vernon \cite{Feynman:1963fq} in their seminal work on open quantum systems. 

The form of the coupling between the probe and the bath systems in \eqref{eq:SpsiO} suggests 
that we can interpret the source  $J$ of the operator $\mathcal{O}$ as the probe field $\Psi$ itself. Further assuming that the dynamics of our probe system is slow, we can then integrate out the `fast' degrees of freedom which make up the environment. This will obtain for us an effective SK action for this probe, which was termed as the influence functional of the probe in \cite{Feynman:1963fq}. 
The Schwinger-Keldysh generating functional of the CFT $\mathcal{Z}_\text{SK}[J_R, J_L]$ can hence be given an alternate interpretation as the influence functional of the probe $S_\text{IF}[\Psi_\skR, \Psi_\skL]$.

Various structural features we described above for the generating functional can then be re-interpreted as necessary features for the probe influence functional.  As described in \cref{sec:intro} if the environment is sufficiently forgetful, one would expect that a local influence functional could be written down for the probe. We will indeed see that the holographic influence functionals derived this way do satisfy this property. Given the difficulty of constructing local influence functionals from perturbative methods (see \cref{sec:intro}), this is indeed a fortunate circumstance. We will show that   a variety of open EFTs can be derived this way by using grSK geometry and holography.

The KMS conditions and Schwinger-Keldysh collapse rule then have a definite counterpart from the influence functional perspective. One can first of all construct a stochastic field theory that is dual to the influence functional of the open system. In this stochastic field theory, the structural features described above get re-interpreted as non-linear generalizations of fluctuation-dissipation theorems (FDTs). Our holographic  construction naturally leads to these non-linear FDTs via the physics of the Hawking radiation and its interaction with the ingoing modes. As far as the authors are aware, this is the first instance where these non-linear FDTs are derived within a field theoretic setup by integrating out bath degrees of freedom.\footnote{ The particle analogues of non-linear FDTs for a Brownian particle have for instance appeared in \cite{Chaudhuri:2018ihk,Chakrabarty:2018dov,Chakrabarty:2019qcp,Chakrabarty:2019aeu}. }

The stochastic description for the  dynamics of the open system degree of freedom which follows from the  influence functionals $S_\text{IF}[\Psi_\skR, \Psi_\skL]$ can be obtained in the following manner. We work in the average/difference basis $\Psi_a $ and $\Psi_d$ defined analogously to \eqref{eq:avdif} for the system variables. The idea is to write the dynamics of  the average field $\Psi_a$ as a Langevin equation after eliminating the difference or fluctuation field $\Psi_d$. We recall that the average field is the `classical' variable, where the difference field is the `stochastic/fluctuation' variable.  One starts with the following ansatz for the Langevin dynamics
\begin{equation}\label{eq:Langevin}
\begin{split}
\mathcal{E}[\Psi_a,\eta]  & = f\, \eta\,, \\
\mathcal{E}[\Psi_a,\eta]  &\equiv 
	\left(-K\,  \partial_t^2 + D \,\nabla^2 + \gamma \,\partial_t \right) \Psi_a\\
& \qquad + \sum_{k=1}^{n-1} \left( \theta_{k} \frac{\eta^{k-1}}{k!} \frac{\Psi_a^{n-k}}{(n-k)!} +  \bar{\theta}_{k} \frac{\eta^{k-1}}{k!} \frac{\Psi_a^{n-k-1}}{(n-k-1)!} \partial_t \Psi_a \right).
\end{split}
\end{equation}
The parameters $\{K,\, D,\, \gamma, \, \theta_k ,\, \bar{\theta}_k \}$ are coupling constants, to be determined in terms of the influence phase parameters. The variable $\eta$ here is the thermal/stochastic noise, with  strength $f$; it is drawn from a non-Gaussian probability distribution 
\begin{equation}
\mathcal{P}[\eta] \sim \exp\left(-\int d^dx \left(\frac{f}{2!}\,\eta^2 + \frac{\theta_n}{n!} \,\eta^n \right) \right) \,.
\end{equation}

To relate the Langevin ansatz \eqref{eq:Langevin} to an effective action arising from the influence functionals (see Eq.~\eqref{eq:Spsin} for an explicit action) we follow the Martin-Siggia-Rose (MSR) trick \cite{Martin:1973zz}, wherein one converts the stochastic Langevin equation into an effective action using a Lagrange multiplier field $\Psi_d$
and the functional integral identity:
\begin{equation}\label{eq:msr}
\begin{split}
1 &= \int \mathcal{D}\Psi_a\, \mathcal{D}\Psi_d\,\mathcal{D}\eta \, e^{-i\int d^dx  \, \left(\mathcal{E}[\Psi_a,\eta]\,-f \, \eta\right) \Psi_d} \, \mathcal{P}[\eta] \,.
\end{split}
\end{equation}

Integrating out the noise field $\eta$ one obtains the SK effective action for $\Psi_a$ and $\Psi_d$. To capture the leading order influence phase we only need to shift $\eta \to \eta + i\, \Psi_d$ and take the limit  $\eta \to 0$. We then arrive at the Schwinger-Keldysh effective action for our probe system
\begin{equation}\label{eq:SLang}
\begin{split}
S_\Psi &= 
	-\int d^dx  \, \Psi_d \left[ -K \partial_t^2  + D\,\nabla^2  + \gamma \,\partial_t \right] \Psi_a \\
& 
	\qquad + i \, \int d^dx\,
	\Bigg[  - f\, \frac{ (i\Psi_d)^2 }{2!}+ 
	 \sum_{k=1}^{n}\, \frac{(i\Psi_d)^{k}}{k!}  \left( \theta_{k}  + \bar{\theta}_{k}  \, \partial_t \right)\frac{\Psi_a^{n-k}}{(n-k)!}  \Bigg] .
\end{split}
\end{equation}
Recall that the standard (linear) FDT relates the friction term $\gamma$ and stochastic noise $f$ via 
\begin{equation}\label{eq:}
\gamma = \frac{\beta\, f}{2}\,.
\end{equation}	
We find that one obtains non-linear FDTs which relate the non-Gaussian couplings  $\theta_k$ and $\bar{\theta}_k$ amongst each other in the form.
\begin{equation}\label{eq:genFD}
\frac{2}{\beta}\, \bar{\theta}_k +  \theta_{k+1} + \frac{1}{4}\theta_{k-1} =0\,.
\end{equation}	
This is the advertised FDT  for which we will give a derivation once we have derived the influence functionals from holography in \cref{sec:stochastic}.

\section{Scalar propagation in grSK geometries}
\label{sec:sprops}

We now turn to the solutions of the  wave equations in diverse dimensions. We will first describe the general set-up, identifying various Green's functions of interest. Our goal is to determine the full solution for the scalar field with prescribed sources $J_{\skR}$ and $J_{\skL}$ on the two boundary segments of the grSK geometry. It will turn out that a useful way to proceed is to first identify the ingoing propagator $G^+$, which solves the wave equation with infalling boundary conditions, and thence use time reversal to obtain the outgoing propagator $G^-$ (or a linear combination, the Hawking propagator, $G^H$, that we introduce below).

 Once we have the general framework we will exploit the relative simplicity in $d=2$ where the BTZ black hole is a quotient of \AdS{3} to find an explicit expression for our propagators. For $d> 2$ the \SAdS{d+1} geometries do not admit closed form solutions to the scalar wave equation. However, one can solve for the propagators explicitly  in a gradient expansion, i.e., order by order perturbatively in frequencies and momenta.  We further elaborate on this gradient expansion scheme in \cref{sec:eogradient}  exposing some useful technical tricks to organize the solution and extract the propagators. 

 In what follows we impose Dirichlet (standard) boundary conditions for our scalar field asymptotically (for simplicity), so the conformal dimension of the CFT operator and the mass of the field propagating in the grSK geometry are related by $\Delta = \frac{d}{2} + \sqrt{ \frac{d^2}{4} + m^2\, \lads^2 }$.

\subsection{Scalar boundary to bulk propagators in grSK geometry}
\label{sec:}

The classical equation of motion for a minimally coupled scalar field 
\begin{equation}\label{eq:phieom}
\partial_A \left( \sqrt{-g}\, g^{AB} \, \partial_B \Phi\right) - m^2 \,\Phi = 0 \,,
\end{equation}	
in the  grSK geometry \eqref{eq:sadsct}. Written out explicitly in Fourier decomposition $\Phi(v,\ctor, {\bf x}) = \int_k\, \Phi_k(\zeta) \, e^{i\,k\,x}$ we find:
\begin{equation}\label{eq:kgPhi}
\pdv{}{\ctor} \left(r^{d-1} \pdv{\Phi_k}{\ctor}\right) + \frac{\beta\omega}{2}\left(r^{d-1}\, \pdv{\Phi_k}{\ctor} + \pdv{}{\ctor} \left(r^{d-1}\Phi_k \right)  \right) + \frac{\beta^2}{4}  r^{d-1} \, f(r) \left(\abs{\bf k}^2 + r^2\, m^2 \right) \Phi_k    =0 \,.
\end{equation}
We can now proceed to solve this problem, but it is useful to understand some structural aspects first. 

We wish to identify the boundary to bulk propagators for the wave equation \eqref{eq:kgPhi}, which allows us to evolve a non-normalizable source at the boundary of the spacetime into a field value at a bulk locale. We will start by first identifying the retarded or ingoing bulk to boundary Green's function $G^+$ and subsequently use information about the time reversal symmetry to extract the outgoing or advanced Green's function.  

Let us first introduce a new pair of  radial derivatives \cite{Chakrabarty:2019aeu}
\begin{equation}\label{eq:Dtorpm}
D_\ctor^\pm = \pdv{}{\ctor} \pm \frac{\beta\,\omega}{2} \,, 
\end{equation}	
which conjugate to each other in the form:
\begin{equation}\label{eq:Dpmconj}
e^{\beta\omega\,\ctor} \, D_\ctor^+ \, e^{-\beta\omega\,\ctor} = D_\ctor^-\,.
\end{equation}	
Notice that $D_\ctor^\pm$ are related to each other by time reversal. The rationale for introducing them is that these derivatives allow us to absorb the odd powers of $\omega$ in the wave equation into themselves. Indeed, in  terms of these derivations the scalar equation of motion takes the form: 
\begin{equation}\label{eq:Dpkg}
D_\ctor^+ \left( r^{d-1} \, D_\ctor^+ \Phi_k\right) + \frac{\beta^2}{4} \, r^{d-1} \, \left( f\, \abs{\bf k}^2 + r^2\, f\, m^2 - \omega^2 \right) \Phi_k = 0\,.
\end{equation}	 

\paragraph{Ingoing boundary to bulk propagator:} The ingoing Green's function $G_{in} \equiv G^+$ is a solution to \eqref{eq:Dpkg} satisfying a regularity condition at the horizon and normalized to unity at the cut-off boundary of the spacetime.
\begin{equation}\label{eq:}
G^+\big|_{r_c } =1 \,, \qquad \dv{G^+}{\ctor} \bigg|_{r_h}  =0 \,.
\end{equation}	
The choice of boundary conditions is such that we are looking at infalling modes across the future horizon, which isolates for us the quasinormal modes (general solution being a superposition of these modes in a linear theory). As such the ingoing Green's function will the one that was obtained in \cite{Son:2002sd} who argue for the ingoing boundary conditions to compute the retarded propagator.
We will shortly demonstrate how to obtain $G^+$ perturbatively in $\omega$ and $|{\bf k}|$, i.e., in a gradient expansion in general $d$ and also obtain an explicit analytic form in $d=2$.

\paragraph{Outgoing boundary to bulk propagator:} Once one knows  ingoing Green's function $ G^+$  the advanced or outgoing Green function should be obtained by suitably time reversing it. As argued at the end of \cref{sec:gravgeom}, while our coodinatization of the geometry is not time reversal invariant, there indeed an involution realized by the diffeomorphism \eqref{eq:trevA}. Let us see how this acts on the equation of motion \eqref{eq:Dpkg}. First, we note that after reversing the frequency dependence we obtain $G^-(\omega, {\bf k}) 
\equiv G^+(-\omega, {\bf k})$. Using the conjugation relation \eqref{eq:Dpmconj} we can then infer that the function $G^{-}(\omega,{\bf k}) e^{-\beta\,\omega\,\ctor}$ satisfies the wave equation provided  
\begin{equation}\label{eq:Dpkgm}
D_\ctor^- \left( r^{d-1} \, D_\ctor^- G^-\right) + \frac{\beta^2}{4} \, r^{d-1} \, \left( f\,\abs{\bf k}^2 + r^2\, f\,m^2 - \omega^2 \right) G^- = 0\,.
\end{equation}	 
Note that  \eqref{eq:Dpkgm} differs from \eqref{eq:Dpkg} only in the signs of the temporal derivatives i.e., through $\omega \to -\omega$.
It therefore follows that the outgoing Green's function can be obtained in Fourier domain as
\begin{equation}
G_{out}(\omega, |{\bf k}]) \equiv G^+(-\omega, |{\bf k}|) e^{-\beta\,\omega\,\ctor} \equiv G^-(\omega, |{\bf k}|) e^{-\beta\,\omega\,\ctor}\,.
\end{equation}

\paragraph{Full solution and boundary conditions:} Now that we have the formal expressions for the ingoing and outgoing Green's functions, we can take a suitable superposition to write down the general solution for the linear wave equation. The explicit form of the full solution takes the form 
\begin{equation}\label{eq:solKG}
\Phi(\ctor,\omega,{\bf k}) = C_+(\omega,{\bf k})\,G^+(\ctor,\omega,{\bf k}) + C_-(\omega,{\bf k})\,G^-(\ctor,\omega,{\bf k})e^{-\beta\,\omega\,\ctor} \,.
\end{equation}

We can now impose boundary conditions at the conformal boundary $r=r_c\pm i\,\varepsilon$ of the grSK geometry. We demand:
\begin{equation}
\Phi_k\big|_{\ctor=0} = J_{\skL}(\omega, {\bf k}) \,, \qquad \Phi_k \big|_{\ctor=1} = J_{\skR}(\omega,{\bf k})\,.
\end{equation}
Using the fact that $G^+$ and $G^-$ are normalized to unity at these boundaries, we find 
\begin{equation}
C_+(\omega,{\bf k}) + C_-(\omega,{\bf k}) = J_{\skL} \,, \qquad C_+(\omega,{\bf k}) + C_-(\omega,{\bf k})e^{-\beta\omega} = J_{\skR} \,.
\end{equation}
which results in the solution
\begin{equation}\label{eq:}
C_-(\omega, {\bf k})  = -(1+ n_\omega) \left(J_{\skR} - J_{\skL} \right)\,,
\qquad 
C_+(\omega,{\bf k}) = (1+n_\omega)\, J_{\skR} - n_\omega \, J_{\skL} \,.
\end{equation}	
Hence the general solution to the scalar wave equation \eqref{eq:solKG} takes the form  
\begin{equation}\label{eq:phiLR}
\Phi_k(\ctor,\omega,{\bf k}) = G^+(\ctor,\omega,{\bf k}) \,  \bigg( (1+n_\omega)\, J_{\skR} -n_\omega \, J_{\skL} \bigg) - G^-(\omega,{\bf k}) \, e^{\beta\, \omega\, (1-\ctor)}\,n_\omega \bigg( J_{\skR} - J_{\skL} \bigg) .
\end{equation}
where we have used the Bose-Einstein identity:
\begin{equation}\label{eq:beid}
1 + n_\omega = e^{\beta \omega}\, n_\omega\,.
\end{equation}	

It is helpful, before proceeding further, to rewrite the result in terms of  linear combinations of the L/R Schwinger-Keldysh sources. For instance, in the retarded-advanced basis \eqref{eq:advret} we find
\begin{equation}\label{eq:phiadvret}
\Phi(\ctor,\omega,{\bf k}) = -G^+(\ctor,\omega,{\bf k}) \, \JF 
+ G ^-(\ctor,\omega,{\bf k}) \, e^{\beta\omega(1-\ctor)} \, \JP\,.
\end{equation}

More interesting to us is the solution in the average-difference basis:
\begin{equation}\label{eq:phiad}
\begin{split}
\Phi(\ctor,\omega,{\bf k}) &=
 G^+(\ctor,\omega,{\bf k}) \prn{ 
	J_a(\omega, {\bf k}) + \left(n_\omega+ \frac{1}{2} \right)  J_d(\omega, {\bf k}) } \\
& \qquad \quad 	- \,n_\omega \,e^{\beta\omega(1-\ctor)} \, G^-(\ctor,\omega,{\bf k}) \,J_d(\omega, {\bf k})\,,\\
& \equiv G^+  \, J_a + \frac{1}{2}\, G^H  \, J_d\,.
\end{split}
\end{equation}
In the last line we have isolated the contribution from  the average and difference sources. The coefficient of the latter is a suitable admixture of the ingoing and outgoing modes which in fact deserves to be called the \emph{Hawking Green's function}  which is a solution to the wave equation with the boundary conditions 
\begin{equation}
\lim_{\ctor\rightarrow 0} G^H = 1\qquad \text{and} \qquad \lim_{\ctor\rightarrow 1}G^H = -1 \,.
\end{equation}
Explicitly, it is given by:
\begin{equation}
G^H \equiv \coth\left(\frac{\beta\omega}{2}\right)\,  G^+-  e^{\frac{\beta\omega}{2} (1-2\, \ctor)} \, \csch\left( \frac{\beta\omega}{2}
\right)   G^-\,.
\end{equation}
%

\subsection{Propagators in $d=2$}
\label{sec:d2props}

As a warm up we start with the BTZ geometry where $d=2$. The metric we recall is 
\begin{equation}\label{eq:btz}
\begin{split}
ds^2 &= -r^2 (1-\frac{r_h^2}{r^2}) \, dv^2 + 2\, dv dr + r^2\, dx^2 \\
& = -r_h^2\, \sinh^2 \rho\, dv^2  + 2 r_h\, \sinh\rho \,dv d\rho + r_h^2 \, \cosh^2\rho\, dx^2 \,,
\end{split}
\end{equation}	
where we have written the metric in ingoing coordinates both in the standard AdS radial coordinate as well as in the BTZ adapted global coordinate $r = r_h \cosh\rho$. We have either by direct integration or by simplifying \eqref{eq:ctoradsd} the following expression for the mock tortoise coordinate
\begin{equation}\label{eq:ctorbtz}
\begin{split}
\dv{\ctor}{r} = \frac{r_h}{i\pi}\, \frac{1}{r^2 - r_h^2} 
 \;\; &\Longrightarrow \;\;
\sqrt{\frac{r-r_h}{r+r_h} } = \tanh\frac{\rho}{2} = e^{i\pi(\ctor+\ctor_c)} \,,
\end{split}
\end{equation}	
where accounted for the cut-off surface where we are imposing our boundary conditions. 

We can solve the massive, minimally coupled, scalar wave equation \eqref{eq:kgPhi} in terms of hypergeometric functions
\begin{equation}\label{eq:btzsol}
\begin{split}
& (\sech \rho)^\Delta \left(1+ \tanh^2 \frac{\rho}{2}\right)^{\Delta -\bpt_+ - \bpt_-}  \ 
{}_2F_1\left( \bpt_+\,,  \bpt_-\,, 1+ \bpt_+ + \bpt_- -\Delta  \,;\tanh^2\rho\right) \!,\\
 &  
  (\sech\rho)^\Delta \,\left(\frac{\tanh^2 \frac{\rho}{2}}{1+ \tanh^2 \frac{\rho}{2}}\right)^{\Delta -\bpt_+ - \bpt_-}
 \;  {}_2F_1\left(  \Delta-\bpt_- \,,   \Delta- \bpt_+ \,, 1+ \Delta  -\bpt_+ -\bpt_-  
\,; \tanh ^2\rho\right)  \!.
\end{split}
\end{equation}	
The first of these is the solution that satisfies ingoing boundary conditions and is regular at the future horizon (near $\rho \sim 0$, we see that the linearly independent solutions are $\phi(\rho) = c_1 + c_2 \, \rho^{ i\,\beta \omega }$ of which the constant behaviour is the correct ingoing mode).  To keep expressions compact, we have introduced the lightcone like dimensionless combination of  frequencies and momenta, including contributions from the dimension which will appear in the solutions  below:
\begin{equation}\label{eq:ppm}
\begin{split}
\bpt_+ & =  i\, \frac{\beta}{4\pi} (k-\omega) +\frac{\Delta}{2}  \,, \qquad  
\bpt_-  = -i\, \frac{\beta}{4\pi} (k+\omega) +\frac{\Delta}{2}  \,.
\end{split}
\end{equation}	

The ingoing Green's function of interest, normalized to unit at the AdS boundary $\zeta =0$ or $\rho_c + i\,\epsilon$ can then be immediately inferred to be:
\begin{equation}\label{eq:btzGp}
\begin{split}
G^+(\zeta, \omega, k) &= 
 \left( \frac{ 1 -e^{2\pi i\,(\ctor+\ctor_c)}}{ 1 + e^{2\pi i\,(\ctor+\ctor_c)}}\right)^\Delta\,  \left( \frac{ 1 -e^{2\pi i\,\ctor_c}}{ 1 + e^{2\pi i\,\ctor_c}}\right)^{-\Delta} \;
 \left(\frac{1 + e^{2\pi i\,\left(\ctor+\ctor_c\right)}}{1 + e^{2\pi i\,\ctor_c}}\right)^{\Delta -\bpt_+ - \bpt_-}  \\ 
& \hspace{2.5cm}\times  \frac{{}_2F_1\left( \bpt_+\,,  \bpt_- \,, \bpt_+ + \bpt_- -\Delta +1 \,;\sec^2\pi(\ctor+\ctor_c)\right)}{
	{}_2F_1\left(  \bpt_+\,, \bpt_-\,, \bpt_+ + \bpt_- -\Delta +1 \,; \sec^2\pi\ctor_c\right) } \,.
\end{split}
\end{equation}	
We have written the answer in the mock tortoise coordinate which makes clear that the solution is continuous across the horizon-cap in the grSK geometry. The knowledge of the retarded Green's function is sufficient  to obtain the full solution for the field $\Phi$ on the Schwinger-Keldysh contour using \eqref{eq:phiad}.

\subsection{Propagators in $d>2$: gradient expansion}
\label{sec:dgenprops}

In dimensions $d>2$ the scalar wave equation \eqref{eq:kgPhi} does not admit a simple closed form solution in the \SAdS{d+1} backgrounds. One can however make progress by solving the equations order by order in a  low-energy, long-wavelength limit, i.e.,  we can expand our Green's function in the limit where $\beta\omega, \beta |{\bf k}| \ll 1$. We consider 
\begin{equation}\label{eq:Gpgrad}
G^+(\ctor, \omega, |{\bf k}|) =   \sum_{n,m=0}^\infty \, G^+_{m,n}(\ctor)\, \left(\frac{\beta \omega}{2}\right)^m\, \left(\frac{\beta |{\bf k}|}{2}\right)^n \,,
\end{equation}	
where we continue to work with the dimensionless frequency and momenta. We have isolated the leading order term in the gradient expansion with some hindsight to simplify the computations.

To determine the solution for the scalar field, we first solve the equation \eqref{eq:kgPhi} with  ingoing boundary conditions. This amounts to imposing the following boundary conditions on the Green's function $G^+$:
\begin{equation}\label{eq:ingbc}
G^+\bigg|_{r=r_c} =1\,, \qquad \dv{G^+}{\ctor}\bigg|_{r=r_h} =0\,.
\end{equation}	
For the series coefficients in the gradient expansion this translates to the requirement
\begin{equation}\label{eq:Gpbcs}
\begin{split}
G^+_{0,0}\bigg|_{r=r_c} =1\,, \qquad G^+_{n,m}\bigg|_{r=r_c} = \dv{G^+_{n,m}}{\ctor}\bigg|_{r=r_h} =0\,, \quad \forall\; n ,m \in \mathbb{Z}_+ \,.
\end{split}
\end{equation}	
Some of the coefficients above are trivial, spatial reflection symmetry sets $G^+_{n,2m+1} =0 $, so we do not encounter any terms which are odd in momenta.

It transpires that the knowledge of $G^+_{n,m}$ suffices to obtain the solution to the scalar wave equation in the average-difference basis in the long-wavelength gradient expansion. The final expression can be compactly summarized as
\begin{equation}\label{eq:Phigrad}
\begin{split}
 \Phi &=  \sum_{n,m=0}^\infty\, G^+_{n,2m} \left(\frac{\beta^2|{\bf k}|^2}{4}\right)^{m} (D_\ctor^+)^n \bigg\{J_a -\frac{1}{2}J_d \left[ (1-2\ctor) + \frac{\beta \omega}{2\times2!}\, \left((1-2\ctor)^2-1\right) \cdots \right] \bigg\}.
\end{split}
\end{equation}
An explicit derivation of the above as is outlined in \cref{sec:eogradient}, but the basic strategy is easy to describe. We essentially introduce the bulk analog of the retarded-advanced sources and carry out the gradient expansion both for the Green's function and for the statistical factors that enter in the construction \eqref{eq:advret}.  We then use the time-reflection symmetry \eqref{eq:trevA} to determine the outgoing Green's function, and employ \eqref{eq:phiad} to assemble the pieces to give the solution to the wave equation.

Closed form solutions for the leading order terms in the gradient expansion can be obtained (see \cref{sec:gradexpmass}). We present below of the basic data that we will use in the rest of the discussion. 
\begin{equation}\label{eq:Gpmngen}
\begin{split}
G^+_{0,0}  &= \frac{P_{-\frac{\Delta}{d}} \left(2\,\frac{r^d}{r_h^d}-1\right)}{P_{-\frac{\Delta}{d}} \left(2\,\frac{r_c^d}{r_h^d}-1\right)}\,,\\
G^+_{1,0}  &=    
	 -G_{0,0}^+ \, \int_0^{\ctor} \;  d\ctor'  \left(1 - \left(\frac{G_{0,0}^+(\ctor_h) }{G_{0,0}^+(\ctor') }\right)^2 \left(\frac{r_h}{r'}\right)^{d-1} \right) , \\
G^+_{0,1}  &= 0\,.
\end{split}
\end{equation}
Higher order terms can similarly be obtained and we give some explicit expressions in \cref{sec:gradexpmass}; see for instance \eqref{eq:Gpexpd} for a massless scalar in arbitrary dimensions and  \eqref{eq:Gpexp2} for an arbitrary scalar in $d=2$.

\section{Influence functionals}
\label{sec:influence}

We  now  have all the pieces necessary to compute the influence functionals of interest. We first outline the computation of the quadratic effective action  in \cref{sec:s2eff}. As is usual in AdS/CFT this is obtained as a boundary term in the on-shell action computation. For the higher order influence functionals we need to employ the standard Witten diagram technology on the grSK geometry (justified in \cref{sec:wittendia}). Armed with these results we compute the $n$-point contact influence functions in \cref{sec:sneff}. Along the way we will argue that the non-Gaussian contributions are well defined after a renormalization of the sources, and determine the appropriate counterterm action necessary to obtain the physical influence functionals. This will turn out to be crucial when our system couples to a marginal operator of the environment theory.  We will give some explicit results for cubic and quartic self-interactions in \cref{sec:sneff}.  
In \cref{sec:2dcubic,sec:counterterns} we compile various technical details underlying the results we present in this section.

\subsection{Quadratic effective action}
\label{sec:s2eff}

Let us begin with the evaluation of the quadratic part of the influence functional. Since we have solved for the field $\Phi$ on the grSK contour, it follows that result should be given by a boundary term. This is indeed the case, for starting with \eqref{eq:sphi4} with $\lambda =0$ we have upon passing to momentum space
\begin{equation}\label{eq:S2bdy}
\begin{split}
S_{(2)} &= 
	\frac{i}{\beta} \,\int \frac{d^{d}k_1}{(2\pi)^{d}}\int\frac{d^{d}k_2}{(2\pi)^{d}} \, (2\pi)^{d}\delta^{d}(k_1+k_2)  \\
 &\qquad 
 	\times\oint d\ctor\; r^{d-1} \left[ D_\ctor^+ \Phi(k_1)D_\ctor^+ \Phi(k_2) - \Phi(k_1) \frac{\beta^2}{4}\left(f\, |{\bf k_2}|^2 + r^2\, f\, m^2 -\omega_2^2 \right) \Phi(k_2)\right]\, \\
&= 
	\frac{i}{\beta} \int \frac{d^dk_1}{(2\pi)^{d}}\frac{d^{d}k_2}{(2\pi)^{d}}(2\pi)^{d} \delta^{d}(k_1+k_2)\int_{0}^1 d\ctor \frac{d}{d\ctor}\left[ r^{d-1} \Phi(k_1) D^+_\ctor \Phi(k_2) \right] \\
&= 
	\frac{i}{\beta} \int \frac{d^dk_1}{(2\pi)^{d}}\frac{d^{d}k_2}{(2\pi)^{d}}(2\pi)^{d} \delta^{d}(k_1+k_2)\,\bigg[ r^{d-1} \Phi(k_1) D^+_\ctor \Phi(k_2) \bigg]_{\ctor=0}^{\ctor=1}  \,.
\end{split}
\end{equation}
In the second line we substituted the equation of motion \eqref{eq:Dpkg} and performed the $\ctor$ contour integral in the final line, expressing the result as a pure boundary term on the grSK contour. 

The general structure of the answer from evaluating the boundary term takes the following form in the average-difference basis:
\begin{equation}\label{eq:}
S_{(2)} = 
	 \int \frac{d^dk}{(2\pi)^d} 
	\bigg[ \ifad_{ad}(\omega,k) \;  J_a(\omega,k) \, J_d(-\omega,-k) + \ifad_{dd}(\omega,k)\;   J_d(\omega,k) \, J_d(-\omega,-k)
	\bigg] .
\end{equation}	
We see that $\ifad_{aa} = 0$ a consequence of Schwinger-Keldysh unitarity of the microscopic theory. This is because the coefficient of the average source is the ingoing Green's function which is manifestly regular on the grSK contour. 

\paragraph{On-shell action in $d=2$:} 	Given our solution in the BTZ geometry it is straightforward to evaluate the boundary term. 
Using the normalized wavefunction  built from \eqref{eq:btzsol}, the boundary term contribution evaluates to 
\begin{equation}\label{eq:}
\begin{split}
\mathfrak{I}_{ad}(\omega,k) &= \frac{4 \pi^2}{\beta^2} \bigg\{
	  \bpt_+ + \bpt_- -\Delta \, \frac{r_c^2}{r_h^2}   \\
 & \qquad \quad \;\;+ \left(1-\frac{r_h^2}{r_c^2}\right)	\frac{ 2\,\bpt_+ \, \bpt_-\;  {}_2\widetilde{F}_1\left( 1+\bpt_+, 1+ \bpt_- , 2 - i\,  \frac{\beta\omega}{2\pi} \,; 1-\frac{r_h^2}{r_c^2} \right)}{ {}_2\widetilde{F}_1\left(\bpt_+, \bpt_- , 1 - i\,  \frac{\beta\omega}{2\pi} \,; 1-\frac{r_h^2}{r_c^2}  \right)} \bigg\}_{\text{cf} \;r_c^{4-2\Delta}} \!.
\end{split}
\end{equation}
where ${}_2\widetilde{F}_1(a,b,c,\xi)$ is the regularized hypergeometric function ${}_2\widetilde{F}_1(a,b,c,\xi) = \frac{1}{\Gamma(c)}\, {}_2F_1(a,b,c,\xi) $.

The subscript at the end instructs us to extract the coefficient of $r_c^{4-2\Delta}$, which is the end result of carrying out a counterterm subtraction using standard holographic renormalization methods.\footnote{  We assume, for simplicity,  that the field $\Phi$ satisfies standard (Dirichlet) boundary condition at infinity. While this restricts us to  $\Delta \geq \frac{d}{2}$, the result is unchanged for $\Delta \in (\frac{d}{2}-1,\frac{d}{2})$ once we include additional boundary terms to account for the alternate (Neumann) boundary conditions.} To understand this recall that we have normalized $G^+(\ctor_c,\omega,k) = 1$ which means that the two point function is obtained from the term that scales like $\left(\frac{1}{r_c}\right)^{2\Delta -4}$ leading to  the aforementioned prescription. A short calculation results in:
\begin{equation}\label{eq:Iad2}
\begin{split}
\ifad_{ad}(\omega,k)     
&= 
	2 \,r_h^{2\Delta-2} \,\frac{\Gamma(\bpt_+) \, \Gamma(\bpt_-) \, \Gamma(2-\Delta)}{\Gamma(\Delta -1) \, \Gamma(\bpt_+ +1 -\Delta ) \, \Gamma(\bpt_- +1 - \Delta)}\\
&=
	 -\frac{2}{\pi}\, r_h^{2\Delta-2} \frac{1}{\Gamma(\Delta-1)^2} \bigg| \Gamma\left( \bpt_+\right)\,\Gamma\left( \bpt_-\right)\bigg|^2 \, 
	 \frac{\sin\left(\pi (\bpt_- -  \Delta )\right)  \sin\left(\pi (\bpt_+ - \Delta )\right)}{\sin(\pi \Delta)} \\
&= 
	-\frac{1}{\pi}\, r_h^{2\Delta-2} \frac{1}{\Gamma(\Delta-1)^2} \bigg| \Gamma\left(\frac{\Delta}{2} - i\,\frac{\beta (k +\omega)}{4\pi}\right)\,\Gamma\left(\frac{\Delta}{2} + i\,\frac{\beta(k-\omega)}{4\pi}\right)\bigg|^2 \\
& 
	\qquad \times  \left( \csc\left(\pi \Delta\right)\, \cosh\left( \frac{\beta k}	{2}\right) - \cot\left(\pi\Delta\right) \, \cosh(\frac{\beta\omega}{2}) 
	+i\, \sinh\left(\frac{\beta\omega}{2}\right)\right).
\end{split}
\end{equation}

This result is indeed the correct expression for the retarded Green's function  $G_R(\omega,k)$ for a 2d CFT on the infinite line.   One can obtain it directly by starting from the conformal 2-point function on the plane, conformally mapping it to the cylinder to obtain the (Euclidean) thermal correlator, and thence take the discontinuity across the lightcone branch cut while analytically continuing it to the timelike Lorentzian domain (using appropriate $i\epsilon$ prescription to do so).  The result has been obtained in various places in the literature before. In \cite{Gubser:1997cm} the computation of the Fourier transform was described and the imaginary part of the Green's function obtained (in the context of computing the absorption cross-section of D-branes). The first, principled, holographic derivation of the Green's function was given in \cite{Son:2002sd}.

The retarded Green's function $G_R(\omega,k)$ has poles in the lower half  complex $\omega$ plane, at 
\begin{equation}\label{eq:}
\omega_\text{qn} = \pm  k\,- 2\pi i \,T\, \left(\Delta + 2n\right) \,, \qquad n \in \mathbb{Z}_{\geq 0}\,.
\end{equation}	
These poles of course correspond to the BTZ quasinormal modes and set the characteristic scale for the decay of the response function in the time domain \cite{Horowitz:1999jd,Birmingham:2001pj}.  Finally, we also note that the expression can be written in a form that is symmetric between the operator $\mathcal{O}$ of dimension $\Delta$  and its shadow  $\mathcal{O}_s$ of dimension $\widetilde{\Delta} =d-\Delta$ (with $d=2$ here). To make  the operator dimension's contribution to $\bpt_\pm $  defined in \eqref{eq:ppm} explicit we redefine the light-cone momenta:
\begin{equation}\label{eq:lpmdef}
\bkt_+ = i\,\frac{\beta}{4\pi} \,(k-\omega)= \bpt_+ -\frac{\Delta}{2} \,,
\qquad 
\bkt_- = -i\frac{\beta}{4\pi}\,(k+\omega) = \bpt_- -\frac{\Delta}{2} \,.
\end{equation}	
We can then write the 2-point influence functional in terms of the function  
\begin{equation}\label{eq:Gfn2pdef}
 \mathfrak{G}(\bkt_+,\bkt_-,\Delta)  \equiv  \Gamma(\bkt_+  +\tfrac{\Delta}{2}) \, \Gamma(\bkt_- +\tfrac{\Delta}{2}) \,\Gamma(1-\Delta) \,,
\end{equation}	
  as
\begin{equation}\label{eq:}
\ifad_{ad}(\omega, k) = \frac{2}{2-\Delta} \,  r_h^{2\Delta-2} \, \frac{\mathfrak{G}(\bkt_+,\bkt_-,\Delta)}{\mathfrak{G}(\bkt_+,\bkt_-, \widetilde{\Delta}) } \,.
\end{equation}	
We will find this notation useful in simplifying the analysis of the 3-point influence functional.

Having understood the computation of the influence functional $\ifad_{ad}$ we next can compute $\ifad_{dd}$. The computation proceeds along similar fashion lines and we obtain 
\begin{equation}\label{eq:Idd2}
\begin{split}
\mathfrak{I}_{dd}(\omega, k) 
&= 
	\frac{i}{4} \, \frac{\cos\left(\pi (\bpt_+ + \bpt_- - \Delta)\right) \, \sin\left(\pi\Delta\right)}{\sin\left(\pi (\bpt_- -  \Delta )\right)  \sin\left(\pi (\bpt_+ - \Delta )\right) } \, \mathfrak{I}_{ad}(\omega, k)  \\
&= 
	 \frac{r_h^{2\Delta-2} }{2\pi i}\,  \frac{\cosh\left(\frac{\beta\omega}{2}\right)}{\Gamma(\Delta-1)^2} 
	 \bigg| \Gamma\left(\frac{\Delta}{2} - i\,\frac{\beta (k +\omega)}{4\pi}\right)\,\Gamma\left(\frac{\Delta}{2} + i\,\frac{\beta(k-\omega)}{4\pi}\right)\bigg|^2 \\
& = \frac{i}{2} \coth\left(\frac{\beta\omega}{2}\right)\, 	\Im\left( \mathfrak{I}_{ad}(\omega,k) \right) .
\end{split}
\end{equation}	
We have expressed the final result making manifest the fluctuation/dissipation relation. We recall that  $\Im(\ifad_{ad})$ gives us the spectral function at finite temperature, see \cite{Chou:1984es,Haehl:2017eob} for further details.

\paragraph{On-shell action in gradient expansion:} The analysis of the influence functionals in a gradient expansion is straightforward given our explicit expressions from before. It is however convenient in actuality to assemble the pieces somewhat differently and work in a basis that is better adapted to the bulk field $\Phi$. We describe such a basis built from the even and odd parts of the ingoing Green's function in \cref{sec:eogradient}. The expressions of interest are the solution to the wave equation \eqref{eq:Phiev},  the field radial gradient \eqref{eq:Dphiev}, and the expansion of the even and odd sources \eqref{eq:Jeoexp}.  To compute the quadratic influence functional, we need to evaluate the on-shell action as a boundary term \eqref{eq:S2bdy}. This amounts to knowledge of the values of $G_{n,m}^+$ and their derivatives at $\ctor =0$. Explicitly,  letting 
\begin{equation}\label{eq:}
G^+_{n,m}(0) = 	\begin{cases}
	& 1 \,,\; n=m=0\\
	& 0 \,, (n,m) \neq (0,0)
	\end{cases}
\,, \qquad \dv{G^+_{n,m}}{\ctor}\bigg|_{\ctor = 0 } =\dot{g}_{n,m}\,,
\end{equation}	
the quadratic influence functional can be simplified to the form
\begin{equation}\label{eq:Sos2}
\begin{split}
\ifad_{ad}(\omega,{\bf k}) 
 &= 	\frac{i}{\beta} \bigg\{ r_c^{d-1} \, 
 	\left[(1+\dot{g}_{1,0} ) \beta \omega 	+ \frac{(\beta |{\bf k}|)^2}{2} \, \dot{g}_{0,2} + \frac{(\beta\omega)^2}{2} \, \dot{g}_{2,0} \right] \bigg\}_{\text{cf} \ r_c^{2(d-\Delta)}} \,,\\
\ifad_{dd}(\omega,{\bf k}) 
 &= 	\frac{i}{\beta} \bigg\{ r_c^{d-1} \, 
 	\left[(1+\dot{g}_{1,0} )	+ \frac{(\beta\omega)^2}{12} \left( 1+ \dot{g}_{1,0}  +3 \,\dot{g}_{3,0} \right) +  \frac{(\beta|{\bf k}|)^2}{4} \, \dot{g}_{1,2}   \right] \bigg\}_{\text{cf} \ r_c^{2(d-\Delta)}} 	\,.
\end{split}
\end{equation}

We  have been able to evaluate the expressions analytically to leading order in  the gradient expansion. We recall that we have normalized $G_{0,0}^+(0) =1$ and use the solution for $G_{1,0}^+$ given  in \eqref{eq:Gpmngen}, to learn that 
\begin{equation}\label{eq:}
r_c^{d-1} \left(1+ \dot{g}_{1,0}\right) = r_h^{d-1}\,  \left[\frac{P_{-\frac{\Delta}{d}} \left( 1\right)}{P_{-\frac{\Delta}{d}} \left(2\,\frac{r_c^d}{r_h^d}-1\right)} \right]^{\!2} \,.
\end{equation}	
Expanding out the Legendre polynomial we pick up the a coefficient of the desired power of $r_c$, and find the influence functionals to be 
\begin{equation}
\begin{split}
\ifad_{ad}(\omega, {\bf k}) & =i\, \frac{\Gamma\!\left(\frac{\Delta}{d}\right)^4}{\Gamma\!\left(2\frac{\Delta}{d}-1\right)^2} 
\;  r_h^{2\Delta -d-1}\, \omega \,,\\   
\ifad_{dd}(\omega, {\bf k}) &= \frac{i}{\beta}\,  \frac{\Gamma\!\left(\frac{\Delta}{d}\right)^4}{\Gamma\!\left(2\frac{\Delta}{d}-1\right)^2} 
\;  r_h^{2\Delta -d-1}\,.
\end{split}
\label{eq:Infd}
\end{equation}

While we are retaining terms only to leading order in the gradient expansion one can nevertheless see that the linearized version of the fluctuation dissipation relation \eqref{eq:Idd2} continues to hold quite generally (as is in fact readily inferred from \eqref{eq:Sos2}).  

The subleading terms in the influence functional $\ifad_{ad}$ are computable. While they do not seem to be amendable to a closed form  analytic expression, we can easily extract the dependence on the physical parameters.  Using the results for the higher order terms in the gradient expansion given in \cref{sec:gradexpmass} we can show that the quadratic order terms in $\ifad_{ad}$ are given by the following:
\begin{equation}\label{eq:g0220}
\begin{split}
\bigg\{\frac{i}{\beta} \, r_c^{d-1} \, \dot{g}_{0,2} \bigg\}_{\text{cf} \ r_c^{2(d-\Delta)} }
&= 
	-\frac{1}{2\pi\beta}\, \frac{\Gamma\!\left(\frac{\Delta}{d}\right)^4}{\Gamma\!\left(2\frac{\Delta}{d}-1\right)^2} \, r_h^{2\Delta -d-1}  \, \mathfrak{g}_{0,2} \,, \\
\bigg\{\frac{i}{\beta} \, r_c^{d-1} \, \dot{g}_{2,0} \bigg\}_{\text{cf} \ r_c^{2(d-\Delta)} }
&= 
	\frac{1}{2\pi\beta}\, \frac{\Gamma\!\left(\frac{\Delta}{d}\right)^4}{\Gamma\!\left(2\frac{\Delta}{d}-1\right)^2} \, r_h^{2\Delta -d-1}   \, \mathfrak{g}_{2,0} \,.
\end{split}
\end{equation}
Here $\mathfrak{g}_{0,2}$ and $\mathfrak{g}_{2,0}$ are purely numerical coefficients and computed by the integral expressions involving 
Legendre functions; see \eqref{eq:g02dot} and \eqref{eq:g20dot}, respectively. Specifically, they are  
\begin{equation}\label{eq:g0220num}
\begin{split}
\mathfrak{g}_{0,2} 
&= 
	\bigg[ \int_1^{\varrho_c} \, d\varrho \, \varrho^{-\frac{2}{d}} \left(P_{-\frac{\Delta}{d}} (2\varrho-1)\right)^2\bigg]_\text{finite} \\
\mathfrak{g}_{2,0} 
&= 2\pi i\, \frac{G^+_{1,0}(r_h)}{G^+_{0,0}(r_h)} +  \, \mathfrak{g}_{0,2}  +\,
	\bigg[ \int_1^{\varrho_c} \, d\varrho \, \frac{\varrho^{-\frac{1}{d}}}{\varrho-1} \left(\varrho^{-\frac{1}{d}} \, P_{-\frac{\Delta}{d}} (2\varrho-1)^2 -1\right)\bigg]_\text{finite} .
\end{split}
\end{equation}	
We have used  \eqref{eq:G10sol} and expressed the answer as integrals over Legendre functions which can be evaluated numerically. 
For $\Delta \in \left(\frac{d}{2}, \frac{d}{2} +1\right)$ the integrals are absolutely convergent and thus may be determined without need for a detailed counterterm analysis. Representative data for these quantities as a function of  $\Delta$ in various dimensions is plotted in \cref{fig:g0220plots} within this window of conformal dimensions. 
We will exploit this structural form when writing down the effective action for the open quantum degree of freedom in \cref{sec:stochastic}.
\begin{figure}[h]
\centering
\includegraphics[width=.4\textwidth]{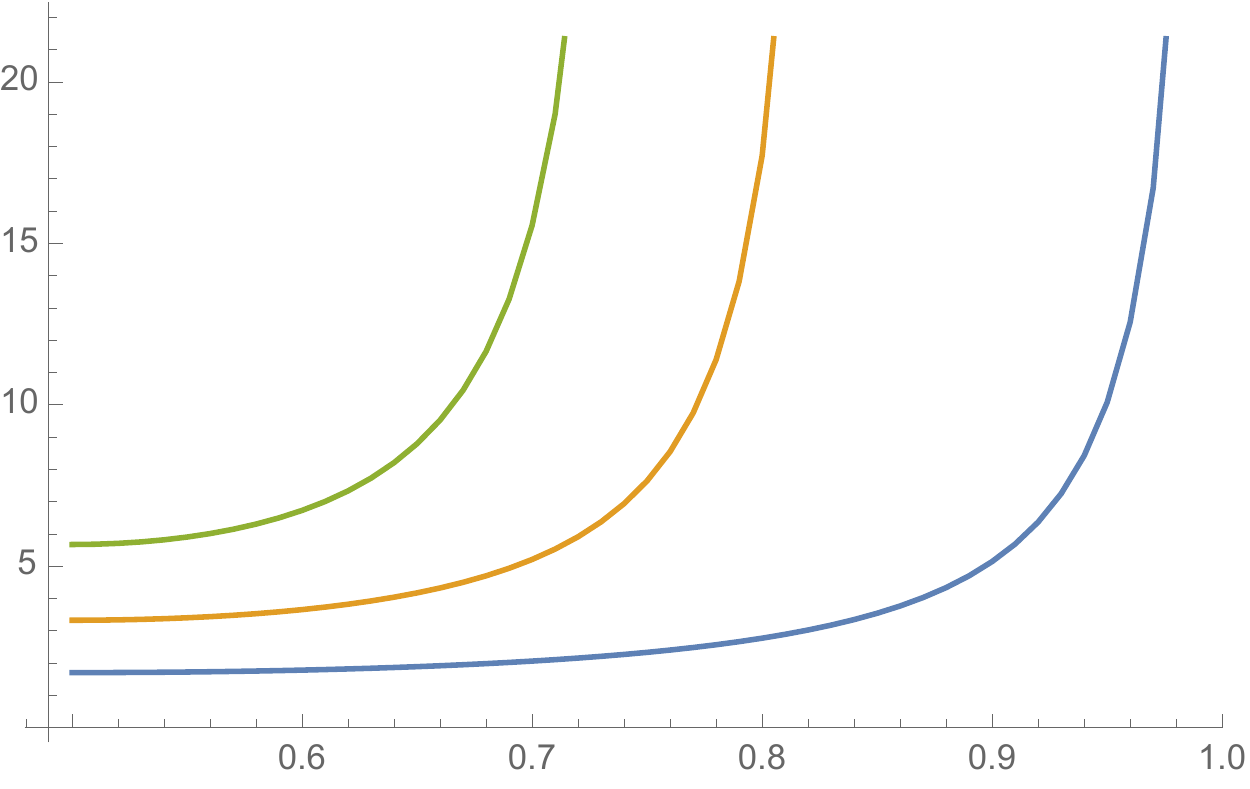}
\hspace{1cm}
\includegraphics[width=.4\textwidth]{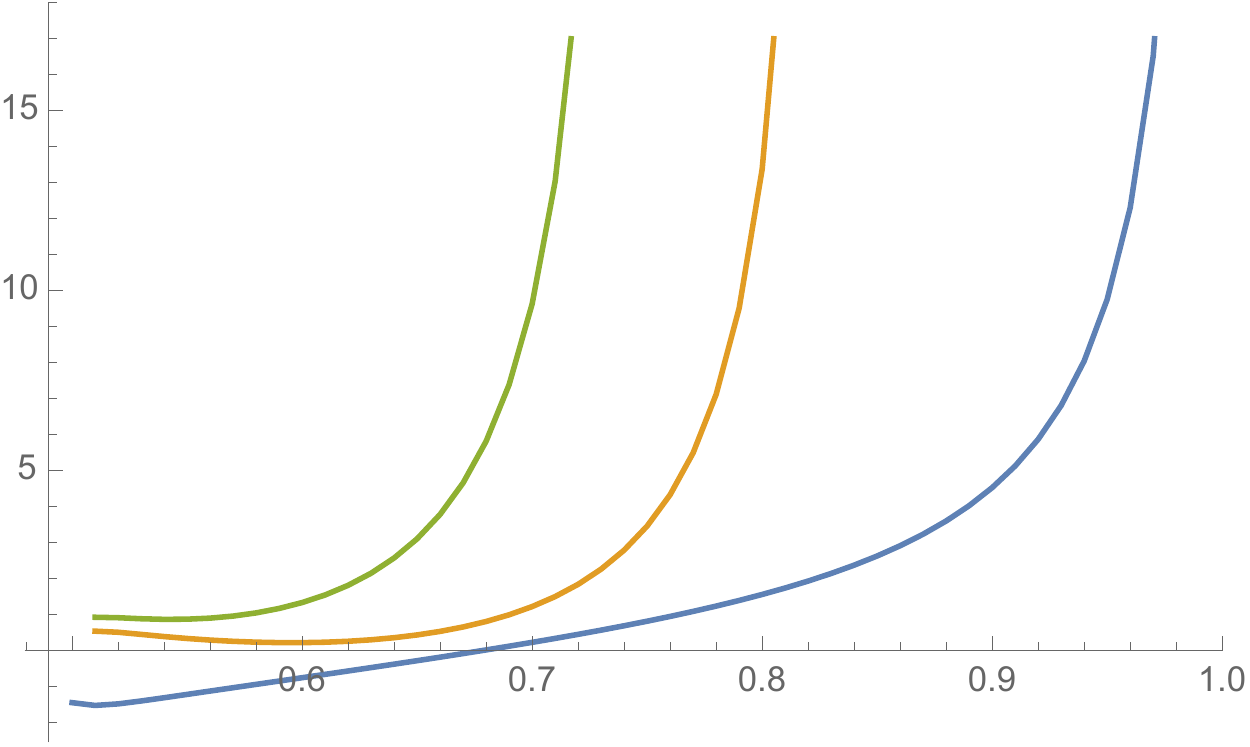}
\begin{picture}(0,0)
\setlength{\unitlength}{1cm}
\put (-0.1,-0.2) {$\frac{\Delta}{d}$}
\put (-7.6,-0.2) {$\frac{\Delta}{d}$}
\put (-14.6,3.2) {$\mathfrak{g}_{0,2}$}
\put (-7.0,3.2) {$\mathfrak{g}_{2,0}$}
\put (-9.0,1.5) {$\scriptscriptstyle{d=2}$}
\put (-11.2,1.5) {$\scriptscriptstyle{d=3}$}
\put (-12.8,1.5) {$\scriptscriptstyle{d=4}$}
\put (-4.3,1.5) {$\scriptscriptstyle{d=4}$}
\put (-3.2,1.5) {$\scriptscriptstyle{d=3}$}
\put (-1.2,1.5) {$\scriptscriptstyle{d=2}$}
\end{picture}
\caption{The numerical values of the quantities $\mathfrak{g}_{0,2}$ and $i\, \mathfrak{g}_{2,0}$ as a function of the conformal dimension $\Delta$ in dimensions $2$, $3$, and $4$, respectively. We have confined attention to the case of relevant operators $\Delta \in (\frac{d}{2},\frac{d}{2}+1)$ when the integrals are convergent without need of additional counterterms.}
\label{fig:g0220plots}
\end{figure}
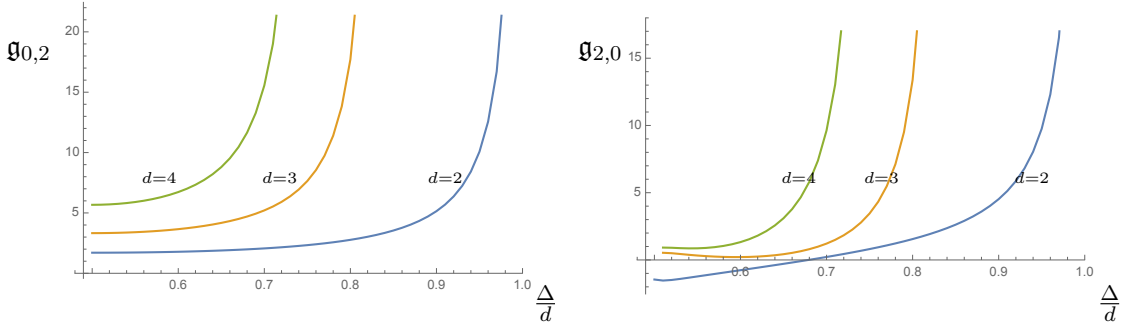
%

\subsection{Interactions: contact self-interaction}
\label{sec:sneff}

Using the standard Witten diagrams on the grSK contour, illustrated in \cref{fig:grskW} (we motivate this briefly in \cref{sec:wittendia}), the influence functionals can be straightforwardly  written down. A contact $n$-point  self-interaction vertex in the bulk leads to the following contribution to the influence functional:
\begin{equation}
\begin{split}
S_{(n)}  =  -\frac{\lambda_n}{n!} \int \prod_{i=1}^n \frac{d^dk_i}{(2\pi)^d} (2\pi)^{d} \delta\left(\sum_{i=1}^{n}k_i\right) \oint d\ctor \, \sqrt{-g}\, \, \prod_{i=1}^n \Phi(\ctor,k_i) 
\end{split}
\end{equation}

We can in general simplify expressions such as the above by using the fact that the contour integral over the mock tortoise coordinate can be done by basically integrating the discontinuity across the branch cut extending from the horizon over the radial coordinate. To wit, for any function on the grSK geometry $\mathfrak{L}(\ctor) $
\begin{equation}\label{eq:skcontourI}
\begin{split}
\oint d\ctor \, \sqrt{-g} \; \mathfrak{L}(\ctor)  
&= 
	\int_{\ctor_h}^{\ctor_c} \, d\ctor \, \sqrt{-g} \,\bigg( \mathfrak{L}(\ctor + 1 ) - \mathfrak{L}(\ctor)\bigg) \\
& = 
	\int_{r_h}^{r_c} \, dr \, r^{d-1} \,\bigg( \mathfrak{L}(\ctor(r) + 1 ) - \mathfrak{L}(\ctor(r))\bigg) \,,
\end{split}
\end{equation}	
where we used $\sqrt{-g}\, \dv{\ctor}{r} = r^{d-1}$. In the second line above, we have assumed that the integrand doesn't have a simple pole at $r=r_h$ and hence the horizon itself gives no contribution to the integral. Consequently, entire contribution arises
from the discontinuity across the branch cut that extends  from the horizon to the conformal boundary.

\begin{figure}[h]
\centering
\includegraphics[width=.37\textwidth]{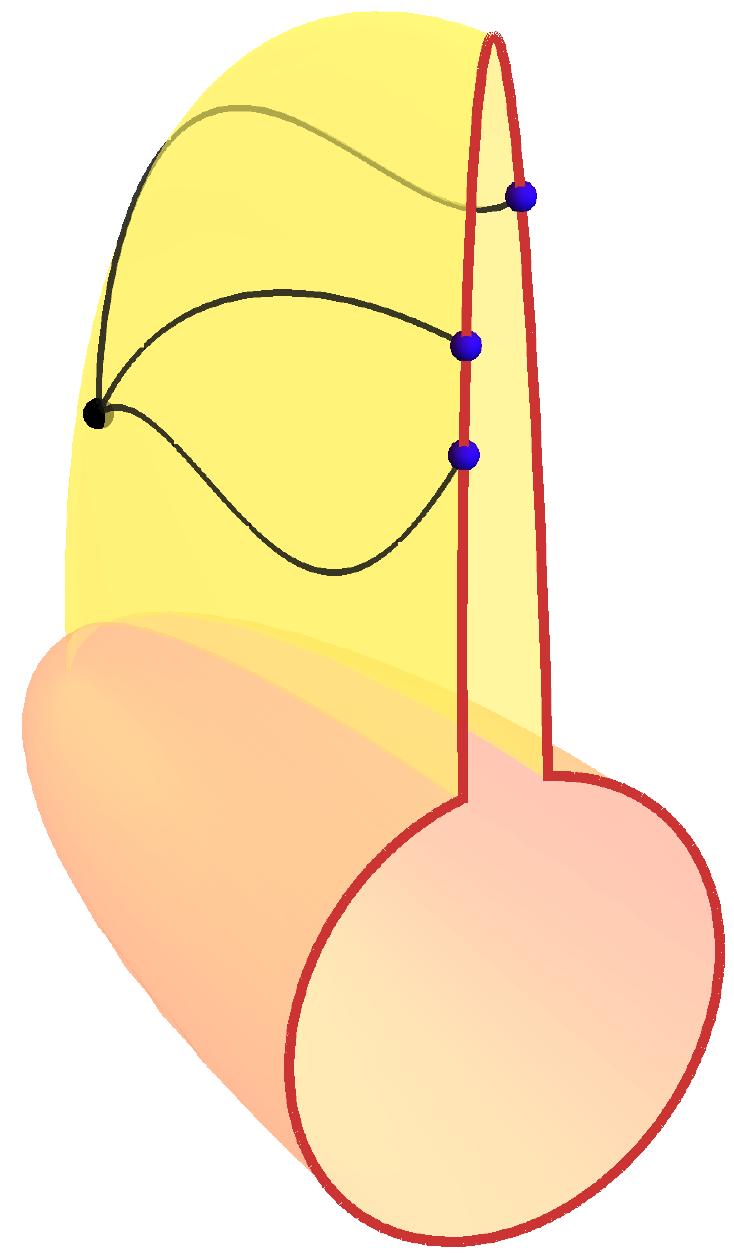}
\hspace{1cm}
\includegraphics[width=.4\textwidth]{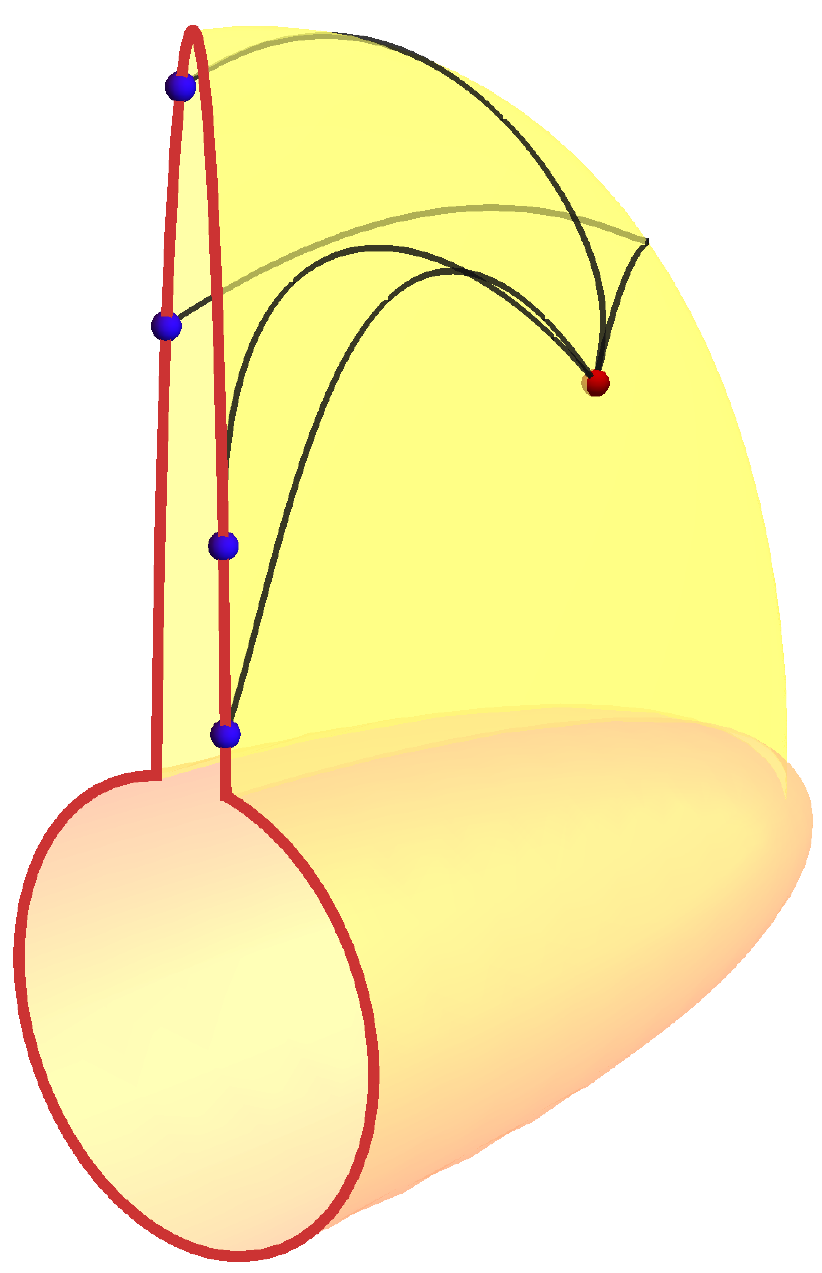}
\caption{Illustration of Witten diagrams on the grSK geometry computing 3 and 4-point influence functions of the boundary field theory. The boundary operator insertions (blue) lie on the thermal SK contour. The bulk field is constructed using the boundary-bulk propagators, and the bulk vertex is integrated over the Lorentzian section of the grSK geometry.  }
\label{fig:grskW}
\end{figure}
%
\subsubsection{Influence functionals in the advanced-retarded basis}
\label{sec:iffpn}

Using the explicit solution on the grSK contour \eqref{eq:phiadvret} we identify from \eqref{eq:Sgen} the influence functionals in the retarded-advanced basis to be 
\begin{align}\label{eq:FPinfgen}
&\ifra_{F\cdots FP\cdots P}(k_1,\cdots,k_n) 
= \text{coeff} \left(\JF(k_1) \cdots \JF(k_p) \JP(k_{p+1}) \cdots \JP(k_n) \right) \nonumber \\
& = 
	-\frac{\lambda_n}{p! (n-p)!} \, \oint d\ctor\, \sqrt{-g}\, (-)^p \, \prod_{i=1}^p  \, G^+(\zeta, \omega_i, {\bf k}_i) \prod_{j=p+1}^{n}\, 
	G^-(\zeta, \omega_j, {\bf k}_j)  \,e^{\beta \omega_j  (1-\ctor)} \\
&= 
	\frac{\lambda_n\,(-)^{p+1} \, }{p!(n-p)!} \, \left(1-e^{\beta \sum_{j=p+1}^n \,\omega_j}\right)\,	 \int_{r_h}^{r_c} dr\, r^{d-1} \, 
	 \prod_{i=1}^p  \, G^+(\zeta, \omega_i, {\bf k}_i) \prod_{j=p+1}^{n}\, e^{-\beta\omega_j \ctor}
	G^+(\zeta, -\omega_j, {\bf k}_j) \,. \nonumber
\end{align}
One can evaluate the integrals directly given the solution to the Green's function on the grSK contour. Note that the causal structure of the influence function is completely manifest. The ingoing Green's function $G^+(\ctor, \omega, {\bf k})$ is manifestly regular in the upper-half $\omega$ plane, while its conjugate $G^-(\ctor, \omega, {\bf k})$ is regular on the lower-half plane by frequency reversal. The time-reversal also correctly incorporates the statistical factor which enters the final expression.

In the holographic setting, we expect the only singularities in $G^+(\ctor, \omega, {\bf k})$ to arise from quasinormal poles, while those of 
$G^-(\ctor, \omega, {\bf k})$ are attributable to the time-reversed anti-quasinormal poles. So the influence functional $\ifra_{F\cdots FP\cdots P}(k_1,\cdots,k_n) $ will have quasinormal poles for all the $\omega_i$ with $i \in F$ and anti-quasinormal poles for $\omega_k$ with $k \in P$. We shall see an explicit illustration of this below.

The radial integral in the expression above, in general needs to be regulated. However, for a certain range of operator dimensions, specifically, for $\Delta < (1-\frac{1}{n}) \, d$ the integrals are absolutely convergent and need no regulating.\footnote{  One can  infer this by examining the asymptotic behaviour of the Green's functions. However, it is also  worth noting that the UV divergences are determined by the  CFT correlation functions in the vacuum, for which the criterion specified is well known, see eg., \cite{Bzowski:2015pba}. } In the discussion below we will focus on the case where the operator coupling to our system is sufficiently relevant to avoid introducing any regulators.

As an illustration consider the result for the 3-point influence function in a 2d CFT  which can be obtained to be (using energy-momentum conservation to eliminate the $\omega_3$ and ${\bf k}_3$)
\begin{equation}\label{eq:2dffp}
\begin{split}
\ifra_{FFP}(k_1,k_2,k_3) &= 
	- \lambda\, r_h^{3\Delta -4}\, 
	\left(1-e^{-\beta \omega_3}\right) 
	\Gamma\left(1 +\tfrac{i\,\bwt_3}{\pi}\right)\, \sum_{\delta_i \in \{\Delta , \widetilde{\Delta}\}} \mathfrak{J}^\delta_{FFP}(k_1,k_2,k_3) 
\end{split}
\end{equation}
where the function $\mathfrak{J}^\delta_{FFP}$ is given as an infinite sum involving generalized hypergeometric functions:
\begin{equation}\label{eq:2dffpa}
\begin{split}
\mathfrak{J}_{FFP}^\delta(k_1,k_2,k_3) 
&=	
	\frac{\mathfrak{G}(\bkt_{1+}, \bkt_{1-},\delta_1)\, \mathfrak{G}(\bkt_{+2}, \bkt_{1-},\delta_2)}{\mathfrak{G}(\bkt_{1+},\bkt_{1-},\widetilde{\Delta}) \, \mathfrak{G}(\bkt_{2+},\bkt_{2-},\widetilde{\Delta})} 
	\Bigg[\sum_{n=0}^\infty \, 
	 \frac{(-)^n\, \Gamma(1-\delta_3-n)}{\Gamma(n+1)\,\Gamma(1-\delta_3-2n)}
	 \\
& 
	\qquad\, \times  \frac{\mathfrak{G}(\bkt_{3+}^*,\bkt_{3-}^*, \delta_3+2n)}{\mathfrak{G}(\bkt_{3+}^*,\bkt_{3-}^*,\widetilde{\Delta})} \,
	\frac{\Gamma(-1+n+ \tfrac{\delta_1+\delta_2+\delta_3}{2})}{\Gamma(n+ \frac{\delta_1+\delta_2+\delta_3}{2} +\tfrac{i\, \bwt_3}{\pi})}\\
&	\qquad  \times \ {}_3F_2\left(
		\begin{array}{c}
		\bkt_{1+} + \tfrac{\delta_1}{2}\,, \bkt_{1-} + \tfrac{\delta_1}{2}\,, -1+ \tfrac{\delta_1+\delta_2+\delta_3}{2}+n \\
		 \delta_1\,, \tfrac{\delta_1+\delta_2+\delta_3}{2}+n + \tfrac{i\,\bwt_3}{\pi}
		\end{array}\;; 1
	\right) \\
&	\qquad  \times 
	\ {}_4F_3\left(
		\begin{array}{c}
		 \bkt_{2+} + \tfrac{\delta_2}{2}\,, \bkt_{2-} + \tfrac{\delta_2}{2}\,, -n\,,1-n- \delta_3  \\
		 \delta_2\,,  1-\bkt_{3+}^*-\tfrac{\delta_3}{2} - n \,,  1-\bkt_{3-}^*-\tfrac{\delta_3}{2} - n 
		\end{array}\;; 1
	\right) \Bigg] .
\end{split}
\end{equation}
This result is valid for $1< \Delta < \frac{4}{3}$, chosen so that the radial integral was absolutely convergent. 
We have employed a notational contrivance to keep the expression simple -- the above is actually a sum over eight terms with similar structure, where each of the three operators enters either as itself, or its shadow. This is indicated by using $\delta$ to sample the operator and shadow operator dimension as indicated in the summation. The lightcone momentum of the outgoing mode is conjugated because of time-reversal. We can use energy-momentum conservation to rewrite $\bkt_{3\pm}^* =   \bkt_{12\pm} $ as a function of $\omega_1+ \omega_2$ and $k_1+k_2$.  Finally, we have chosen to write the final expression as a infinite sum, but one can equivalently have presented the result as set of contour integrals (which is where the sum is obtained from).  We give a detailed derivation of this result in  \cref{sec:2dcubic}.  

Let us focus on some of the features that the readily visible from the above parameterization of the influence functional. Firstly, the result is factorized into right and left movers. This is not manifest in when we consider the AdS radial integral representation, but is expected on grounds that the Euclidean thermal 3-point correlator factorizes into holomorphic and anti-holomorphic parts. Secondly, the influence functional, correctly captures the causality requirements noted above. 
The only singularities which can arise in the complex frequency plane are from the Gamma functions for $\omega_1$ and $\omega_2$. One can also check that the generalized hypergeometric functions do not contribute any singularities (this is easier to see directly from the infinite sum or contour integral representation \eqref{eq:2dffpsum}).

 Of the eight possible choices of $\delta_i$ we note that there can only be poles when $\delta_i  = \Delta$. When $\delta_i = 2-\Delta$ the Gamma function in the denominator also a pole which cancels out the behaviour of the numerator leaving a finite answer. We conclude that the correlator is analytic in the upper half of the complex $\omega_1$ and $\omega_2$ planes and encounters the usual quasinormal type poles in the lower half-planes. So the influence functional only has simple poles at  are at the following locations determined by the quasinormal modes.
\begin{equation}\label{eq:ffppoles}
\begin{split}
\omega_1 = \pm k_1 - 2\pi i\, T (2n_1&+ 2m_1 + \Delta) \,, \qquad \omega_2 = \pm k_2 - 2\pi i\, T (2n_2+ 2m_2 + \Delta) \,,\\ 
\omega_1+\omega_2 &= \pm (k_1+k_2) - 2\pi i\, T (2n_3+ 2m_3 + \Delta)\,.
\end{split}
\end{equation}	
The last set translates upon using energy-momentum conservation to the anti-quasinormal modes of the advanced operator with frequency $\omega_3$. 

We note in closing that one can give a closed form expression for the thermal real-time 3-point functions in 2d CFT in terms of Meijer G-functions. This has been derived by Fourier transforming the cylinder correlator of a 2d CFT to momentum space with an appropriate $i\epsilon$ prescription in \cite{Becker:2014jla}.\footnote{ We thank Sean Colin-Ellerin for alerting us to this result.} We have not attempted to simplify our expression \eqref{eq:2dffp} to this form, though we note that the analytic structure does match between the two expressions (as it must).

\subsubsection{Influence functionals in the average-difference basis}
\label{sec:ifadn}

If we wish to evaluate the influence functionals in the average-difference basis we can simply implement the basis transformation to go from  \eqref{eq:advret} to \eqref{eq:avdif}. While working in the long-wavelength gradient expansion however, one must also account for the statistical factor, which has been folded into \eqref{eq:Phigrad}. We will now use this expression to evaluate the influence functionals in the average-difference basis.

As noted above, depending on the nature of the operator $\mathcal{O}$ we may need to incorporate counterterms to evaluate the influence functional which enter into the effective field theory of open system degrees of freedom. For the influence functionals to respect the microscopic unitarity, these counterterms must be both state independent and suitably factorized across the SK contour. We see that this is indeed possible, provided that we suitably renormalize the sources for the holographic operator $\mathcal{O}$ obtain finite correlation functions. Operationally, from our discussion in \cref{sec:holopen} this means that we must renormalize our system degrees of freedom while constructing the open effective field theory.\footnote{  We will continue here to write the expressions in terms of the boundary sources $J$ of the holographic environment, leaving implicit the identification $\Psi_a \sim J_a$ and $\Psi_d \sim J_d$. }

We will first describe the structure of the divergences encountered in the computation of the influence functionals themselves. These will appear from the bulk calculation in UV divergent terms from the radial integral, scaling with our radial cut-off $r_c$.  We confine our attention to the leading order terms in the gradient expansion, so the results will be accurate to $\order{\beta \omega}$, for convenience. A more thorough analysis at arbitrary orders in the gradient expansion can be carried out along similar lines, and the details will appear elsewhere. 

To begin with let us make the following two assertions about the non-linear influence functional. 
\begin{enumerate}
\item  First, the anharmonic influence phase to linear order in the coupling $\lambda_n$  takes the form:
\begin{equation}
\begin{split}\label{eq:BareInf}
S_{(n)}^{\text{bare}}
&= 
	 -  \lambda_n \int d^dx  \sum_{k=1}^n\frac{1}{(n-k)!}\left(\Jb_a+\frac{i}{8}\beta\partial_t \Jb_d\right)^{n-k} \\
& \qquad \qquad  
	\times \left[\ifIb_{n,k} \frac{( \Jb_d)^k}{k!}
-\ifIb_{n,k+1} \frac{( \Jb_d)^{k-1}}{(k-1)!}\frac{i}{2} \beta\partial_t \Jb_d\right]\ .  
\end{split}
\end{equation}
 Here $\ifIb_{n,k}$ denotes the integral
\begin{equation}\label{eq:ifIk0}
\begin{split}
\ifIb_{n,k}  \equiv \oint d\ctor \, \sqrt{-g}\, \left(G^+_{0,0}\right)^n \left( \ctor-\frac{1}{2} + \frac{G^+_{1,0}}{ G^+_{0,0}} \right)^k ,
\end{split}
\end{equation}
where $G^+_{m,n}$ to the desired order are given by the expressions in \eqref{eq:Gpmngen}. We have anticipated the need for regulating the sources and the integrals and used the superscript ${\sf b}$ to denote that the above expressions are the bare results, prior to any renormalization prescription. 
\item Second, the integrals  $\ifIb_{n,k} $ appearing in the computation could be  divergent. This depends on the conformal dimension $\Delta$  of the operator $\mathcal{O}$. We find:
\begin{itemize}
\item For $\Delta < (1-\frac{1}{n}) d$ the  integrals involved in evaluating $\ifIb_{n,k}$ are convergent as noted in \cref{sec:iffpn}.
\item For relevant operators with $\Delta \in [(1-\frac{1}{n}) d,d)$ we need to regulate some of the integrals. In particular, the integrals $\ifIb_{n,2k+1}$  with an odd argument are divergent. We can estimate them to behave as follows:
\begin{equation}\label{eq:}
\ifIb_{n,2k+1}  = \ifIr_{2k+1}+\frac{\Lambda_\Delta}{4^k} \,, \qquad \Lambda_\Delta =  \frac{r_c^{n\Delta - (n-1) d }}{n \Delta - (n-1) \,d} \,, 
\end{equation}	
where  $\ifIr_{n,k}$ denote the \emph{renomalized} integrals which are completely finite and well-defined as $r_c\to \infty$. 
\item Finally, for a marginal operator $\Delta =d$ all the integrals are divergent. The functions $\ifIb_{n,2k+1} $ are power-law divergent, but now for  even argument $\ifIb_{n,2k}$ one encounters a logarithmic divergence.
\begin{equation}
\begin{split}
\ifIb_{n,2k+1}  = \ifIr_{2k+1}+\frac{\Lambda_d}{4^k}\ ,\qquad  \ifIb_{n,2k}  = \ifIr_{2k}+k\frac{\Lambda_l}{4^{k-1}}\,.
\end{split}
\end{equation}
The divergences $\Lambda_d $ and $\Lambda_l$ are given by 
\begin{equation}\label{eq:Lambdam0}
\begin{split}
\Lambda_d\equiv \frac{r_c^d}{d} ,\qquad  \Lambda_l\equiv  \frac{i}{\pi}\, r_h^d \log\frac{r_c}{r_h}  \ .
\end{split}
\end{equation}
\end{itemize}
\end{enumerate}
We prove the aforementioned assertions regarding the structure of the bare influence functional and the divergences of the integrals appearing therein in  \cref{sec:counterterns}.

Armed with these statements we can now assert the following result: define the \emph{renomalized} probes $\Jr_a$ and $\Jr_d$ via
\begin{equation}
\begin{aligned}
\Jb_a &\equiv \Jr_a \, &\quad  &  \Jb_d \equiv  \Jr_d   \,,  &\quad &  \frac{d}{2} \leq \Delta < d \,,\\ 
\Jb_a &\equiv \Jr_a -\frac{\Lambda_l}{2\Lambda_d} \, \Jr_d \ ,  & \quad&  \Jb_d \equiv  \Jr_d +\frac{\Lambda_l}{2\Lambda_d} \, 
i\, \beta\partial_t \Jr_d   \,, &\quad & \Delta = d .
\end{aligned}
\end{equation}
While the integrals are divergent for some range of relevant operators, we will argue below that the divergences can be canceled by a standard counterterm. However, we see that there is a need for a non-trivial renormalization of the sources in the case of a marginal operator, arising primarily from the logarithmic divergence encountered in $\ifIb_{n,2k}$. Note that as $r_c\to \infty$ we have  $\lim_{r_c\to\infty}\frac{\Lambda_l}{\Lambda_d}=0$, i.e., the bare and the renormalized sources agree when the cutoff is removed. At finite cutoff however we need to renormalize the sources slightly; this is achieved by mixing them with difference probes in a temperature dependent manner.  If the influence functional was completely finite, this deformation would disappear as we take $r_c\to \infty$ limit. But given the UV divergences in the case under consideration,  this small renormalization, when ignored, can lead to new divergences which look like temperature-dependent divergences which cannot be canceled by unitary, state-independent counterterms factorized across the SK contour.

However, with respect to the renormalized sources there is no ambiguity and the result is consistent with the  requirements delineated above. To see this, define the following counterterm action in terms of the renormalized probes that is unitary, state-independent, and factorized across the SK contour
\begin{equation}
\begin{split}\label{eq:ctInf}
S_{(n)}^{\text{c.t.}}=\int d^dx \, \frac{\lambda_n}{n!}  \left[\left(\Jr_a+\frac{1}{2}\, \Jr_d\right)^n
-\left( \Jr_a-\frac{1}{2}\, \Jr_d\right)^n\right] .  
\end{split}
\end{equation}
With this choice we obtain a manifestly finite influence functional  to linear order in couplings $\lambda_n$ and to linear order in derivatives, given by the expression
\begin{equation}\label{eq:snfin}
\begin{split}
S_{(n)} &\equiv - \lim_{r_c\to\infty} \big(S_{(n)}^{\text{bare}}+S_{(n)}^{\text{c.t.}} \big) \\
&= -  \lambda_n \, \int d^dx   \sum_{k=1}^n\frac{1}{(n-k)!}\left(\Jr_a+\frac{i}{8}\beta\partial_t \Jr_d\right)^{n-k} 
\left[\ifIr_{n,k} \, \frac{(\Jr_d)^k}{k!}
-\ifIr_{n,k+1}\,\frac{(\Jr_d)^{k-1}}{(k-1)!}\frac{i}{2} \beta\partial_t\Jr_d\right]\ .  
\end{split}
\end{equation}

Once we have performed the renormalization described, we find that we have to evaluate the renormalized integrals $\ifIr_{n,k}$. We have been able to obtain closed form expressions for a marginal operator with $\Delta =d$  (see \cref{sec:mass0}). For the quartic influence functional, passing to momentum space and letting $\delta(k) =  (2\pi)^d \delta^d\left(\sum_{i=1}^{4}k_i\right)  $, we find:
\begin{equation}
\begin{split}
\ifad_{aaaa} &= 
	0 \,,\\
\ifad_{aaad} &= 
	\frac{i\lambda_4}{3!} \, \frac{r_h^d}{d} \, \delta(k)\,, \\
\ifad_{aadd} &=
	 \frac{-i\lambda_4}{4}\, \frac{r_h^d}{d}\, \delta(k)\, \frac{\beta\omega_4}{4} \,, \\ 
\ifad_{ddda} &=
	\frac{i\lambda_4}{3!} \, \frac{r_h^d}{d} \, \delta(k) \left( \frac{1}{2} + \frac{3i\zeta(3)}{2\pi^3}\, \beta\omega_4 \right)\,, \\ 
\ifad_{dddd} &=
	 \frac{i\lambda_4}{4!} \, \frac{r_h^d}{d}\, \delta(k)\, \left( \frac{3i\zeta(3)}{\pi^3} \right)\,. 
\end{split}
\end{equation}

On the other hand closed form expressions for $\Delta <d$ are not easy to come by. One can however  extract the overall dependence on the dimensionful parameters ($r_h$ which sets the thermal scale) and obtain a result up to a numerical factor which is a function of the scaling dimension. These coefficients are given in terms of integrals of Legendre functions. More specifically, we find  that the influence functionals are given in terms of the renormalized integrals
\begin{equation}\label{eq:Fnkscaling}
\begin{split}
\ifIr_{n,k} &= 
	\frac{r_h^{n\Delta - (n-1)d}}{d} \,
 	\frac{\Gamma\left(\frac{\Delta}{d}\right)^{2n}}{\Gamma\left(\frac{2\Delta}{d}-1\right)^n} \; \mathfrak{F}_{n,k}(\Delta) 
\end{split}
\end{equation}	
where 
\begin{equation}\label{eq:Fnknumbers}
\begin{split}
\Fnk_{n,k}(\Delta) & =   
	\int_1^\infty\, d\varrho\, \left[P_{-\frac{\Delta}{d}}(2\varrho-1)\right]^n\, 
	\Bigg[\left(\frac{i}{\pi }\,  \frac{Q_{-\frac{\Delta}{d}}(2\varrho-1)}{P_{-\frac{\Delta}{d}}(2\varrho-1)} + i\cot\left(\frac{\pi \, \Delta}{d}\right)+ \frac{1}{2}\right)^k \\ 	
& 
\hspace{5cm} - \left(\frac{i}{\pi}\,  \frac{Q_{-\frac{\Delta}{d}}(2\varrho-1)}{P_{-\frac{\Delta}{d}}(2\varrho-1)} + i\cot\left(\frac{\pi\,\Delta}{d}\right) - \frac{1}{2}\right)^k \Bigg]\,.
\end{split}
\end{equation}	
To derive this expression we used the fact that one can solve for the combination $\ctor+ \frac{G^+_{1,0}}{G^+_{0,0}}$ explicitly in terms of Legendre functions as we describe in \eqref{eq:G01zeta}. One can easily numerically estimate these integrals entering the effective action. For the range of relevant  operators where the integrals are convergent we quote results from a simple numerical integration in \cref{fig:Fnkplots}.

\begin{figure}[h]
\centering
\includegraphics[width=.4\textwidth]{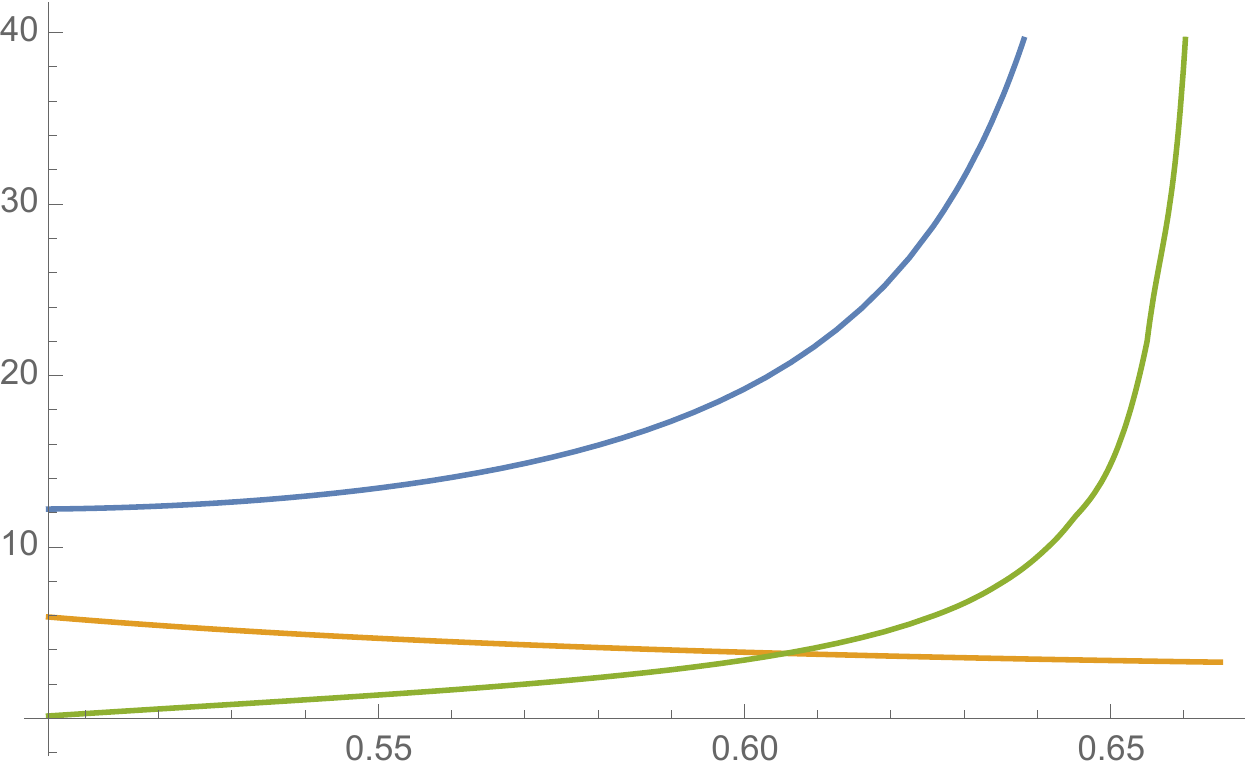}
\hspace{1cm}
\includegraphics[width=.4\textwidth]{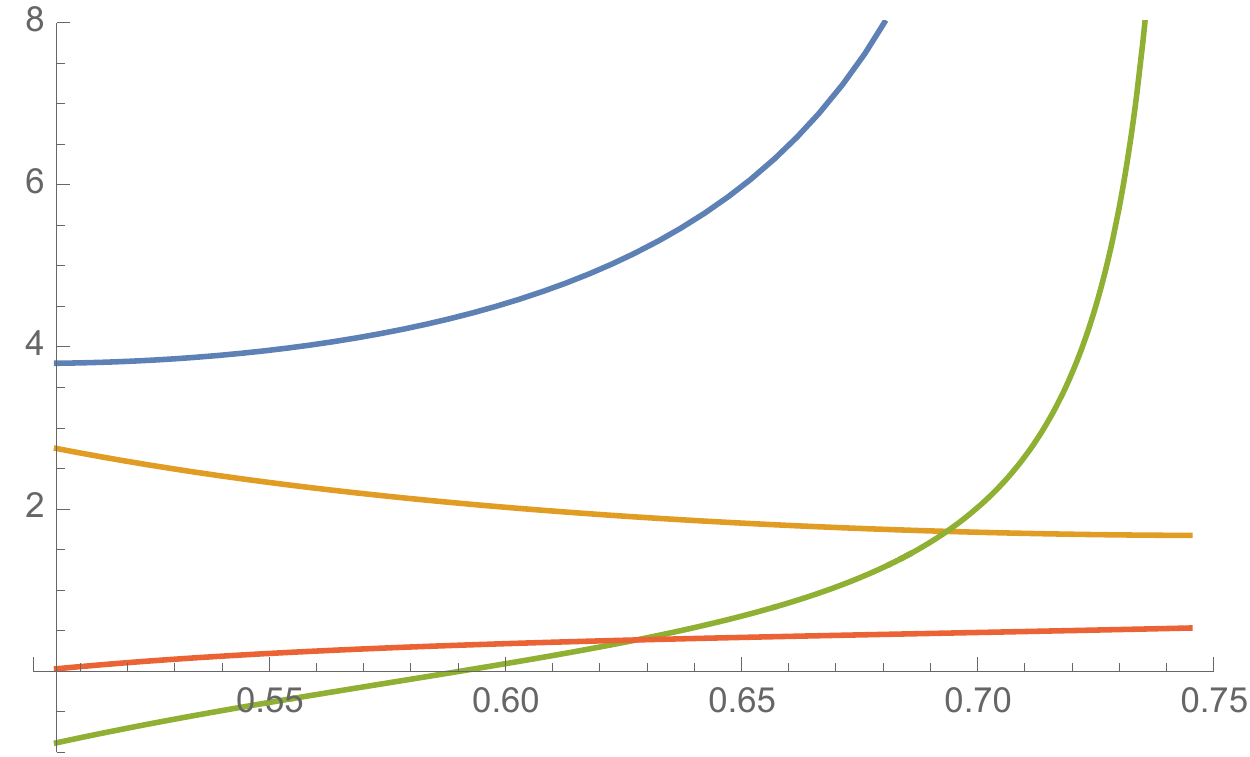}
\begin{picture}(0,0)
\setlength{\unitlength}{1cm}
\put (-0.1,-0.1) {$\frac{\Delta}{d}$}
\put (-7.6,-0.1) {$\frac{\Delta}{d}$}
\put (-14.6,3.2) {$\Fnk_{3,k}$}
\put (-7.0,3.2) {$\Fnk_{4,k}$}
\put (-12.8,1.5) {$\scriptscriptstyle{\Fnk_{3,1}}$}
\put (-8.9,1.5) {$\scriptscriptstyle{\Fnk_{3,3}}$}
\put (-11.9,0.8) {$\scriptscriptstyle{\Im(\Fnk_{3,2})}$}
\put (-4.9,2.2) {$\scriptscriptstyle{\Fnk_{4,1}}$}
\put (-1.5,2) {$\scriptscriptstyle{\Fnk_{4,3}}$}
\put (-4.9,1.4) {$\scriptscriptstyle{\Im(\Fnk_{4,2})}$}
\put (-5.5,0.7) {$\scriptscriptstyle{\Im(\Fnk_{4,4})}$}
\end{picture}
\caption{The numerical values of the quantities $\Fnk_{3,k}$ and $\Fnk_{4,k}$ as a function of the conformal dimension $\Delta$ (normalized by dimension). We have confined attention to the case of relevant operators $\Delta \in (\frac{d}{2},\frac{n-1}{n}\,d)$ when the integrals are convergent without need of additional counterterms. We note that $\Fnk_{n,2k}$ are purely imaginary and  $\Fnk_{n,2k+1}$ are real -- we have thus plotted just the imaginary or real  parts in the corresponding case. }
\label{fig:Fnkplots}
\end{figure}
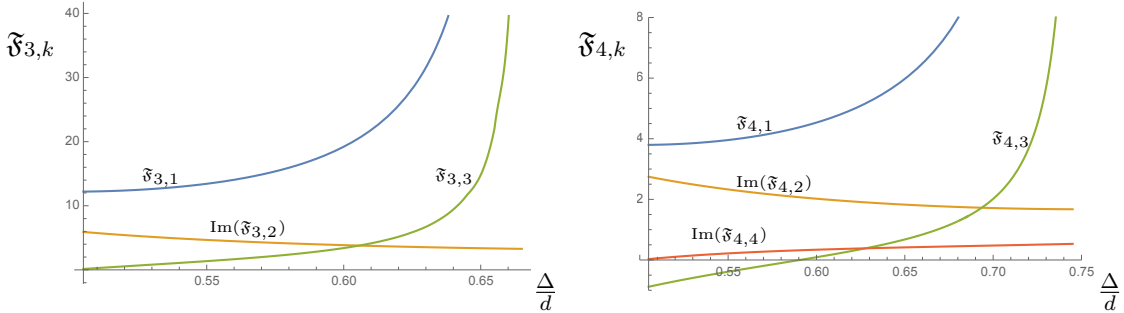
%

\section{Stochastic description of the open effective field theory}
\label{sec:stochastic}

Given that we have computed the influence functionals we can return to the problem of constructing the open effective field theory for our system degree of freedom $\Psi(x)$. As remarked in \cref{sec:holopen} this is easy, since all we need to do at the linear order is replace the sources $J_a$ and $J_d$ by the average and difference field $\Psi_a(x)$ and $\Psi_d(x)$, respectively. The renormalized effective action for when the operator $\Psi$ couples to has a $n$-point contact interaction in the holographic set-up is then immediate to write down, using  the results of \cref{sec:s2eff} and \cref{sec:sneff}. Collecting the results from \eqref{eq:Sos2}, and \eqref{eq:snfin} the effective action in position space reads:
\begin{equation}
\begin{split}\label{eq:Spsin}
S_\Psi & =	 
	\mathcal{N}_2 \, \int d^dx  \left( - \partial_t \Psi_a \,\Psi_d + \frac{i}{\beta} \Psi_d^2  -\frac{\beta}{4\pi}\, \mathfrak{g}_{2,0} \partial_t\Psi_a \partial_t \Psi_d +\frac{\beta}{4\pi}\, \mathfrak{g}_{0,2} \nabla_i\Psi_a \nabla_i \Psi_d \right) \\
& 
	\qquad - \ \lambda_n\, \mathcal{N}_n\, \int d^dx   \Bigg[ \sum_{k=1}^n\,
	\mathfrak{F}_{n,k}(\Delta)\, 	\frac{\Psi_a^{n-k}}{(n-k)!}\, \frac{\Psi_d^k}{k!}   \\
& 	
	\hspace{3.5cm} - 	\frac{i\beta}{2} \, 	
	\sum_{k=1}^{n-1} \left(\mathfrak{F}_{n,k+1}\, - \frac{1}{4}\,\mathfrak{F}_{n,k-1}\, \right)
		\frac{\Psi_a^{n-k}}{(n-k)!}\, \frac{\Psi_d^{k-1}}{(k-1)!}\, 	\partial_t\Psi_d\Bigg]\, .  
\end{split}
\end{equation}
where  
\begin{equation}\label{eq:}
\mathcal{N}_2 =  r_h^{2\Delta -d-1}  \;\frac{\Gamma\!\left(\frac{\Delta}{d}\right)^4}{\Gamma\!\left(2\frac{\Delta}{d}-1\right)^2} 
\,,
\qquad 
\mathcal{N}_n = \frac{r_h^{n\Delta - (n-1)d}}{d} \,
 \frac{\Gamma\left(\frac{\Delta}{d}\right)^{2n}}{\Gamma\left(\frac{2\Delta}{d}-1\right)^n} \,.
\end{equation}	
The quantities $\mathfrak{g}_{0,2}$ and $\mathfrak{g}_{2,0}$ are finite numbers (they are functions of $\Delta$ and $d$) obtained from integral expressions in \eqref{eq:g0220}, as are $\mathfrak{F}_{n,k}(\Delta)$ which is given in \eqref{eq:Fnknumbers}. Sample values of these are also plotted in \cref{fig:g0220plots} and \cref{fig:Fnkplots}, respectively.

We can compare the effective action we have derived with the general expectation based on stochastic dynamics of the open system \eqref{eq:Langevin}.  The coupling constants of appearing in the Langevin effective action obtained earlier in \eqref{eq:SLang} can be read off from  \eqref{eq:Spsin}. At the Gaussian order we have the relations
\begin{equation}\label{eq:Langpars1}
\begin{split}
 \gamma &= \mathcal{N} _2 \,,\hspace{2.69cm}  f = \frac{2\,\gamma}{\beta} \,, \\
 K &=  \frac{\beta}{4\pi}\, \mathfrak{g}_{2,0}\,\gamma\,,  \hspace{1.5cm}   D = \frac{\beta}{4\pi}\, \mathfrak{g}_{0,2}\,\gamma
 \end{split}
\end{equation}
while the non-Gaussian terms lead to 
\begin{equation}\label{eq:Langpars2}
\begin{split}
\theta_k &= - \frac{\lambda_n}{i^{k+1}} \, \mathcal{N}_n \, \mathfrak{F}_{n,k}(\Delta) \,,\hspace{4.1cm} \text{for} \; k\in \{1,\cdots\,, n\}\,,\\
\bar{\theta}_k & =  -\frac{\beta}{2}\, \frac{\lambda_n}{i^{k}} \, \mathcal{N}_n \left( \mathfrak{F}_{n,k+1}(\Delta) - \frac{1}{4}\, \mathfrak{F}_{n,k-1}(\Delta) \right) \,,\qquad  \text{for} \; k\in \{1,\cdots\,, n-1\}\,.
\end{split}
\end{equation}

The first set of these relations \eqref{eq:Langpars1} capture the standard fluctuation dissipation relations expected from a Gaussian noise source. The fact that $\theta_k$ and $\bar{\theta}_k$ also owe their origins to the same set of influence functionals (and thus to the same underlying microscopic dynamics of the bath fields we integrated out) leads to a set of generalized FDTs quoted earlier in \eqref{eq:genFD}, which we reproduce here for convenience:
\begin{equation}\label{eq:genFDrep}
\frac{2}{\beta}\, \bar{\theta}_k + \theta_{k+1} + \frac{1}{4}\theta_{k-1} =0\,.
\end{equation}	
%

\section{Discussion}
\label{sec:discuss}

We have initiated the study of open quantum systems, where the environment/bath is modeled by a strongly correlated thermal medium with short relaxation times. We modeled the latter using holography, and used standard arguments to relate the real-time thermal observables of the bath to the influence functionals of the open effective field theory. For simplicity, our focus was on systems comprising of a single scalar degree of freedom, coupled to a gauge invariant operator of the bath theory of scaling dimension $\Delta$. 

As we have endeavored to explain, it hitherto has been an unanswered question, at least within the remit of perturbative techniques, whether the dynamics of the system is described by a local influence functional, even assuming that the environment is sufficiently scrambling. As one might suspect, holographic systems, dual to black holes are maximally efficient in scrambling, and thus ought to be able to obtain local influence functions. This is indeed borne out by our explicit analysis, whereby the holographic influence functionals are manifestly local, and provide a proof-of-existence of such local open EFTs. We have furthermore shown that this open EFT can be given a stochastic interpretation, satisfying the required non-linear fluctuation-dissipation relations. 

From a gravitational standpoint, the discussion also sheds further light on the semiclassical geometries that compute real-time observables in black hole backgrounds, providing further evidence to the proposal of \cite{Glorioso:2018mmw}. In this context, we capture in the influence functionals, not only the known dissipative physics contained in response functions, gravitationally encoded in the quasinormal modes, but also at the same time, learn about their interactions with the outgoing Hawking quanta. While earlier works, \cite{Chakrabarty:2019aeu}, demonstrated this in the context of a single Brownian particle degree of freedom, the present discussion generalizes this to scalar probes of the black hole environment. We should note that earlier attempts to understand fluctuations of the outgoing Hawking quanta,  as  for example in  \cite{Leahy:1983vb}, where the authors studied an interacting scalar model in an ersatz two-dimensional black hole were plagued by infra-red divergences. These are due to the fact that probe field has a large effect on the background in the solution chosen, as well as, the fact that the horizon boundary conditions are not imposed appropriately.\footnote{ We thank Alok Laddha for bringing this work to our attention and for useful discussions on this issue.} The calculation we have described in the preceding sections suffers no such ambiguities and indeed reproduces results that are consistent with expectation of the dual boundary field theory.

From the viewpoint of holography, the grSK geometries we have employed provide a satisfactory answer to an old conundrum. How does one efficiently compute higher-point correlation functions in black hole backgrounds, while manifestly respecting causality properties of response functions and fluctuations thereof? One question of particular interest was whether one was required to integrate the interaction vertex of the Witten diagrams throughout the maximal Kruskal extension of the black hole geometry, including regions near the singularity.   The grSK geometries give a satisfactory answer: observables are computed in smooth two-sheeted geometries, each extending only up to the future horizon and joined together across a smooth horizon-cap, without any information about the black hole interior, per se. To be clear however, our analysis probes a stationary thermal medium with external sources, which are not themselves backreacting on the medium. We have utilized this to allow the real-time part of the Schwinger-Keldysh contour to extend all the way to $t\to \infty$, despite the fact that evolution future of the latest operator insertion in Lorentzian time is redundant (by the collapse rules). We are as yet unaware of a geometric construction that makes this redundancy manifest, an issue that will be important when we discuss certain generalizations below.\footnote{ We thank Rob Myers for raising this question.}

There are several straightforward generalizations of our analysis such as discussing open systems with spin degrees of freedom coupled to holographic matter \cite{Loganayagam:2020aa}. More interesting are situation where the dynamics involves coupling the open system to conserved currents of the holographic theory. For instance,  the open system could be coupled to a conserved $R$-current of the holographic field theory, or to the energy-momentum tensor. In either case, one has to examine the dynamics of gauge fields in the grSK geometries and compute the Schwinger-Keldysh correlators of these conserved currents. As mentioned in \S\ref{sec:intro}, there already exists an analysis in \cite{Glorioso:2018mmw} for the dynamics of a Maxwell field in the grSK Schwarzschild-AdS geometry. However, there are several peculiarities of the gauge dynamics which deserve clarification (as already noted in the reference cited). Understanding probe gauge field dynamics is but a precursor to the more interesting question: how does one account for gravitational backreaction and consider holographic environments with non-trivial temporal evolution. A non-trivial example which would be worth analyzing is the couping of our system to a holographic plasma which is spatially inhomogeneous and temporally evolving towards equilibrium. This would require us to understand the gravitational Schwinger-Keldysh construction for the fluid/gravity spacetimes \cite{Hubeny:2011hd}.

While we have focused on understanding the influence functionals that are computable by the Schwinger-Keldysh temporal ordering, some of these thermal observables are in turn related by the general KMS relations to certain-out-of-time-order (OTO) observables 
\cite{Haehl:2017eob}.  For instance, all OTO three-point correlation functions can be generated from two Schwinger-Keldysh correlators, while many of the OTO four-point functions are related to ones with Schwinger-Keldysh ordering (though not the oft studied chaos correlator, which lies in a separate KMS orbit). A natural question is to encode these into the influence functionals, and use the system to probe OTO correlation functions of the holographic bath. This requires a suitable generalization of the grSK geometry to an grOTO saddle, which would be interesting to analyze. Note that the imprints of OTO observables on a Brownian particle probe has been analyzed hitherto in \cite{Chaudhuri:2018ihk}.

Finally, in the holographic setting, we have argued that one needs to carry out a suitable renormalization of the open system's degrees of freedom, in order to derive a local effective field theory, whilst respecting microscopic unitarity, state-independence, and factorization across the two legs of the Schwinger-Keldysh contour. This was starkly visible at leading order in the gradient expansion for the coupling of a bosonic degree of freedom to a marginal operator of the holographic thermal field theory. While we have analyzed the renormalization effects at leading order, developing a system holographic  renormalization procedure, and understanding the requisite counter-terms, along the lines of \cite{Skenderis:2008dg}, would be very helpful. For one these would be invaluable in attempting to construct a Wilsonian effective field theory for open systems (analogous to \cite{Heemskerk:2010hk,Faulkner:2010jy}). Of particular interest in this context is to understand how strongly correlated thermal environments evade the issue of non-trivial (infra-red) divergences that appear to arise in perturbative open field theory computations \cite{Avinash:2017asn,Agon:2017oia,Avinash:2019qga,Chatterjee:2020aa}. 

\section*{Acknowledgements}
It is a pleasure to thank Bidisha Chakrabarty, Soumyadeep Chaudhuri, Sean Colin-Ellerin,  Guglielmo Grimaldi, Veronika Hubeny, Juan Maldacena, Shiraz Minwalla, Rob Myers, Akhil Sivakumar,  and Spenta Wadia for helpful discussions. We would also like to thank Yogesh Dandekar for initial collaboration on the project. 
CJ and RL would like to thank the hospitality of Indian Institute of Science Education and Research (Bhopal) during National Strings Meeting 2019 (NSM2019) where this work was presented. RL would like to thank the organizers of the ``Workshop on Holography, Entanglement and Complexity'' at  Ashoka University and the workshop on ``Recent advances in QFT and gravity, Amplitudes and CFT correlators'' at Saha Institute of Nuclear Physics. CJ and RL would also like to acknowledge our debt to the people of India for their steady and generous support to research in the basic sciences.
MR would like to thank KITP, UCSB for hospitality during the workshop ``Gravitational Holography'', where the research was supported in part by the National Science Foundation under Grant No. NSF PHY1748958 to the KITP. 
MR was supported  by U.S.\ Department of Energy grant DE-SC0009999 and by funds from the University of California. 

\appendix
\section{Gradient expansion on the grSK contour}
\label{sec:eogradient}

In the main text we defined the retarded-advanced basis for the sources on the boundary. It is useful to extend this to the bulk and define
two related combinations:
\begin{equation}\label{eq:Jevenodd}
\begin{split}
J_{even} \equiv -\JF + e^{\beta\omega(1-\ctor)} \, \JP = \left[ J_{a} - \left( \frac{e^{\beta\omega(1-\ctor)}-1}{e^{\beta\omega}-1} - \frac{1}{2} \right) J_d \right]\\ = \left(1-e^{-\beta\omega\ctor} \right) (n_\omega +1) J_{\skR} + \left( e^{\beta\omega(1-\ctor)}-1 \right)n_\omega J_{\skL} \,, \\
J_{odd} \equiv -\JF - e^{\beta\omega(1-\ctor)} \JP = \left[ J_{a} + \left( \frac{e^{\beta\omega(1-\ctor)}+1}{e^{\beta\omega}-1} + \frac{1}{2} \right) J_d \right]\\ = \left(1+e^{-\beta\omega\ctor} \right) (n_\omega +1) J_{\skR} - \left( e^{\beta\omega(1-\ctor)}+ 1 \right)n_\omega J_{\skL} \,.
\end{split}
\end{equation}
The solution for $\Phi$ can then be written as with this simple change of basis as
\begin{equation}\label{eq:Phieo}
\begin{split}
\Phi = G_{even}\, J_{even} + G_{odd}\, J_{odd}\,, 
\end{split}
\end{equation}
where
\begin{equation}\label{eq:Gpeo}
\begin{split}
G_{even} \equiv \frac{1}{2}\left( G^+ + G^- \right) \,, \qquad G_{odd} \equiv \frac{1}{2} \left( G^+ - G^- \right)\,.
\end{split}
\end{equation}
Note that $G^+_{even}$, $G^+_{odd}$ are obtained by separating the ingoing Green function $G^+$ into even and odd parts under 
frequency reversal once we recall  that $G^-(\ctor, \omega, {\bf k}) = G^+(\ctor, -\omega, {\bf k})$. 

The advantage of the combinations $J_{even}$ and $J_{odd}$ lies in the following fact: given the ingoing Green function $G^+$ in a boundary gradient expansion, one can obtain the solution on the full grSK contour by a simple substitution. We simply multiply the even frequency part of the Green's function by $J_{even}$ and the odd frequency part  by $J_{odd}$, i.e., 
\begin{equation}\label{eq:}
\Phi(\ctor,\omega, {\bf k}) = \sum_{n,m=0}^\infty \, \bigg( G^+_{2n,m}(\ctor)\;  J_{even} + G^+_{2n+1,m}(\ctor)\;   \bwt \, J_{odd} \bigg)  \bwt^{2n}\, \bqt^m \,, 
\end{equation}	
where we resort to the truncated notation:
\begin{equation}\label{eq:wqdef}
\bwt = \frac{\beta\omega}{2} \,, \qquad \bqt = \frac{\beta |{\bf k|}}{2}\,.
\end{equation}	

There is another useful property obeyed by the even and odd combinations which is worth highlighting:
\begin{equation} 
\label{eq:Jevod}
\begin{split}{}
 D_\ctor^+ J_{even} = \bwt\,J_{odd}\ ,\quad  D_\ctor^+ J_{odd} = \bwt\,J_{even}\, .
\end{split}
\end{equation}
This can be established by a direct differentiation of the definitions above. It follows that $ J_{even} $ and $ J_{odd}$ are solutions of the 
time-reversal invariant differential equation 
\begin{equation}
\bigg(\left(D_\ctor^+\right)^2- \bwt^2\bigg) \Phi=0\ .
\end{equation}
This equation involves no spatial boundary derivatives and hence should be thought of as a differential equation on each ingoing Eddington-Finkelstein tube in  the grSK contour (the terminology comes from the fluid/gravity correspondence, cf., \cite{Hubeny:2011hd}). We can then define $ J_{even} $ as the solution that interpolates between $J_{\skL}$ at $\ctor=0$ to $J_{\skR}$ at $\ctor=1$. The odd combination $J_{odd}$ is obtained by differentiation $D_\ctor^+ J_{even} =\bwt\, J_{odd}$ as given from \eqref{eq:Jevod}. More explicitly, we have
\begin{equation}\label{eq:Phiev}
\begin{split}
 \Phi  = \sum_{n,m=0}^\infty G_{n,m}^+ \bqt^m \, (D_\ctor^+)^n J_{even}\,.
\end{split}
\end{equation}

For computations involving the full solution written down in a derivative expansion we can use the following gradient expansions of the even and odd sources: 
\begin{equation}\label{eq:Jeoexp}
\begin{split}
J_{even} 
&=
	 J_a -\frac{1}{2}J_d \left[ y + \frac{\bwt}{2!}\, (y^2-1) + \frac{\bwt^2}{3!} \,  y\,(y^2-1) + \frac{\bwt^3}{4!}\, (y^2-1)^2 + \cdots \right] \\
2\, \bwt\, J_{odd}  
&=
	2\, \bwt\, J_a + J_d \Big[2+ \bwt\, y + \frac{1}{2!} \left( \bwt  \right)^2 \left(y^2 + \frac{1}{3}\right)  + \frac{1}{3!} \left( \bwt  \right)^3 y(y^2-1)\\
&
	\hspace{3cm} + \frac{1}{4!} \left( \bwt  \right)^4 \left(y^4 -2y^2 -\frac{1}{15}\right) +\cdots\Big]\,.
\end{split}
\end{equation}
Here we have introduced a new bulk variable $y$ defined as 
\begin{equation}\label{eq:}
y = 1 - 2\,\ctor  \,,
\end{equation}	
to  simplify the expressions somewhat.  Using the gradient expansion of $J_{even}$ given above, in \eqref{eq:Phiev} we end up the result \eqref{eq:Phigrad} quoted in the main text. For certain computations it is helpful to have the solution in the gradient expansion directly in position space. One can verify that \eqref{eq:Phigrad} can be simplified to the following form, accurate to linear order in the derivative expansion  
\begin{equation} \label{eq:Phitime}
 \begin{split}
\Phi
&=
	\left(G_{0,0}^+ +\frac{i}{2}  G^+_{1,0} \,\beta\partial_t \right) 
	\bigg\{ J_a + \frac{i}{8}\,\beta\partial_t J_d+ J_d\left(\ctor-\frac{1}{2}  +\frac{G^+_{1,0}}{G^+_{0,0}} \right) \\
& 
	\qquad \qquad -\;\frac{i}{2} \,\beta\partial_t J_d \left(\ctor-\frac{1}{2} + \frac{G^+_{1,0}}{ G^+_{0,0}} \right)^2\bigg\} 
	+  \order{\left(\beta\partial_ t\right)^2} .
\end{split}
\end{equation}

It is also useful to record here a simple expression for the $D_\ctor$ derivative of the field which enters into the computation of the quadratic influence functional. It is
\begin{equation}\label{eq:Dphiev}
\begin{split}
D_\ctor^+ \Phi 
&= 
\sum_{n=0}^\infty\, \sum_{m=0}^\infty  \bqt^m\, \left(\dv{G_{n,m}^+}{\ctor}  + G_{n-1,m}^+ \right)  (D_\ctor^+)^n J_{even}\,,
\end{split}
\end{equation}
with $G_{-1,m}^+ \equiv 0$.

Note that, by construction, $J_{odd}$ occurs in the full solution with at least one $\beta\omega$ factor multiplying it. It follows that if $G^+$ has a derivative expansion, so does the full solution on the grSK contour written in \emph{average-difference}/Keldysh basis. Note also that when the even part of the ingoing solution is lifted to the holographic SK contour, the order of the derivative expansion is maintained. In contrast, when the odd part of the ingoing solution is lifted to the holographic SK contour, the source $J_d$ occurs with one less time derivative.

\section{Gradient expansion of the Green's functions}
\label{sec:gradexpmass}

For a massive scalar field $m^2 = \Delta (\Delta -d) \neq 0$  on the grSK geometry we now analyze the wave equation \eqref{eq:phieom} in a gradient expansion \eqref{eq:Gpgrad}. It is helpful to introduce a new radial coordinate 
\begin{equation}\label{eq:}
 \varrho \equiv \frac{r^d}{r_h^d}  \;\; \Longrightarrow \;\; \dv{}{\ctor} = 2\pi i\, \varrho^{\frac{1}{d}} (\varrho-1)\, \dv{}{\varrho}\,,
\end{equation}	
and re-express \eqref{eq:phieom} as 
\begin{equation}\label{eq:phieomrho}
\left(\dv{}{\varrho}\left( \varrho (\varrho -1) \, \dv{}{\varrho}\right)  -\nu (\nu-1)\right) \Phi  = \bwt\, \mathcal{S}_1[\Phi] + \bqt^2 \, \mathcal{S}_2[\Phi]\,,
\end{equation}	
in terms of  the following operators:
\begin{equation}\label{eq:DSops}
\begin{split}
\mathcal{S}_1[\Phi]&= 
	-\frac{1}{2\pi\,i} \, \frac{1}{\Phi} \, \dv{}{\varrho}\left(\varrho^{1-\frac{1}{d}}\, \Phi^2\right)
\,,\\ 
\mathcal{S}_2[\Phi] &=
	\frac{1}{(2\pi)^2} \, \varrho^{-\frac{2}{d}}\, \Phi\,.
\end{split}
\end{equation}	
We have also defined a rescaled conformal dimension
\begin{equation}\label{eq:nudef}
\nu = \frac{\Delta}{d}
\end{equation}	
which will be the only parameter entering our expressions.  Upon plugging in the expansion \eqref{eq:Gpgrad} we immediately find the recursion relation:
\begin{equation}\label{eq:recurse1}
\left(\dv{}{\varrho}\left( \varrho (\varrho -1) \, \dv{}{\varrho}\right)  -\nu(\nu-1)\right)  G_{m,n}^+  =  \mathcal{S}_1[G_{m-1,n}^+] + \mathcal{S}_2[G_{m,n-2}^+]\,,
\end{equation}	
with the understanding that $G_{m,n}^+ =0$ for either $m<0$ or $n<0$.  We will now solve this order by order sequentially.

We first note that the leading order term in the expansion, the zero-mode or the DC part, $G_{0,0}^+$ satisfies the homogeneous equation with no sources: 
\begin{equation}\label{eq:G00eqn}
\left(\dv{}{\varrho}\left( \varrho (\varrho -1) \, \dv{}{\varrho}\right)  -\nu (\nu-1)\right)  G_{0,0}^+  =0\,.
\end{equation}	
We can exploit this fact to simplify our recursion relation, remove the contribution arising from the mass, and rewrite the differential part of the expression as a total derivative, amenable to integration by quadratures. Define the ratio
\begin{equation}\label{eq:Gprdef}
\Gpr{m,n} =  \frac{G_{m,n}^+}{G_{0,0}^+}\,,
\end{equation}	
which one can check satisfies the following equation
\begin{equation}\label{eq:recurse2}
\dv{}{\varrho}\left( \varrho (\varrho -1) \, \left(G_{0,0}^+\right)^2 \dv{\Gpr{m,n}}{\varrho}\right)  
= G_{0,0}^+ \left(\mathcal{S}_1[G_{m-1,n}^+] + \mathcal{S}_2[G_{m,n-2}^+]\right) ,
\end{equation}	
where we have exploited the fact that $G_{0,0}^+$ is in the kernel of the differential operator.
We impose regular boundary condition at the horizon and normalization at the conformal boundary at each order in perturbation theory as described in \eqref{eq:Gpbcs}. The general solution can then be immediately written down:
\begin{equation}\label{eq:Gmnsol}
\Gpr{m,n} = \int_{\varrho_c}^\varrho \frac{ d\varrho' }{\varrho' (\varrho'-1) \left(G_{0,0}^+(\varrho')\right)^2} \int_1^{\varrho'} d\bar{\varrho}
\, G_{0,0}^+(\bar{\varrho})\left(\mathcal{S}_1[G_{m-1,n}^+] +   \mathcal{S}_2[G_{m,n-2}^+]\right)  ,
\end{equation}	
where the inner integral has its constant of integration chosen to ensure that the pole at $\varrho'=1$ is canceled.

\paragraph{Zeroth order solution:} 
Let us examine some of the leading order terms in the gradient expansion explicitly.  The differential operator appearing on the l.h.s.\ of \eqref{eq:phieomrho} is the Legendre differential operator. Consequently,  $G_{0,0}^+$ satisfies the homogeneous Legendre differential equation \eqref{eq:G00eqn} which has the general solution\footnote{ We define the associated Legendre function of the second kind as in \cite[Eq.~14.3.7]{NIST:DLMF} which disambiguates the various definition for this function employed in the literature. In particular, we define it in terms of the regularized hypergeometric function as
\begin{equation}\label{eq:}
Q_{\nu}(x) = \sqrt{\pi} \frac{\Gamma(\nu+1)}{(2x)^{\nu+1}} \, {}_2\widetilde{F}_1\left(\frac{\nu}{2}+\frac{1}{2},\frac{\nu}{2}+1, \nu+\frac{3}{2},\frac{1}{x^2} \right)\,.
\end{equation}	
This expression is well defined for $x\in (1,\infty)$ which is the  domain of interest to us. 
} 
\begin{equation}
G^+_{0,0} = c_1 \, P_{-\nu} (2 \varrho-1) + c_2 \, Q_{-\nu} (2\, \varrho-1)\,,
\end{equation}
 It is easy to check that the Legendre function of second kind $Q_{-\nu} (2\varrho-1)$ diverges at the horizon, $\varrho=1$.  Hence the normalized solution obeying our boundary conditions is simply
\begin{equation}\label{eq:G00sol}
G^+_{0,0} = \frac{P_{-\nu} \left(2\,\varrho-1\right)}{P_{-\nu} \left(2\,\varrho_c-1\right)}\,.
\end{equation}	
This is the result quoted in \eqref{eq:Gpmngen}. 

\paragraph{First order solution:} At the next order it is immediate to see that $G_{0,1}^+$ (or for that matter any $G_{m,2n+1}^+$) vanishes. We then have to solve for $G_{1,0}^+$ which can be ascertained to satisfy the inhomogeneous equation
\begin{equation}\label{eq:}
\begin{split}
\dv{}{\varrho}\left( (\varrho^2 - \varrho) \, (G_{0,0}^+)^2 \, \dv{\Gpr{1,0}}{\varrho}\right) 
& = -  \frac{1}{2\pi i} \, \dv{}{\varrho} \left( \varrho^{1-\frac{1}{d}} \, \left(G_{0,0}^+\right)^2\right),
\end{split}
\end{equation}	
which gives a solution by quadratures 
\begin{equation}\label{eq:G10sol}
\begin{split}
\Gpr{1,0} =  -\frac{1}{2\pi i} \, \int^{\varrho}_{\varrho_c}  d\varrho' \;
\left(  \frac{1}{\varrho'^{\frac{1}{d}} \, (\varrho' - 1)}  -  \frac{1}{(\varrho'^2 - \varrho') \,P_{-\nu}(2\varrho'-1)^2 } \right) ,
\end{split}
\end{equation}	
where constants of integration have been chosen to ensure regularity at the horizon $\varrho =1$ and vanishing of the function at the cut-off surface. We have in addition also made use \eqref{eq:G00sol} to write the result as an integral over Legendre functions.  It is useful to massage this to the form:
\begin{equation}\label{eq:G01zeta0}
\begin{split}
\ctor + \Gpr{1,0}  
&=  
	 \frac{1}{2\pi i}\, \int_{\varrho_c}^{\varrho} \, \frac{d\varrho'}{\varrho' \,( \varrho'-1)} \left( \frac{1}{P_{-\nu} \left(2\varrho'-1 \right)}\right)^2 \,, 
\end{split}
\end{equation}	
which will enter in the computation of the influence functionals in the gradient expansion. This integral can be explicitly evaluated in terms of Legendre functions of the second kind. In terms of $x = 2\varrho -1$ we find after a small algebraic manipulation, the following simple relation
\begin{equation}\label{eq:G01zetaint}
\begin{split}
 P_{-\nu}(x)^2 \, \dv{x} (\ctor + \Gpr{1,0}) 
 &= 
 	\frac{1}{i\pi} \, \frac{1}{x^2-1} \\
 &= 
 	- \frac{1}{i\pi} \, \text{Wr}\{P_{-\nu}(x), Q_{-\nu}(x)\}	 \\
 &= 
 	-\frac{1}{i\pi}\, P_{-\nu}(x)^2\, \dv{x}\left(\frac{Q_{-\nu}(x)}{P_{-\nu}(x)}\right)	,
 \end{split} 
\end{equation}	
where we have identified the Wronskian of the Legendre functions, see \cite[Eq.~14.2.10]{NIST:DLMF}. It then follows that 
\begin{equation}\label{eq:G01zeta}
\ctor + \Gpr{1,0}  = \frac{i}{\pi}\, \frac{Q_{-\nu}(2\varrho-1)}{P_{-\nu}(2\varrho-1)} + i\, \cot \pi \nu \,, \qquad \nu \notin \mathbb{Z}_+
\end{equation}	
where  we fixed the constant by matching the asymptotics of the integral and the Legendre functions. Note that the integral is convergent for $\nu > \frac{1}{2}$, though in the main text we will often restrict attention to $\nu \in (\frac{1}{2},1]$ . The bound on $\nu$ arises from our focus on relevant and marginal operators (a detailed analysis of convergence properties is given in \cref{sec:counterterns}).

\paragraph{Second order solution:} At the next order in the gradient expansion we first solve for $G_{0,2}^+$ which satisfies the inhomogeneous equation
\begin{equation}\label{eq:G02eqn}
\dv{}{\varrho}\left( (\varrho^2 - \varrho) \, (G_{0,0}^+)^2 \, \dv{ \Gpr{0,2} }{\varrho}\right) 
 = \frac{1}{(2\pi)^2}\, \varrho^{-\frac{2}{d}} \, \left( G_{0,0}^+\right)^2 \,.
\end{equation}	
Integrating this and imposing our boundary conditions \eqref{eq:Gpbcs} we find the integral expression
\begin{equation}\label{eq:G02sol}
\Gpr{0,2}(\varrho) = \frac{1}{(2\pi)^2} \int_{\varrho_c}^\varrho \, \frac{d\varrho'}{\varrho' (\varrho'-1) \left(G_{0,0}^+(\varrho')\right)^2} \int_1^{\varrho'} d\bar{\varrho} \; 
\bar{\varrho}^{-\frac{2}{d}} \, \left( G_{0,0}^+(\bar{\varrho})\right)^2  .
\end{equation}	
Similarly, one finds for the temporal component $G_{2,0}^+$ the following expression:
\begin{equation}\label{eq:G20sol}
\Gpr{2,0}(\varrho) = -\frac{1}{2\pi i} \, \int_{\varrho_c}^\varrho \, \frac{d\varrho'}{\varrho' (\varrho'-1) \left(G_{0,0}^+(\varrho')\right)^2} 
\int_1^{\varrho'} d\bar{\varrho} \; \frac{G_{0,0}^+(\bar{\varrho})}{G_{1,0}^+(\bar{\varrho})} \dv{}{\bar{\varrho}} \left( \bar{\varrho}^{1-\frac{1}{d}} \left(G_{1,0}^+(\bar{\varrho})\right)^2\right) .
\end{equation}	
For the computation of the quadratic influence functional, we do not need the precise form of these functions. Having their derivatives at the cut-off surface suffices to determine the quantities $\dot{g}_{0,2}$ and $\dot{g}_{2,0}$ entering $\ifad_{ad}(\omega,{\bf k})$ in \eqref{eq:Sos2}. These can be computed straightforwardly as we have:
\begin{equation}\label{eq:g02dot}
\begin{split}
\dot{g}_{0,2} =
	\dv{G_{0,2}^+}{\ctor} \bigg|_{\ctor =0} 
&= 
	\frac{2\pi i}{(2\pi)^2}\,  \frac{\varrho_c^{\frac{1}{d}-1}}{G_{0,0}^+(\varrho_c)} \,
	\int_1^{\rho_c} \, d\varrho\, \varrho^{-\frac{2}{d}} \, \left( G_{0,0}^+(\varrho)\right)^2 \\
&=  
	\frac{i}{2\pi}\,  \frac{\varrho_c^{\frac{1}{d}-1}}{P_{-\nu}(2\varrho_c-1)^2}\, 
	\int_1^{\rho_c} \, d\varrho\,  \varrho^{-\frac{2}{d}} \, P_{-\nu}(2\varrho-1)^2 ,
\end{split}
\end{equation}
and 
\begin{equation}\label{eq:g20dot}
\begin{split}
\dot{g}_{2,0} =
	\dv{G_{2,0}^+}{\ctor} \bigg|_{\ctor =0} 
&= 
	-  \varrho_c^{\frac{1}{d}-1} \,
	\int_1^{\rho_c} \, d\varrho\;
	 \frac{1}{\Gpr{1,0}(\varrho)} \dv{}{\varrho} \left( \varrho^{1-\frac{1}{d}} \, \Gpr{1,0}(\varrho)^2 \, G_{0,0}^+(\varrho)^2\right)  \\
&= 
	-\frac{\varrho_c^{\frac{1}{d}-1}}{P_{-\nu}(2\varrho_c-1)^2}\,
	 \int_1^{\varrho_c} \, d\varrho \, 
	\bigg( \dv{\varrho}\left(\varrho^{1-\frac{1}{d}} \, \Gpr{1,0} \,P_{-\nu}(2\varrho-1)^2 \right) \\
&\hspace{5cm}	
	 + \varrho^{1-\frac{1}{d}}\,
		P_{-\nu}(2\varrho-1)^2\, \dv{\Gpr{1,0}}{\varrho}	\bigg)\\
&= -\frac{\varrho_c^{\frac{1}{d}-1}}{P_{-\nu}(2\varrho_c-1)^2}\, 
	\bigg[
	-\Gpr{1,0}(1) -\frac{1}{2\pi i} \, \int_1^{\varrho_c} \, d\varrho \, 
		\varrho^{-\frac{2}{d}}\, P_{-\nu}(2\varrho-1)^2\, \\
&		\hspace{3cm} 
		-\frac{1}{2\pi i} \, \int_1^{\varrho_c} \, d\varrho \,  \frac{\varrho^{-\frac{1}{d}}}{\varrho-1} 
		\left(\varrho^{-\frac{1}{d}} \, P_{-\nu}(2\varrho-1)^2-1\right)
	\bigg] ,
\end{split}
\end{equation}
where we have integrated by parts and used \eqref{eq:G10sol} to simplify the integral. These are the expressions that are compiled in \eqref{eq:g0220} and \eqref{eq:g0220num}.

\paragraph{Explicit solutions for massless fields:}
The expressions are in fact simplest for $d|\Delta$, i.e., $\nu \in \mathbb{Z}_+$, since then the Legendre functions are simple polynomials (note that $P_{-\nu}(x) = P_{\nu-1}(x)$ from the Legendre differential equation).  For instance, carrying out the exercise we find for a massless, minimally coupled field $m^2 =0$  or $\Delta =d$:
\begin{equation}\label{eq:Gpexpd}
\begin{split}
G^+_{0,0} &=1\\
G^+_{1,0} &= - \int_0^\ctor\, d\ctor' \bigg( 1-\left( \frac{r_h}{r'}\right)^{d-1}\bigg) \\
G^+_{0,1} & =0 \\
G^+_{2,0} &= \int_0^\ctor d\ctor' \int_{\ctor_h}^{\ctor'} d\ctor'' \bigg(1+ \left(\frac{{r''}}{r'}\right)^{d-1} \bigg)
 \bigg(1- \left(\frac{r_h}{r''}\right)^{d-1} \bigg)
 \\
& \qquad+ \;\int_0^{\ctor_h} d\ctor' \bigg(1- \left(\frac{r_h}{r'}\right)^{d-1}\bigg)
 \int_0^\ctor d\ctor'' \bigg(1- \left(\frac{r_h}{r''}\right)^{d-1} \bigg)
 \,
\\
G^+_{0,2} &= - \int_0^\ctor d\ctor' \int_{\ctor_h}^{\ctor'} d\ctor'' \left(\frac{{r''}}{r'}\right)^{d-1} f(r'')\,.
\end{split}
\end{equation}

\paragraph{Solutions in the BTZ geometry:} 
It would not be a surprise to reader to know that the above expressions can be integrated in $d=2$. We can alternately use the knowledge of the hypergeometric series  to aid in expanding out \eqref{eq:btzGp}. 
\begin{equation}\label{eq:hypser}
{}_2F_1(a,b,c;z) = \sum_{n=0}^\infty\, \frac{(a)_n\, (b)_n}{(c)_n\, n!} \, z^n  
\end{equation}
to aid the expansion, but note that we have resum the terms to get the perturbative contributions. The resummation can be done in terms of polylogarithms. For example:
\begin{equation}\label{eq:hyppol}
{}_2F_1(\epsilon \,a, \epsilon \, b, 1+ \epsilon \,c;z) =1+\epsilon^2\, a\, b\, \text{Li}_2(z) + \order{\epsilon^3}
\end{equation}
Using this we can expand $G^+(\omega, k)$ to the desired order, for 
\begin{equation}\label{eq:btzGpexp}
\begin{split}
G^+(\zeta, \omega, k) & =  
	\bigg[1+ i \, \frac{1}{\pi} \, \bwt \log \left(\frac{1 + e^{2\pi i\left(\ctor+\ctor_c\right)}}{1 + e^{2\pi i\,\ctor_c}}\right)  - \frac{1}{2\,\pi^2} \, \bwt^2 \, \left(\log \left(\frac{1 + e^{2\pi i (\ctor+\ctor_c)}}{1 + e^{2\pi i\,\ctor_c}}\right) \right)^2 +\cdots \bigg] \\
	& \qquad  \times
 	\bigg[ 1+ \frac{1}{4\,\pi^2} \, \left(\mathfrak{q}^2 - \bwt^2\right)  \, \left(\text{Li}_2\left(\sec^2\pi (\ctor+\ctor_c)\right)- \text{Li}_2\left(\sec^2\pi\ctor_c\right)\right)  + \cdots\bigg]
\end{split}
\end{equation}
where we have kept terms to $\order{\bwt^2}$ and $\order{\mathfrak{q}^2}$, respectively.  In other words, the analog of \eqref{eq:Gpexpd} now reads to quadratic order:
\begin{equation}\label{eq:Gpexp2}
\begin{split}
G^+_{0,0} & =1 \\
G^+_{1,0} & =\frac{i}{\pi} \,  \log \left(\frac{1 + e^{2\pi i\left(\ctor+\ctor_c\right)}}{1 + e^{2\pi i\,\ctor_c}}\right)  \\ 
G^+_{2,0} & = -\frac{1}{2\pi^2}  \left[\log \left(\frac{1 + e^{2\pi i\left(\ctor+\ctor_c\right)}}{1 + e^{2\pi i\,\ctor_c}}\right)\right]^2   - \frac{1}{4\pi^2} \left(\text{Li}_2\left(\sec^2\pi (\ctor+\ctor_c)\right)- \text{Li}_2\left(\sec^2\pi\ctor_c\right)\right)  \\
G^+_{0,2} & = \frac{1}{4\pi^2} \left(\text{Li}_2\left(\sec^2\pi (\ctor+\ctor_c)\right)- \text{Li}_2\left(\sec^2\pi\ctor_c\right)\right) .
\end{split}
\end{equation}
%

\section{Witten diagrams on the grSK contour}
\label{sec:wittendia}

In this Appendix we give a short argument in favour of using Witten diagrams on the grSK contour. Per se the argument is no different from the one used in the standard AdS/CFT context on a single sheeted geometry, but for completeness we give a brief account.
Consider the action \eqref{eq:sphi4} whose equation of motion is:
\begin{equation}
\frac{1}{\sqrt{-g}} \partial_A \left[\sqrt{-g} g^{AB} \partial_B \Phi \right] - m^2\, \Phi - \frac{\lambda_n}{(n-1)!} \Phi^{n-1} = 0\,. {}
\end{equation}
We can solve this equation perturbatively with $\lambda$ as a perturbation parameter, viz., 
\begin{equation}
\Phi = \Phi_0 + \lambda\, \Phi_1 + \cdots \,.
\end{equation}
Here $\Phi_0$ solves \eqref{eq:kgPhi} while $\Phi_1$ satisfies the following inhomogeneous equation
\begin{equation}
\frac{1}{\sqrt{-g}} \partial_A \left[\sqrt{-g} g^{AB} \partial_B \Phi_1 \right]  - m^2\, \Phi_1= \lambda_n \frac{1}{(n-1)!} \Phi_0^{n-1} \,. 
\end{equation}

It will be helpful for our analysis to fix the boundary conditions in a manner that is well adapted to this perturbation theory, so that a minimal number of terms contribute in the evaluation of the action. We pick:
\begin{equation}\label{eq:bclam}
\begin{split}
\Phi_0(\ctor=0,k) = J_{\skL}(k)\,, \qquad \Phi_0(\ctor=1,k) = J_{\skR}(k)\,, \qquad \Phi_i(\ctor=0,k) = 0 \,\, \forall i>0\,.
\end{split}
\end{equation}

Assuming that we have the solution to  a desired order we can then evaluate the on-shell action. For instance, if we compute the action  up to $\order{\lambda_n}$, as appropriate for a contact bulk interaction, we will simply find the contribution from the zeroth order solution, for:
\begin{equation}
\begin{split}
S_{os} 
&= 
	- \int d^{d}x \oint d\ctor  \, \sqrt{-g}\,\left[\frac{ g^{AB}}{2}\Big( \partial_{A}\Phi_0\, \partial_{B}\Phi_0 + 2\lambda_n \partial_A \Phi_1 \, \partial_B \Phi_0 \Big) + \frac{\lambda_n}{n!}\Phi_0^n \right] + \order{\lambda_n^2}\,, \\
&= 
	-\int d^d x  \oint d\ctor\, \,\Bigg[\frac{1}{2}\partial_A \Big(\sqrt{-g}\, g^{AB}\left(\Phi_0 + 2\lambda_n\Phi_1 \right)  \partial_{B}\Phi_0 \Big)\\
&
\hspace{2cm} 
	 - \lambda_n\, \Phi_1
	  \left[ \partial_B \left( \sqrt{-g} g^{AB} \partial_A \Phi_0 \right)  - \sqrt{-g}\, m^2\, \Phi_0 \right]
	   + \frac{\lambda_n}{n!}\sqrt{-g}\,\Phi_0^n  \Bigg] + \order{\lambda_n^2}  \\
&= - \int d^{d}x \oint d\ctor\, \,\Bigg[\frac{1}{2}\partial_A \Big(\sqrt{-g}\, g^{AB} \Phi_0 \partial_{B}\Phi_0 \Big) + \frac{\lambda_n}{n!} \, \sqrt{-g} \, \Phi_0^n  \Bigg] \,.
\end{split}
\end{equation}
In passing from the first to the second line we plugged in the equation of motion. We then see that as expected the first term in the second line is a total derivative, and thus a boundary term. Since our boundary conditions \eqref{eq:Gpbcs} set  $\Phi_1=0$ at the conformal boundary, so its contribution to the boundary term vanishes.  The other term in the expression is the $\Phi_0$ equation of motion which vanishes on-shell.  The action with the surviving terms is  the expression which only cares about  $\Phi_0$ and indeed the contribution to the interaction can be obtained by  writing out $\Phi_0(\ctor, x)$ in terms of the boundary values using boundary-bulk propagators. Given this, we have chosen drop the subscript `0' from $\Phi_0$ in the main text.

\section{Cubic influence functionals in 2d CFTs}
\label{sec:2dcubic}

The computation of the influence functional in the RA basis in a two-dimensional CFT is quite general, since conformal invariance on the plane fixes the 3-point function in Euclidean space. One can obtain the thermal correlation function by conformally mapping the complex plane onto the cylinder. A further analytic continuation with suitable $i\epsilon$ prescription gives the result in Lorentz signature. For a certain choice of operator ordering the result is given in momentum space in \cite{Becker:2014jla}. We will now derive the general influence functions in momentum space from the bulk BTZ geometry. It is worth emphasizing that while holography provides a simple way to get the answer, the result is not particularly holographic, since it is constrained entirely by the underlying conformal symmetry.

For definiteness we will compute the result for three scalar primary operators $\mathcal{O}_i$ with conformal dimensions $\Delta_i$ with $i=1,2,3$. We will assume that the OPE coefficient  is  $C_{123}$ which sets the strength of the bulk vertex (the result presented in the main text corresponds to the special case $\Delta_1 = \Delta_2 = \Delta_3 = \Delta$). We take the bulk action to be given by 
\begin{equation}
\begin{split}
S  &=  
	- \sum_{i=1}^3 \, \oint d\ctor \int d^3 x \, \sqrt{-g}\,\left[ \frac{1}{2} \, g^{AB} \, \partial_{A}\Phi_i \partial_{B}\Phi_i +\frac{1}{2}\, \Delta_i \,  (\Delta_i -2) \,\Phi_i^2 \right]\\
& 
	\qquad \qquad  -C_{123} \int \prod_{i=1}^2 \frac{d^3k_i}{(2\pi)^3} (2\pi)^{3} \delta\left(\sum_{i=1}^{3}k_i\right) \oint d\ctor \, \sqrt{-g}\, \, \prod_{i=1}^3 \Phi_i(\ctor,k_i) \,.
\end{split}
\end{equation}
The cubic influence functional $\ifra_{FFP}$ is then given by
\begin{equation}\label{eq:FFP2d}
\begin{split}
& \ifra_{FFP} (k_1,k_2,k_3) 
 = 
	C_{123}\, \oint d\ctor\, \sqrt{-g}\,   G^+_{\Delta_1}(\zeta, \omega_1, k_1) \, G^+_{\Delta_2}(\zeta, \omega_2, k_2) \,
	G^-_{\Delta_3}(\zeta, \omega_3, k_3) \,e^{\beta \omega_3 (1-\ctor)} \\
&= C_{123} \, \left(1- e^{\beta\, \omega_3}\right)  \int_{r_h}^{r_c} dr\, r \, 
	G^+_{\Delta_1}(r, \omega_1, k_1) \, G^+_{\Delta_2}(r, \omega_2, k_2) \, 
	G^-_{\Delta_3}(r, \omega_3, k_3) \, \left(\frac{r-r_h}{r+r_h}\right)^{i\, \frac{\beta \omega_3}{2\pi}}\,,
\end{split}
\end{equation}
where the BTZ Green's function in radial coordinates is given by
\begin{equation}\label{eq:}
\begin{split}
G^+_{\Delta}(r,\omega, k) &=	\mathcal{N}\, 
	\left( \frac{r_h}{r}\right)^\Delta \,\left( \frac{r}{r+r_h}\right)^{\Delta - \bpt_+ - \bpt_-} \,  {}_2\widetilde{F}_1\left(\bpt_+\,, \bpt_- \,, \bpt_+ + \bpt_- +1 - \Delta \,; 1-\frac{r_h^2}{r^2}\right) \\
	\mathcal{N} &\equiv \frac{\Gamma(\bpt_+) \, \Gamma(\bpt_-)}{\Gamma(\Delta -1) } \,,
\end{split}
\end{equation}
where we have included the normalization factors which set the source term to unit on the boundary of the spacetime.\footnote{ This normalization is different from what is used in the main part of the discussion where we have left the cut-off explicit. Here we choose to normalize the propagator by demanding that $\lim_{r\to \infty} r^{d-\Delta} \, G^+_\Delta =1$. If one uses this normalization then the computation of the two point function proceeds as described in \cref{sec:s2eff} with some minor changes. Firstly, we should pick up  the constant term in the asymptotic expansion to compute $S_{(2)}$. Secondly, there is a renormalization of the operator -- the 2-point functions computed in this fashion are rescaled by a factor of $\frac{\Delta}{2(\Delta-1)}$ relative to the result in \eqref{eq:Iad2}. 
For the computation of the  3-point function, we should supply a factor $r_h^{\Delta_1+\Delta_2+ \Delta_3 -6}$ at the end to ensure that we have the correct scaling of the  correlator.\label{fn:normalizations}}
In the second line we have written out the discontinuity across the branch cut which includes a factor of $e^{-2\beta\omega_3\zeta}$ in terms of the standard BTZ radial coordinate.

The non-normalizable mode in $G^\pm_{\Delta}$ leads to a  $r^{\Delta -2}$  fall-off. This can be seen from the fact that (with $\epsilon= \frac{r_h}{r}$)
\begin{equation}\label{eq:2F1asym}
\begin{split}
{}_2\widetilde{F}_1\left(\bpt_+\,,\bpt_-\,, 1-i\,\frac{\bwt}{\pi} \,; 1-\epsilon^2\right) 
 \;\;\overset{\mathop{\longrightarrow}}{\scriptscriptstyle{\epsilon\to 0}} \;\;
	\frac{1}{\mathcal{N}}\, \epsilon^{2-2\Delta}    
	+ \frac{ \Gamma(1-\Delta)}{\Gamma(\,\bpt_++1-\Delta) \, \Gamma(\,\bpt_- +1-\Delta)} + \cdots
\end{split}
\end{equation}	

This implies that the radial integral is absolutely convergent\footnote{ In general dimensions the conformal three-point function in momentum space has a convergent integral representation for $\Delta_1 + \Delta_2 + \Delta_3 < 2d$, see \cite{Bzowski:2015pba}.} 	if $\Delta_1 + \Delta_2 + \Delta_3 < 4$. We will restrict to this range of conformal dimensions to avoid introducing UV regulators in the computation for the present (a discussion of the regulators can be found in \cref{sec:counterterns}). 

On way to proceed is to use the contour integral representation of the hypergeometric function. A variant of the Barnes integral representation \cite[15.6.7]{NIST:DLMF} gives
\begin{equation}\label{eq:ifIdef}
\begin{split}
{}_2\widetilde{F}_1\left(a\,, b\,, c\,; z\right) 
&= 
	\mathcal{M}(a,b,c)\, 
	 \int_{\mathcal{C}} \, \frac{ds}{2\pi i} \,\Gamma(s) \, \Gamma(c-a-b+s) \, \Gamma(a-s) \, \Gamma(b-s) (1-z)^{-s}\,,  \\
\mathcal{M}(a,b,c) &=  \frac{1}{\Gamma(a)\, \Gamma(b)} \, \frac{1}{\Gamma(c-a)\,\Gamma(c-b)} \,.
\end{split}
\end{equation}
This is valid for $|\arg(1-z)| < \pi$ and the choice of the contour $\mathcal{C}$ is as follows. One separates the poles of the Gamma functions in the integrand into two sets. The first set includes the poles of $\Gamma(s)$ and $\Gamma(c-a-b+s)$ i.e., 
$s \in \mathcal{P}_\text{left} = \{-n  \,,  -n +a+b-c\,| \, n=0,1,2, \cdots\}$. The second set comprises the poles of $\Gamma(a-s)$ and $\Gamma(b-s)$ i.e.,  $s \in \mathcal{P}_\text{right} = \{a-n  \,,  b-n \,| \, n=0,1,2, \cdots\}$. The contour $\mathcal{C}$ is chosen as a separatrix between $\mathcal{P}_\text{left}$ and $\mathcal{P}_\text{right}$. Depending on the range of $z$ and the parameters $a,b,c$ one can either take it to be a vertical contour between$\mathcal{P}_\text{left}$ and $\mathcal{P}_\text{right}$, or a contour that encircles the poles of one of the sets. 

For the BTZ Green's function we can write the ingoing Green's function  using \eqref{eq:lpmdef} and \eqref{eq:Gfn2pdef} as
\begin{equation}\label{eq:}
\begin{split}
G^+_{\Delta}(r,\omega, k) &=	
	\frac{1}{\mathfrak{G}(\bkt_\pm, \widetilde{\Delta})} \,\left( \frac{r}{r+r_h}\right)^{\frac{i}{\pi}\, \bwt} 
	\int_{\mathcal{C}} \,\frac{ds}{2\pi i}\  \mathcal{G}_\Delta(\bkt,\Delta, s)
	\left( \frac{r_h}{r}\right)^{\Delta-2s} \,, \\
\mathcal{G}_\Delta(\bkt_\pm, s)	 
& = 
	\Gamma(s) \, \Gamma(1-\Delta + s ) \, \Gamma(\bkt_++\tfrac{\Delta}{2} - s)\, \Gamma(\bkt_-+\tfrac{\Delta}{2} -s) \,.
\end{split}
\end{equation}
We also have shortened the notation for functions of both  $\bkt_+$ and $\bkt_-$ and written the dependence into simply $\bkt_\pm$
to allow for more compact formulae below. We will do likewise for $\mathfrak{G}(\bkt_\pm, \Delta)$.

The contour we want to pick is one which encircles the poles in the left set 
\begin{equation}\label{eq:}
\mathcal{P}_\text{left} = \{-n  \,,  -n +\Delta -1\,| \, n=0,1,2, \cdots\}\,,
\end{equation}	
and is anchored at $-\infty$. As expected the leading divergence comes from the rightmost pole located at  $s= \Delta -1$ which gives us the  asymptotic behaviour in \eqref{eq:2F1asym}. Taking then $s_i \leq \Delta_i -1$ for $i=1,2,3$ we  see that the influence functional receives contribution from the following radial integral 
\begin{equation}\label{eq:}
\begin{split}
\int_{r_h}^{\infty} \, r\, dr \,& 
	\left( \frac{r_h}{r}\right)^{\sum_{i=1}^3\, (\Delta_i-2s_i)} \,\left( \frac{\,r}{r+r_h}\right)^{\frac{i}{\pi}\, (\bwt_1+ \bwt_2 - \bwt_3)}  \, 
	\left(\frac{r-r_h}{r+r_h}\right)^{\frac{i\,\bwt_3}{\pi}} \\
&  =
	  r_h^2\; \Gamma\left(1+\frac{i\,\bwt_3}{\pi}\right)\;  
	  \frac{\Gamma\left(-1+ \sum_{i=1}^3 \frac{\Delta_i}{2} - s_i\right)}{2\,\Gamma\left(\frac{i\,\bwt_3}{\pi}+\sum_{i=1}^3 \frac{\Delta_i}{2} - s_i \right) } \,. 
 \end{split}
\end{equation}	
We have explicitly used the fact that the integral is absolutely convergent when
\begin{equation}\label{eq:}
\text{Re} \left(\sum_{i=1}^3  (2 s_i- \Delta_i) \right)  < -2  \;\; \Longrightarrow \;\; \Delta_1 + \Delta_2 + \Delta_3 < 4 \,,
\end{equation}	
as argued earlier and have  employed momentum conservation to eliminate $\bwt_2$. The integral as defined appears to have poles when $\bwt_3 = i\,\pi$, but note that the prefactor $(1-e^{\beta\omega_3})$ also vanishes at that point rendering this harmless.

Putting all the pieces together we find:
\begin{equation}\label{eq:}
\begin{split}
\ifra_{FFP}(k_1,k_2,k_3) &= \frac{C_{123}\, r_h^2\, 
	\left(1-e^{\beta \omega_3}\right) \Gamma\left(1+\frac{i\,\bwt_3}{\pi}\right)}{\mathfrak{G}(\bkt_{1\pm},\widetilde{\Delta}_1)
	\, \mathfrak{G}(\bkt_{2\pm},\widetilde{\Delta}_2)\, \mathfrak{G}(\bkt_{3\pm}^*,\widetilde{\Delta}_3)}
	 \left[\prod_{i=1}^3 \int_{\mathcal{C}_i} \frac{ds_i}{2\pi i}\,    K(s_1,s_2,s_3)\right] , \\
	 K(s_1,s_2,s_3) & =  
	\mathcal{G}_{\Delta_1} (\bkt_{1\pm},s_1) \, \mathcal{G}_{\Delta_2} (\bkt_{2\pm},s_2)\, \mathcal{G}_{\Delta_3} (\bkt_{3\pm}^*,s_3) \, 
	\frac{\Gamma\left(-1+ \sum_{i=1}^3 \frac{\Delta_i}{2} - s_i\right)}{2\,\Gamma\left(\frac{i\,\bwt_3}{\pi}+\sum_{i=1}^3 \frac{\Delta_i}{2} - s_i \right) } .
\end{split}
\end{equation}
In writing the above expression we have used the fact that the sign reversal of $\omega_3$ owing to the contribution coming from outgoing Green's function $G^-(r,\omega_3,k_3)$ can be expressed by complex conjugation of the lightcone momenta, for  $\bkt_{\pm}^*(\omega,k) =\bkt_{\mp}(-\omega,k)$.

The main point to note is that the choice of parameters we have made is such that the contribution from the hypergeometric function arising from the radial integral is completely regular in $s_i$. This means that we can close the contours $\mathcal{C}_i$ to the left picking up the contributions from the poles in $\mathcal{P}_+^i$ for $i=1,2,3$. This implies that we can write the final result as a triple sum:
\begin{equation}\label{eq:J123}
\begin{split}
&\mathfrak{J}(k_1,k_2,k_3) \equiv \left(\prod_{i=1}^3 \int_{\mathcal{C}_i} \frac{ds_i}{2\pi i}\, \right)   K(s_1,s_2,s_3) 
\\
&\hspace{2cm} =  	  
\sum_{n_1,n_2,n_3}^\infty \,	\left(\prod_{i=1}^3 \frac{(-)^{n_i} \Gamma(1 - \delta_i -n_i)}{\Gamma(1+n_i) \, \Gamma(1-\delta_i-2\,n_i)} \right) \; 
	 \frac{\Gamma\left(-1+\sum_{i=1}^3  \frac{\delta_i+2n_i}{2}\right)}{2\,\Gamma\left(\sum_{i=1}^3  \frac{\delta_i +2n_i}{2} + \frac{i\, \bwt_3}{\pi}\right) }\\
&
	\hspace{4cm} \times
		  \mathfrak{G} \left(\bkt_{1\pm}, \delta_1+2n_1\right) \, \mathfrak{G}\left(\bkt_{2\pm} ,\delta_2+2n_2\right) \, 
		  \mathfrak{G}\left( \bkt_{3\pm}^*, \delta_3+2n_3\right) \,.
\end{split}
\end{equation}
We have written this result in a shorthand notation: accounting for the poles from $s_i = -n_i$ and from $s_i = -n_i + \Delta_i-1$ is tantamount to picking contributions involving either the operator dimension or that of its shadow for each $i$. This is encoded in binary choice of our parameter $\delta_i$, viz.,  
\begin{equation}\label{eq:}
\delta_i \in \{\Delta_i, 2-\Delta_i\}  \,.
\end{equation}	
The result is expressed in terms of the function $\mathfrak{G}$ introduced during the analysis of the 2-point function, Eq.~\eqref{eq:Gfn2pdef}, for brevity. To be explicit
\begin{equation}\label{eq:}
 \mathfrak{G} \left(\bkt_\pm, \delta+2n\right) = \Gamma\left(\frac{i(\bqt-\bwt)}{2\pi} + \frac{\delta +2n}{2}\right) \, \Gamma\left(-\frac{i(\bqt
 +\bwt)}{2\pi} + \frac{\delta +2n}{2}\right) \, \Gamma\left(1- \delta -2n\right).
\end{equation}	
Note that the first term in the parenthesis in the second line of \eqref{eq:J123} contains the residues from the poles and includes a ratio arising from our having expressed the result in terms of the functions $\mathfrak{G}$ defined above. 

To understand the analytic structure it is helpful to reinstate $\omega_2$ and use energy-momentum conservation to eliminate 
$\omega_3$ instead. Noting that 
\begin{equation}\label{eq:}
\bkt_{3\pm}^*  =   \mp \,i\, \frac{\bqt_3 \mp \bwt_3}{2\pi} = \pm \,i\,\frac{\bqt_1 + \bqt_2 \mp (\bwt_1+ \bwt_2)}{2\pi} 
\equiv \bkt_{12\pm}\,,
\end{equation}	
we can write the three-point influence functional by replacing  $\mathfrak{G}\left( \bkt_{3+}^*, \bkt_{3-}^*,\delta_3+2n_3\right)$ with the simpler expression $\mathfrak{G}\left( \bkt_{12+},  \bkt_{12-},\delta_3+2n_3\right)$. 

The full influence functional then can be expressed as (after supplying the factors of $r_h$ for dimensional reasons, see \cref{fn:normalizations})
\begin{equation}\label{eq:2dffpsum}
\begin{split}
\ifra_{FFP}&(k_1,k_2) = 
	 C_{123}\, r_h^{\Delta_1+ \Delta_2+ \Delta_3 -4}\, 
	\left(1-e^{-\beta (\omega_1+\omega_2)}\right) 
	\Gamma\left(1-\frac{i\,(\bwt_1+\bwt_2)}{\pi}\right)\, \\
& 	\qquad \quad \Bigg[\sum_{n_1,n_2,n_3}^\infty \,	\left(\prod_{i=1}^3 \frac{(-)^{n_i} \Gamma(1 - \delta_i -n_i)}{\Gamma(1+n_i) \, \Gamma(1-\delta_i-2\,n_i)} \right) \; 
	 \frac{\Gamma\left(-1+\sum_{i=1}^3  \frac{\delta_i+2n_i}{2}\right)}{2\,\Gamma\left(\sum_{i=1}^3  \frac{\delta_i +2n_i}{2} + \frac{i\, \bwt_3}{\pi}\right) }\\
&
	\hspace{2.5cm} \times
		  \frac{\mathfrak{G} \left(\bkt_{1\pm}, \delta_1+2n_1\right) \, \mathfrak{G}\left(\bkt_{2\pm}, \delta_2+2n_2\right) \, 
		  \mathfrak{G}\left( \bkt_{12\pm}, \delta_3+2n_3\right)}{\mathfrak{G}(\bkt_{1\pm},\widetilde{\Delta}_1)	\, \mathfrak{G}(\bkt_{2\pm},\widetilde{\Delta}_2)\, \mathfrak{G}(\bkt_{12\pm},\widetilde{\Delta}_3)} \Bigg].
\end{split}
\end{equation}

The analytic structure of the influence function can be read off without further effort. The terms on the second line of 
$\ifra_{FFP}(k_1,k_2)$ involving the sum of $n_i$ are regular as a function of $\omega_1$ and $\omega_2$. The only singularities are from the Gamma functions containing $\bpt_{k\pm} = \bkt_{k\pm}  +\tfrac{\Delta}{2} $, i.e., the pieces that occur already in the 2-point function. Of the eight possible choices of $\delta_i$ we note that there can only be poles when $\delta_i  = \Delta_i$. When $\delta_i = 2-\Delta_i$ the Gamma functions in the denominator factor also have a pole which cancels against that of the numerator leaving a finite answer. We conclude that the correlator is analytic in the upper half of the complex $\omega_1$ and $\omega_2$ planes and encounters the usual quasinormal type poles in the lower half-planes. 

One can in fact carry out the two of the three sums by realizing that some of the sums are the defining expressions for the generalized hypergeometric function. For instance, the sum over $n_1$ leads to ${}_3F_2$ with arguments comprising of the combination $n_2+n_3$:
\begin{align}\label{eq:}
\sum_{n_1=0}^\infty \frac{(-)^{n_1}}{\Gamma(n_1+1)}\, & \Gamma(1-\delta_1-n_1) \,\Gamma(\bkt_{1+}  + \tfrac{\delta_1}{2} + n_1) \,\Gamma(\bkt_{1-}  + \tfrac{\delta_1}{2} + n_1)\,
\frac{\Gamma\left(-1+\sum_{i=1}^3  \frac{\delta_i+2n_i}{2}\right)}{2\,\Gamma\left(\sum_{i=1}^3  \frac{\delta_i +2n_i}{2} + \frac{i\, \bwt_3}{\pi}\right) } \nonumber \\
& = \mathfrak{G}(\bkt_{1\pm}, \delta_1) \, \frac{\Gamma\left(-1+ \tfrac{\delta_1+\delta_2+\delta_3}{2}+n_2+n_3\right)}{2\,\Gamma\left(\tfrac{\delta_1+\delta_2+\delta_3}{2}+n_2+n_3 + \tfrac{i\,\bwt_3}{\pi}\right) }  \\
& \qquad \times \ {}_3F_2\left(
		\begin{array}{c}
		\bkt_{1+} + \tfrac{\delta_1}{2}\,, \bkt_{1-} + \tfrac{\delta_1}{2}\,, -1+ \tfrac{\delta_1+\delta_2+\delta_3}{2}+n_2+n_3 \\
		 \delta_1\,, \tfrac{\delta_1+\delta_2+\delta_3}{2}+n_2+n_3 + \tfrac{i\,\bwt_3}{\pi}
		\end{array}\;; 1 
\right) .\nonumber
\end{align}	
One can then carry out the sum over $n_2$ after a  shift $n_3 +n_2= n$, which allows performing the $n_2$ sum. The result is given as 
\begin{align}\label{eq:}
\mathfrak{J}(k_1,k_2,k_3) 
&= \mathfrak{G}(\bkt_{1\pm}, \delta_1)\, \mathfrak{G}(\bkt_{2\pm}, \delta_2) \, 
	\sum_{n=0}^\infty \, \frac{(-)^n\,\Gamma(1-\delta_3-n) }{\Gamma(n+1)\,\Gamma(1-\delta_3-2n)} \, 
	\frac{\Gamma(-1+n+ \tfrac{\delta_1+\delta_2+\delta_3}{2})}{2\,\Gamma(n+ \frac{\delta_1+\delta_2+\delta_3}{2} +\tfrac{i\, \bwt_3}{\pi})}
	\nonumber \\
& \qquad \times
		\mathfrak{G}(\bkt_{3\pm}^*, \delta_3+2n)\ {}_3F_2\left(
		\begin{array}{c}
		\bkt_{1+} + \tfrac{\delta_1}{2}\,, \bkt_{1-} + \tfrac{\delta_1}{2}\,, -1+ \tfrac{\delta_1+\delta_2+\delta_3}{2}+n \\
		 \delta_1\,, \tfrac{\delta_1+\delta_2+\delta_3}{2}+n + \tfrac{i\,\bwt_3}{\pi}
		\end{array}\;; 1
	\right) \nonumber \\
& \qquad\qquad  \times 
	\ {}_4F_3\left(
		\begin{array}{c}
		 \bkt_{2+} + \tfrac{\delta_2}{2}\,, \bkt_{2-} + \tfrac{\delta_2}{2}\,, -n\,,1-n- \delta_3  \\
		 \delta_2\,,  1-\bkt_{3+}^*-\tfrac{\delta_3}{2} - n \,,  1-\bkt_{3-}^*-\tfrac{\delta_3}{2} - n 
		\end{array}\;; 1
	\right) .
\end{align}	
This is the expression quoted in the main text in \eqref{eq:2dffp} and \eqref{eq:2dffpa}, modulo a reversion to our standard notation and accounting for a symmetry factor from the $k_1 \leftrightarrow k_2$ swap of the $F$-type sources.

\section{Counterterm analysis for influence functionals}
\label{sec:counterterns}

We prove the statements made in \cref{sec:sneff} regarding the renormalization of the nonlinear influence functional. We will demonstrate the divergence of the integrals noted in the main text and then argue that a suitable temperature dependent mixing of the difference source into the average source serves to give a counterterm action that is consistent with microscopic unitarity.

The staring point of our analysis is the time-domain solution of the free massless scalar equation on the gravitational SK contour \eqref{eq:Phitime} which we reproduce here with the bare sources explicitly marked.  
\begin{equation} \label{eq:PhitimeE}
 \begin{split}
\Phi
&=
	\left(G_{0,0}^+ +\frac{i}{2}  G^+_{1,0} \,\beta\partial_t \right) 
	\bigg\{ \Jb_a + \frac{i}{8}\,\beta\partial_t \Jb_d+ \Jb_d\left(\ctor-\frac{1}{2}  +\Gpr{1,0} \right) \\
& 
	\qquad \qquad -\;\frac{i}{2} \,\beta\partial_t \Jb_d \left(\ctor-\frac{1}{2} + \Gpr{1,0} \right)^2\bigg\} 
	+  \order{\left(\beta\partial_ t\right)^2} 
\end{split}
\end{equation}
where we are using \eqref{eq:Gprdef}.  We  work to linear order in the gradients, leaving a more detailed analysis of the gradient expansion for the future.
 
 To compute the influence phase to linear order in perturbation theory, this unperturbed solution is sufficient. Taking the $n^{\rm th}$ power of this solution, we obtain 
 \begin{equation}
 \begin{split}
\frac{\Phi^n}{n!}&= 
	\Bigg\{\sum_{k=0}^n \frac{1}{(n-k)!} \, \left(G_{0,0}^+\right)^n \left(\Jb_a+\frac{i}{8}\beta\partial_t \Jb_d\right)^{n-k}\\
&\qquad
	 \times\frac{1}{k!} \left[ \Jb_d\left( \ctor-\frac{1}{2} + \Gpr{1,0}\right)-
	\frac{i}{2} \beta\partial_t \Jb_d \left( \ctor-\frac{1}{2} +\Gpr{1,0} \right)^2\right]^k\Bigg\}\\
&
	\qquad \quad+\; n\, \frac{i}{2} \beta\partial_t \left[\left( G^+_{0,0}\right)^{n-1} \, G^+_{1,0} \left(\Jb_a+ \Jb_d\left( \ctor-\frac{1}{2} +\Gpr{1,0}\right)\right)\right]+ \order{\left(\beta\partial_ t\right)^2} \ .  
\end{split}
\end{equation}
Note that in the above expression we have expanded the solution accurately to linear order in time-derivatives and we have combined all the contributions coming from $\frac{i}{2}  G^+_{1,0} \beta\partial_t$ into a total derivative using Leibniz rule.

We now compute the integral over the bulk  gravitational SK contour. Only the terms with branch cuts can contribute to the radial integral: this  implies we can drop the $k=0$ term in the sum above since it is analytic. As expected we get no contribution involving only the average sources. We can  also drop the total time derivative in the last line since it  gives a boundary contribution to the influence functional (this is partially the reason for working in coordinate space). Thus, we have
\begin{equation}
 \begin{split}
\int d^dx \oint d\ctor\ \sqrt{-g}\, \frac{\Phi^n}{n!}&=
	\int d^dx \oint d\ctor\  \sqrt{-g}\, 
		\Bigg\{\sum_{k=1}^n \frac{1}{(n-k)!}\, \left(G^+_{0,0}\right)^n  \left(\Jb_a+\frac{i}{8}\beta\partial_t \Jb_d\right)^{n-k} \\
&\quad  
	\times\left[\frac{( \Jb_d)^k}{k!} \left( \ctor-\frac{1}{2} + \Gpr{1,0} \right)^k-
	\frac{i}{2} \beta\partial_t \Jb_d \frac{( \Jb_d)^{k-1}}{(k-1)!}\left( \ctor-\frac{1}{2} +\Gpr{1,0} \right)^{k+1}\right]\Bigg\}\\
&\quad 
	+ \order{\left(\beta\partial_ t\right)^2} \ .  
\end{split}
\end{equation}
Defining the integrals 
\begin{equation}\label{eq:ifInk}
\begin{split}
\ifIb_{n,k}  &\equiv 
	\oint d\ctor \, \sqrt{-g}\, \left(G^+_{0,0}\right)^n \left( \ctor + \Gpr{1,0} -\frac{1}{2}\right)^k  \\
& =  
	\oint dr\, r^{d-1}\,  \left(G^+_{0,0}\right)^n \left( \ctor  + \Gpr{1,0} -\frac{1}{2}\right)^k \,, 
\end{split}
\end{equation}
we get the following contribution to the influence functional
\begin{equation}
\begin{split}\label{eq:BareInfE}
\int &d^dx \oint d\ctor\ \sqrt{-g}\, \frac{\Phi^n}{n!}\\
&=\int d^dx  \sum_{k=1}^n\frac{1}{(n-k)!}\left(\Jb_a+\frac{i}{8}\beta\partial_t \Jb_d\right)^{n-k} 
\left[\ifIb_{n,k} \frac{( \Jb_d)^k}{k!}
-\ifIb_{n,k+1} \frac{( \Jb_d)^{k-1}}{(k-1)!}\frac{i}{2} \beta\partial_t \Jb_d\right] .  
\end{split}
\end{equation}

To proceed we  need to estimate the integrals.  Since we have explicit expressions for the massless scalar field we will first describe the computation in that case, before outlining the general story.

\subsection{Divergence structure for a marginal operator}
\label{sec:mass0}

For the massless scalar, for which the radial functions appearing in the gradient expansion are given in \eqref{eq:Gpexpd} we can simplify the integrals $\ifI_{n,k}$ defined in \eqref{eq:ifInk} since $G^+_{0,0} =1$. Dropping the subscript $n$, since it is unnecessary, we focus on the integrals $\ifI$ defined in \eqref{eq:ifIk0}, viz.,  
\begin{equation}\label{eq:}
\ifIb_k  = \oint dr \, r^{d-1} \left( \ctor + G_{1,0}^+ - \frac{1}{2} \right)^k
\end{equation}	
We then can use the explicit form of $G^+_{1,0}$ in \eqref{eq:Gpexpd} 
\begin{equation}\label{eq:}
G_{1,0}^+ = - \int_0^\zeta d\zeta' \left( 1- \left(\frac{r_h}{r'}\right)^{d-1}\right)
\end{equation}	
and immediately obtain $\zeta + G_{1,0}^+  \sim \frac{1}{r^d}$. Furthermore, since $G_{1,0}^+$ is continuous across the grSK contour the integrals $\ifIb_k $ are simply:
\begin{equation}\label{eq:}
\ifIb_k  = 
	\int_{r_h}^{r_c} dr \, r^{d-1} \bigg[ \left( \ctor+  G_{1,0}^+ + \frac{1}{2}\right)^k - \left( \ctor+  
	G_{1,0}^+ -\frac{1}{2}\right)^k \bigg].
\end{equation}	
To understand the divergence structure it is sufficient to use the asymptotic expansion, though we will use the explicit form. 

First consider the situation with the argument $k$ being an odd integer. Then, 
\begin{equation}\label{eq:}
\begin{split}
 \ifI_{2k+1}\sim  \int^{r_c} dr \, r^{d-1} \bigg[ \frac{1}{4^k} + \binom{2k+1}{2} \, \frac{1}{4^{k-1}}   (\ctor + G_{1,0}^+)^2 + \cdots
 \bigg]
 & =  \frac{1}{4^k \, d}\, r_c^d  + \order{r_c^{-1}}
 \end{split}
\end{equation}	
To check that the other terms do not contribute realize that 
\begin{equation}\label{eq:}
\begin{split}
\int^{r_c} dr \,r^{d-1} \, \log^2\left( \frac{1-\left( \frac{r_h}{r}\right)^d}{1-\left( \frac{r_h}{r_c}\right)^d} \right)  &
= \frac{r^{d}}{d} \, \log^2\left( \frac{1-\left( \frac{r_h}{r}\right)^d}{1-\left( \frac{r_h}{r_c}\right)^d} \right)  \Bigg|^{r_c} \\ 
&  - 2 \int^{r_c} dr  \, \frac{ r_h^d}{r^2}  \frac{1}{1-\left( \frac{r_h}{r}\right)^d} \, \log \left( \frac{1-\left( \frac{r_h}{r}\right)^d}{1-\left( \frac{r_h}{r_c}\right)^d} \right)
\end{split}
\end{equation}	
The first term vanishes at the cut-off and the second is a convergent integral. 
Thus we have established the first of the relations given in \eqref{eq:Lambdam0}, for indeed
\begin{equation}\label{eq:}
\ifIb_{2k+1} = \ifIr_{2k+1} + \frac{1}{4^k} \, \frac{r_c^d}{d} 
\end{equation}	

When the argument $k$ is even case we have a  divergence when we pick up the linear term in $\ctor  + G_{1,0}^+$, for integrating by parts we find 
\begin{equation}\label{eq:}
\begin{split}
 \ifIb_{2k}
 & 
 	\sim  \int^{r_c} dr \, r^{d-1} \bigg[ 2\, \binom{2k}{1}\, \frac{1}{2^{2k-1}}   (\ctor + G_{1,0}^+) + \cdots \bigg] \\
 & = 
 	\frac{k}{4^{k-1}\, \pi i } \int^{r_c}\, dr \, r^{d-1}  \log\left( \frac{1-\left( \frac{r_h}{r}\right)^d}{1-\left( \frac{r_h}{r_c}\right)^d} \right)  \\
&=	-	\frac{k}{4^{k-1}\, \pi i } \int^{r_c} \, dr \, \frac{ r_h^d}{r} \, \frac{1}{1-\left( \frac{r_h}{r}\right)^d}  \\
& = - \frac{k}{4^{k-1}\, \pi i }\,r_h^d\, \log \frac{r_c}{r_h}  + \order{1}\,.
 \end{split}
\end{equation}
thus proving the second relation in \eqref{eq:Lambdam0}
\begin{equation}\label{eq:}
\ifIb_{2k} = \ifIr_{2k} + \frac{i}{ \pi\, 4^{k-1}}\, r_h^d\, \log \frac{r_c}{r_h} \,.
\end{equation}	

For completeness let us record the integrals for the massless scalar that enter into the computation of the quartic influence functional. In intermediate steps we define a rescaled radial variable  
\[
	y = \frac{1}{\varrho} = \left( \frac{r_h}{r}\right)^d
\]
which helps simplify the integration.
\begin{equation}
\begin{split}
\ifIb_1 &= \oint dr\, r^{d-1} \, \left( \ctor +G^+_{1,0} -\frac{1}{2}\right) 
= \int_{r_h}^{r_c} dr\, r^{d-1} = \frac{1}{d} \left( r_c^d - r_h^d \right)\,.
\end{split}
\end{equation}
\begin{equation}
\begin{split}
\ifIb_2 &= \oint dr\, r^{d-1} \, \left( \ctor +G^+_{1,0} -\frac{1}{2}\right)^2 = 2 \int_{r_h}^{r_c} dr\, r^{d-1} \left( \ctor + G^+_{1,0} \right) \\
&=  \frac{1}{2\pi i} \frac{r_h^d}{d} \int_{y_c}^1 \frac{dy}{y^2} \, \log \left( \frac{1-y}{1-y_c} \right) = -\frac{r_h^d}{\pi i} \log\frac{r_c}{r_h}\,.
\end{split}
\end{equation}
\begin{equation}
\begin{split}
\ifIb_3 &= \oint dr\, r^{d-1} \, \left( \ctor +G^+_{1,0} -\frac{1}{2}\right)^3  = \int_{r_h}^{r_c} dr\, r^{d-1} \left[ \frac{1}{4} + 3(\ctor + G^+_{1,0})^2 \right] \\
& = \frac{r_c^d}{4d} - \frac{r_h^d}{4d} + \frac{r_h^d}{d}\frac{3}{(2\pi i)^2} \int_{y_c}^{1} \frac{dy}{y^2}\, \log^2\left( \frac{1-y}{1-y_c}  \right)= \frac{r_c^d}{4d} - \frac{r_h^d}{2d}\,.
\end{split}
\end{equation}
\begin{equation}
\begin{split}
\ifIb_4 &= \oint dr\, r^{d-1} \, \left( \ctor +G^+_{1,0} -\frac{1}{2}\right)^4 = \int_{r_h}^{r_c} dr\, r^{d-1} \left[ 4(\ctor + G^+_{1,0})^3 + (\ctor + G^+_{1,0}) \right] \\ 
&= \frac{r_h^d}{d} \left[\frac{4}{(2\pi i)^3} \int_{y_c}^1 \frac{dy}{y^2} \log^3 \frac{1-y}{1-y_c} + \frac{1}{2\pi i} \int_{y_c}^{1} \frac{dy}{y^2} \log\frac{1-y}{1-y_c} \right] \\
&= \frac{1}{2\pi i} \frac{r_h^d}{d}\, \left( \log(y_c) + \frac{6}{\pi^2} \text{Li}_3(1-y_c) \right) \\
&= -\frac{r_h^d}{2\pi i} \log\frac{r_c}{r_h} + \frac{1}{2\pi i} \frac{r_h^d}{d} \frac{6}{\pi^2} \zeta(3) + \order{r_c^{-d}}\,.
\end{split}
\end{equation} 

\subsection{Divergence structure for relevant operators}
\label{sec:massive}

We shall now generalize the discussion of the divergence structure to an arbitrary operator of dimension $\Delta$. We have derived hitherto the general expressions for the functions entering the gradient expansion in \eqref{eq:Gmnsol}.  While there are still integrals to be done, for the purposes of analyzing the divergence structure, it suffices to exploit the standard AdS asymptotics to extract the leading behaviour.  

We recall that,  $G^+_{0,0}$ solves the massive wave equation in the Schwarzschild-\AdS{d+1} geometry and thus has both the non-normalizable $r^{\Delta-d}$ and the normalizable $r^{-\Delta}$ fall-offs. As written in \eqref{eq:G00sol} from the explicit solution we can see the asymptotic behaviour to be given by 
\begin{equation}\label{eq:}
\varrho_c^{\nu-1}\, G_{0,0}^+ =  \varrho^{\nu-1}  \left(1- \frac{\nu-1}{2}\, \frac{1}{\varrho} + \cdots \right) + 
\frac{\Gamma(\nu)^2 \, \Gamma(1-2\nu)}{ \Gamma(1-\nu)^2\,\Gamma(-1+2\nu)} \,\varrho^{-\nu} \left( 1 + \frac{\nu}{2} \, \frac{1}{\varrho} + \cdots\right)
\end{equation}	
where $\nu = \frac{\Delta}{d}$ was defined earlier in \eqref{eq:nudef}. We recognize the leading term as the non-normalizable mode $\varrho^{\nu-1} \sim r^{\Delta -d}$ and the second series as the normalizable mode $\varrho^{-\nu} \sim r^{-\Delta}$.  Often it is convenient to normalize the Green's function to have a unit source,  which would imply:
\begin{equation}\label{eq:}
\begin{split}
G^+_{0,0} &= \frac{1}{r^{d-\Delta}}\left( 1+ \order{r^{-d}} \right) + s_\Delta\,  \frac{1}{r^\Delta} \left( 1+ \order{r^{-d}}\right) \\
s_\Delta &= \frac{\Gamma(\nu)^2 \, \Gamma(1-2\nu)}{ \Gamma(1-\nu)^2\,\Gamma(-1+2\nu)} \, r_h^{2\Delta-d}  \,.
\end{split}
\end{equation}	
Notice that the subleading term in the source expansion appears at the order $r^{\Delta -2d}$ which is faster than the normalizable fall-off of $r^{-\Delta}$ provided $\Delta  < d $. So for all relevant operators we can ignore the subleading term in the expansion of the source. The special case of a marginal operator $\Delta  =d$ was dealt with explicitly above in \cref{sec:mass0}.

To proceed we need estimates for the combination $ \ctor+  \Gpr{1,0}$ which appears in the functions 
$\ifIb_{n,k}$ defined in \eqref{eq:ifInk}. We recall that this is the combination we solved for in \eqref{eq:G01zeta}. Expanding out the integrand for large $\rho$ we discern the asymptotic behaviour, directly from \eqref{eq:G01zeta0},
\begin{equation}\label{eq:}
\ctor+  \Gpr{1,0}\sim   \int \frac{d\bar{\varrho}}{\bar{\varrho}^{\,2\nu}} =   \frac{d}{d-2\Delta} \, \left(\frac{r_h}{r} \right)^{2\Delta-d}+ \cdots\,.
\end{equation}	
In deriving this expression we have used $P_{-\nu}(\varrho =1) = 1$ and also accounted for the normalization factors in the source term of $G_{0,0}^+$. Note that the most delicate case is when $\Delta = \frac{d}{2}$ for then $G_{0,0}^+ \sim \frac{\log r}{r^{\frac{d}{2}}}$, which is nevertheless convergent. Recall that we are sticking to standard AdS boundary conditions, so $\Delta\geq \frac{d}{2}$.

Armed with this information we can proceed to estimate the integrals $\ifIb_{n,k}$ given in \eqref{eq:ifInk}. We again note that the Green's functions $G_{n,m}^+$ are continuous on the grSK contour, so the only contribution to the contour integral comes from the explicit factor of $\ctor$. We can therefore write
\begin{equation}\label{eq:}
\ifIb_{n,k}  = \int  dr\, r^{d-1} \, (G_{0,0}^+)^n \bigg[ \left( \ctor+ \Gpr{1,0}+ \frac{1}{2} \right)^k - 
 \left(  \ctor+ \Gpr{1,0}- \frac{1}{2} \right)^k \bigg] .
\end{equation}	

The leading divergence comes from the bare contribution in the case of odd argument, whence
\begin{equation}\label{eq:}
\begin{split}
\ifIb_{n,2k+1} 
&\sim 
	\frac{1}{4^k} \int^{r_c} \, dr \,r^{d-1} \left( G^+_{0,0}\right)^{n}   \sim \frac{1}{4^k} \int^{r_c} \, dr \,  r^{d-1 - n\,(d-\Delta)} \,.
\end{split}
\end{equation}	
This is absolutely convergent when 
\begin{equation}\label{eq:}
(1-n)\, d  + n\,\Delta -1< -1 \;\; \Longrightarrow \;\; \Delta < \frac{n-1}{n}\, d \,, 
\end{equation}	
but otherwise predicts a divergence 
\begin{equation}\label{eq:}
\ifIb_{n,2k+1} \sim  \frac{1}{4^k}\, \frac{r_c^{n\Delta - (n-1) d }}{n \Delta - (n-1) \,d} + \text{regular}
\end{equation}	
On the other hand for even argument we would estimate:
\begin{equation}\label{eq:}
\begin{split}
\ifIb_{n,2k} 
&\sim 
	\frac{2k}{4^{k-1}} \int^{r_c} \, dr\, r^{d-1} \left(G^+_{0,0} \right)^{n} \left( \ctor+  \Gpr{1,0}\right) 
\sim  \frac{2k}{4^{k-1}} \int^{r_c} \, dr \,  r^{-1+(n-2)(\Delta-d)}
\end{split}
\end{equation}
leading to 
\begin{equation}\label{eq:}
\ifIb_{n,2k} \sim \frac{2k}{4^{k-1}}\, \frac{r_c^{(n-2) (\Delta -  d) }}{(n-2) (\Delta - d)} + \cdots .
\end{equation}	
For $n>3$ we thus end up with a convergent integral for all relevant operators ($\Delta <d$).So the only renormalization necessary for a relevant operator is to remove the power-law divergence in the functions $\ifIb_{n,2k+1}$ for the range of $\Delta$ specified above.

\input{gravsk-refs}

\end{document}

%% file: gravsk-macros.tex
\definecolor{rust}{rgb}{0.8,0.2,0.2}

\newcommand{\prn}[1]{\left ( #1 \right )}

\def\lads{\ell_\text{AdS}}
\def\AdS#1{AdS$_{#1}$}
\def\SAdS#1{Schwarzschild-AdS$_{#1}$}
\def\tE{t_{_\text{E}}}

\def\ctor{\zeta}
\def\bwt{\mathfrak{w}}
\def\bqt{\mathfrak{q}}
\def\bpt{\mathfrak{p}}
\def\bkt{\mathfrak{K}}
\def\ifad{\mathfrak{I}}
\def\ifra{\mathcal{I}}
\def\ifI{F}
\def\ifIb{F^{\sf b}}
\def\ifIr{F^{\sf r}}
\def\Fnk{\mathfrak{F}}
\def\Jb{J^{\sf b}}
\def\Jr{J^{\sf r}}
\def\Gpr#1{\widetilde{G}^+_{#1}}

\newcommand{\skR}{\text{\tiny R}}
\newcommand{\skL}{\text{\tiny L}}

\newcommand{\JF}{J_{_{\bar F}}}
\newcommand{\JP}{J_{_{\bar P}}}

%% file: gravsk-refs.tex
\providecommand{\href}[2]{#2}\begingroup\raggedright\endgroup

%% file: gravSK-spin0-arXiv-v3.bbl
\begin{thebibliography}{10}

\bibitem{Horowitz:1999jd}
G.~T. Horowitz and V.~E. Hubeny, {\it {Quasinormal modes of AdS black holes and
  the approach to thermal equilibrium}},  {\em Phys. Rev.} {\bf D62} (2000)
  024027, [\href{http://arxiv.org/abs/hep-th/9909056}{{\tt hep-th/9909056}}].

\bibitem{Policastro:2002se}
G.~Policastro, D.~T. Son, and A.~O. Starinets, {\it {From AdS / CFT
  correspondence to hydrodynamics}},  {\em JHEP} {\bf 09} (2002) 043,
  [\href{http://arxiv.org/abs/hep-th/0205052}{{\tt hep-th/0205052}}].

\bibitem{Herzog:2007ij}
C.~P. Herzog, P.~Kovtun, S.~Sachdev, and D.~T. Son, {\it {Quantum critical
  transport, duality, and M-theory}},  {\em Phys. Rev.} {\bf D75} (2007)
  085020, [\href{http://arxiv.org/abs/hep-th/0701036}{{\tt hep-th/0701036}}].

\bibitem{Feynman:1963fq}
R.~Feynman and J.~Vernon, F.L., {\it The theory of a general quantum system
  interacting with a linear dissipative system},  {\em Annals Phys.} {\bf 24}
  (1963) 118--173.

\bibitem{Caldeira:1982iu}
A.~Caldeira and A.~Leggett, {\it {Path integral approach to quantum Brownian
  motion}},  {\em Physica A} {\bf 121} (1983) 587--616.

\bibitem{Breuer:2002pc}
H.~Breuer and F.~Petruccione, {\em {The theory of open quantum systems}}.
\newblock Oxford University Press, 2002.

\bibitem{Schlosshauer:2003zy}
M.~Schlosshauer, {\it {Decoherence, the Measurement Problem, and
  Interpretations of Quantum Mechanics}},  {\em Rev.\ Mod.\ Phys.} {\bf 76}
  (2004) 1267--1305, [\href{http://arxiv.org/abs/quant-ph/0312059}{{\tt
  quant-ph/0312059}}].

\bibitem{Sieberer:2015svu}
L.~Sieberer, M.~Buchhold, and S.~Diehl, {\it {Keldysh Field Theory for Driven
  Open Quantum Systems}},  {\em Rept.\ Prog.\ Phys.} {\bf 79} (2016), no.~9
  096001, [\href{http://arxiv.org/abs/1512.00637}{{\tt arXiv:1512.00637}}].

\bibitem{Lombardo:1995fg}
F.~Lombardo and F.~D. Mazzitelli, {\it {Coarse graining and decoherence in
  quantum field theory}},  {\em Phys. Rev.} {\bf D53} (1996) 2001--2011,
  [\href{http://arxiv.org/abs/hep-th/9508052}{{\tt hep-th/9508052}}].

\bibitem{Agon:2014uxa}
C.~Agon, V.~Balasubramanian, S.~Kasko, and A.~Lawrence, {\it {Coarse Grained
  Quantum Dynamics}},  {\em Phys.\ Rev.\ D} {\bf 98} (2018), no.~2 025019,
  [\href{http://arxiv.org/abs/1412.3148}{{\tt arXiv:1412.3148}}].

\bibitem{Avinash:2017asn}
A.~Baidya, C.~Jana, R.~Loganayagam, and A.~Rudra, {\it {Renormalization in open
  quantum field theory. Part I. Scalar field theory}},  {\em JHEP} {\bf 11}
  (2017) 204, [\href{http://arxiv.org/abs/1704.08335}{{\tt arXiv:1704.08335}}].

\bibitem{Agon:2017oia}
C.~Agón and A.~Lawrence, {\it {Divergences in open quantum systems}},  {\em
  JHEP} {\bf 04} (2018) 008, [\href{http://arxiv.org/abs/1709.10095}{{\tt
  arXiv:1709.10095}}].

\bibitem{Gao:2018bxz}
P.~Gao, P.~Glorioso, and H.~Liu, {\it {Ghostbusters: Unitarity and Causality of
  Non-equilibrium Effective Field Theories}},  {\em JHEP} {\bf 03} (2020) 040,
  [\href{http://arxiv.org/abs/1803.10778}{{\tt arXiv:1803.10778}}].

\bibitem{Avinash:2019qga}
Avinash, C.~Jana, and A.~Rudra, {\it {Renormalisation in Open Quantum Field
  theory II: Yukawa theory and PV reduction}},
  \href{http://arxiv.org/abs/1906.10180}{{\tt arXiv:1906.10180}}.

\bibitem{Faulkner:2010tq}
T.~Faulkner and J.~Polchinski, {\it {Semi-Holographic Fermi Liquids}},  {\em
  JHEP} {\bf 06} (2011) 012, [\href{http://arxiv.org/abs/1001.5049}{{\tt
  arXiv:1001.5049}}].

\bibitem{Gibbons:1976ue}
G.~W. Gibbons and S.~W. Hawking, {\it {Action Integrals and Partition Functions
  in Quantum Gravity}},  {\em Phys. Rev.} {\bf D15} (1977) 2752--2756.

\bibitem{Haehl:2016pec}
F.~M. Haehl, R.~Loganayagam, and M.~Rangamani, {\it {Schwinger-Keldysh
  formalism. Part I: BRST symmetries and superspace}},  {\em JHEP} {\bf 06}
  (2017) 069, [\href{http://arxiv.org/abs/1610.01940}{{\tt arXiv:1610.01940}}].

\bibitem{Son:2002sd}
D.~T. Son and A.~O. Starinets, {\it {Minkowski space correlators in AdS / CFT
  correspondence: Recipe and applications}},  {\em JHEP} {\bf 09} (2002) 042,
  [\href{http://arxiv.org/abs/hep-th/0205051}{{\tt hep-th/0205051}}].

\bibitem{Herzog:2002pc}
C.~P. Herzog and D.~T. Son, {\it {Schwinger-Keldysh propagators from AdS/CFT
  correspondence}},  {\em JHEP} {\bf 03} (2003) 046,
  [\href{http://arxiv.org/abs/hep-th/0212072}{{\tt hep-th/0212072}}].

\bibitem{Barnes:2010jp}
E.~Barnes, D.~Vaman, C.~Wu, and P.~Arnold, {\it {Real-time finite-temperature
  correlators from AdS/CFT}},  {\em Phys. Rev.} {\bf D82} (2010) 025019,
  [\href{http://arxiv.org/abs/1004.1179}{{\tt arXiv:1004.1179}}].

\bibitem{Son:2009vu}
D.~T. Son and D.~Teaney, {\it {Thermal Noise and Stochastic Strings in
  AdS/CFT}},  {\em JHEP} {\bf 07} (2009) 021,
  [\href{http://arxiv.org/abs/0901.2338}{{\tt arXiv:0901.2338}}].

\bibitem{CaronHuot:2011dr}
S.~Caron-Huot, P.~M. Chesler, and D.~Teaney, {\it {Fluctuation, dissipation,
  and thermalization in non-equilibrium AdS$_5$ black hole geometries}},  {\em
  Phys. Rev.} {\bf D84} (2011) 026012,
  [\href{http://arxiv.org/abs/1102.1073}{{\tt arXiv:1102.1073}}].

\bibitem{Chesler:2011ds}
P.~M. Chesler and D.~Teaney, {\it {Dynamical Hawking Radiation and Holographic
  Thermalization}},  \href{http://arxiv.org/abs/1112.6196}{{\tt
  arXiv:1112.6196}}.

\bibitem{Botta-Cantcheff:2018brv}
M.~Botta-Cantcheff, P.~J. Martínez, and G.~A. Silva, {\it {The Gravity Dual of
  Real-Time CFT at Finite Temperature}},  {\em JHEP} {\bf 11} (2018) 129,
  [\href{http://arxiv.org/abs/1808.10306}{{\tt arXiv:1808.10306}}].

\bibitem{Botta-Cantcheff:2019apr}
M.~Botta-Cantcheff, P.~J. Martínez, and G.~A. Silva, {\it {Holographic excited
  states in AdS Black Holes}},  {\em JHEP} {\bf 04} (2019) 028,
  [\href{http://arxiv.org/abs/1901.00505}{{\tt arXiv:1901.00505}}].

\bibitem{Witten:1998qj}
E.~Witten, {\it {Anti-de Sitter space and holography}},  {\em Adv.\ Theor.\
  Math.\ Phys.} {\bf 2} (1998) 253--291,
  [\href{http://arxiv.org/abs/hep-th/9802150}{{\tt hep-th/9802150}}].

\bibitem{Skenderis:2008dh}
K.~Skenderis and B.~C. van Rees, {\it {Real-time gauge/gravity duality}},  {\em
  Phys. Rev. Lett.} {\bf 101} (2008) 081601,
  [\href{http://arxiv.org/abs/0805.0150}{{\tt arXiv:0805.0150}}].

\bibitem{Skenderis:2008dg}
K.~Skenderis and B.~C. van Rees, {\it {Real-time gauge/gravity duality:
  Prescription, Renormalization and Examples}},  {\em JHEP} {\bf 05} (2009)
  085, [\href{http://arxiv.org/abs/0812.2909}{{\tt arXiv:0812.2909}}].

\bibitem{vanRees:2009rw}
B.~C. van Rees, {\it {Real-time gauge/gravity duality and ingoing boundary
  conditions}},  {\em Nucl. Phys. Proc. Suppl.} {\bf 192-193} (2009) 193--196,
  [\href{http://arxiv.org/abs/0902.4010}{{\tt arXiv:0902.4010}}].

\bibitem{Dong:2016hjy}
X.~Dong, A.~Lewkowycz, and M.~Rangamani, {\it {Deriving covariant holographic
  entanglement}},  {\em JHEP} {\bf 11} (2016) 028,
  [\href{http://arxiv.org/abs/1607.07506}{{\tt arXiv:1607.07506}}].

\bibitem{Glorioso:2018mmw}
P.~Glorioso, M.~Crossley, and H.~Liu, {\it {A prescription for holographic
  Schwinger-Keldysh contour in non-equilibrium systems}},
  \href{http://arxiv.org/abs/1812.08785}{{\tt arXiv:1812.08785}}.

\bibitem{deBoer:2018qqm}
J.~de~Boer, M.~P. Heller, and N.~Pinzani-Fokeeva, {\it {Holographic
  Schwinger-Keldysh effective field theories}},  {\em JHEP} {\bf 05} (2019)
  188, [\href{http://arxiv.org/abs/1812.06093}{{\tt arXiv:1812.06093}}].

\bibitem{Chakrabarty:2019aeu}
B.~Chakrabarty, J.~Chakravarty, S.~Chaudhuri, C.~Jana, R.~Loganayagam, and
  A.~Sivakumar, {\it {Nonlinear Langevin dynamics via holography}},
  \href{http://arxiv.org/abs/1906.07762}{{\tt arXiv:1906.07762}}.

\bibitem{deBoer:2008gu}
J.~de~Boer, V.~E. Hubeny, M.~Rangamani, and M.~Shigemori, {\it {Brownian motion
  in AdS/CFT}},  {\em JHEP} {\bf 07} (2009) 094,
  [\href{http://arxiv.org/abs/0812.5112}{{\tt arXiv:0812.5112}}].

\bibitem{Chaudhuri:2018ihk}
S.~Chaudhuri and R.~Loganayagam, {\it {Probing Out-of-Time-Order Correlators}},
   {\em JHEP} {\bf 07} (2019) 006, [\href{http://arxiv.org/abs/1807.09731}{{\tt
  arXiv:1807.09731}}].

\bibitem{Chakrabarty:2018dov}
B.~Chakrabarty, S.~Chaudhuri, and R.~Loganayagam, {\it {Out of Time Ordered
  Quantum Dissipation}},  {\em JHEP} {\bf 07} (2019) 102,
  [\href{http://arxiv.org/abs/1811.01513}{{\tt arXiv:1811.01513}}].

\bibitem{Penington:2019npb}
G.~Penington, {\it {Entanglement Wedge Reconstruction and the Information
  Paradox}},  \href{http://arxiv.org/abs/1905.08255}{{\tt arXiv:1905.08255}}.

\bibitem{Almheiri:2019psf}
A.~Almheiri, N.~Engelhardt, D.~Marolf, and H.~Maxfield, {\it {The entropy of
  bulk quantum fields and the entanglement wedge of an evaporating black
  hole}},  {\em JHEP} {\bf 12} (2019) 063,
  [\href{http://arxiv.org/abs/1905.08762}{{\tt arXiv:1905.08762}}].

\bibitem{Rocha:2008fe}
J.~V. Rocha, {\it {Evaporation of large black holes in AdS: Coupling to the
  evaporon}},  {\em JHEP} {\bf 08} (2008) 075,
  [\href{http://arxiv.org/abs/0804.0055}{{\tt arXiv:0804.0055}}].

\bibitem{Penington:2019kki}
G.~Penington, S.~H. Shenker, D.~Stanford, and Z.~Yang, {\it {Replica wormholes
  and the black hole interior}},  \href{http://arxiv.org/abs/1911.11977}{{\tt
  arXiv:1911.11977}}.

\bibitem{Almheiri:2019qdq}
A.~Almheiri, T.~Hartman, J.~Maldacena, E.~Shaghoulian, and A.~Tajdini, {\it
  {Replica Wormholes and the Entropy of Hawking Radiation}},
  \href{http://arxiv.org/abs/1911.12333}{{\tt arXiv:1911.12333}}.

\bibitem{Chou:1984es}
K.-c. Chou, Z.-b. Su, B.-l. Hao, and L.~Yu, {\it {Equilibrium and
  Nonequilibrium Formalisms Made Unified}},  {\em Phys. Rept.} {\bf 118} (1985)
  1--131.

\bibitem{Haehl:2017eob}
F.~M. Haehl, R.~Loganayagam, P.~Narayan, A.~A. Nizami, and M.~Rangamani, {\it
  {Thermal out-of-time-order correlators, KMS relations, and spectral
  functions}},  {\em JHEP} {\bf 12} (2017) 154,
  [\href{http://arxiv.org/abs/1706.08956}{{\tt arXiv:1706.08956}}].

\bibitem{Chaudhuri:2018ymp}
S.~Chaudhuri, C.~Chowdhury, and R.~Loganayagam, {\it {Spectral Representation
  of Thermal OTO Correlators}},  {\em JHEP} {\bf 02} (2019) 018,
  [\href{http://arxiv.org/abs/1810.03118}{{\tt arXiv:1810.03118}}].

\bibitem{Chakrabarty:2019qcp}
B.~Chakrabarty and S.~Chaudhuri, {\it {Out of time ordered effective dynamics
  of a quartic oscillator}},  {\em SciPost Phys.} {\bf 7} (2019) 013,
  [\href{http://arxiv.org/abs/1905.08307}{{\tt arXiv:1905.08307}}].

\bibitem{Martin:1973zz}
P.~Martin, E.~Siggia, and H.~Rose, {\it Statistical dynamics of classical
  systems},  {\em Phys.Rev.A} {\bf 8} (1973) 423--437.

\bibitem{Gubser:1997cm}
S.~S. Gubser, {\it {Absorption of photons and fermions by black holes in
  four-dimensions}},  {\em Phys. Rev.} {\bf D56} (1997) 7854--7868,
  [\href{http://arxiv.org/abs/hep-th/9706100}{{\tt hep-th/9706100}}].

\bibitem{Birmingham:2001pj}
D.~Birmingham, I.~Sachs, and S.~N. Solodukhin, {\it Conformal field theory
  interpretation of black hole quasinormal modes},  {\em Phys.Rev.Lett.} {\bf
  88} (2002) 151301, [\href{http://arxiv.org/abs/hep-th/0112055}{{\tt
  hep-th/0112055}}].

\bibitem{Bzowski:2015pba}
A.~Bzowski, P.~McFadden, and K.~Skenderis, {\it {Scalar 3-point functions in
  CFT: renormalisation, beta functions and anomalies}},  {\em JHEP} {\bf 03}
  (2016) 066, [\href{http://arxiv.org/abs/1510.08442}{{\tt arXiv:1510.08442}}].

\bibitem{Becker:2014jla}
M.~Becker, Y.~Cabrera, and N.~Su, {\it {Finite-temperature three-point function
  in 2D CFT}},  {\em JHEP} {\bf 09} (2014) 157,
  [\href{http://arxiv.org/abs/1407.3415}{{\tt arXiv:1407.3415}}].

\bibitem{Leahy:1983vb}
D.~Leahy and W.~Unruh, {\it {Effects of a lambda phi**4 interaction on black
  hole evaporation in two-dimensions}},  {\em Phys. Rev. D} {\bf 28} (1983)
  694--702.

\bibitem{Loganayagam:2020aa}
R.~Loganayagam, K.~Ray, and A.~Sivakumar, {\it {Fermionic Open EFT from
  Holography}},  {\em to appear} (2020).

\bibitem{Hubeny:2011hd}
V.~E. Hubeny, S.~Minwalla, and M.~Rangamani, {\it {The fluid/gravity
  correspondence}},  in {\em {Black holes in higher dimensions}}, pp.~348--383,
  2012.
\newblock \href{http://arxiv.org/abs/1107.5780}{{\tt arXiv:1107.5780}}.
\newblock [,817(2011)].

\bibitem{Heemskerk:2010hk}
I.~Heemskerk and J.~Polchinski, {\it {Holographic and Wilsonian Renormalization
  Groups}},  {\em JHEP} {\bf 06} (2011) 031,
  [\href{http://arxiv.org/abs/1010.1264}{{\tt arXiv:1010.1264}}].

\bibitem{Faulkner:2010jy}
T.~Faulkner, H.~Liu, and M.~Rangamani, {\it {Integrating out geometry:
  Holographic Wilsonian RG and the membrane paradigm}},  {\em JHEP} {\bf 08}
  (2011) 051, [\href{http://arxiv.org/abs/1010.4036}{{\tt arXiv:1010.4036}}].

\bibitem{Chatterjee:2020aa}
S.~Chatterjee, C.~Jana, R.~Loganayagam, and A.~Rudra, {\it {Renormalisation in
  open quantum field theory III: Non-local divergences}},  {\em to appear}
  (2020).

\bibitem{NIST:DLMF}
``{\it NIST Digital Library of Mathematical Functions}.''
  http://dlmf.nist.gov/, Release 1.0.26 of 2020-03-15.
\newblock F.~W.~J. Olver, A.~B. {Olde Daalhuis}, D.~W. Lozier, B.~I. Schneider,
  R.~F. Boisvert, C.~W. Clark, B.~R. Miller, B.~V. Saunders, H.~S. Cohl, and
  M.~A. McClain, eds.

\end{thebibliography}
